\documentclass[print, amsmath,amssymb, aps,prc,nofootinbib, twocolumn]{revtex4-2}

\usepackage[usenames,dvipsnames]{xcolor}
\usepackage{graphicx}
\usepackage{dcolumn}
\usepackage{bm}
\usepackage{hyperref}
\usepackage[utf8]{inputenc} 
\usepackage{lipsum}
\usepackage{babel}

\allowdisplaybreaks

\usepackage{multirow}

\newcommand{\nn}{\nonumber}
\renewcommand{\vec}[1]{{\bf #1}}

\newcommand{\rmd}{{\rm d}}

\newcommand{\iunit}{{\rm i}}

\newcommand{\rvec}{\vec{r}}
\newcommand{\rvecp}{\vec{r}{\, }'}

\newcommand{\etal}{\emph{et al.}}

\newcommand{\nuc}[2]{$^{#1}$\textrm{#2}}
\newcommand{\vnabla}{\boldsymbol{\mathbf\nabla}}
\newcommand{\vsigma}{\boldsymbol{\mathbf\sigma}}

\renewcommand{\Re}{\text{Re}}
\renewcommand{\Im}{\text{Im}}

\usepackage{dsfont}

\definecolor{mbscolor}{rgb}{0.60, 0.0, 0.65}

\definecolor{mbscolorp}{rgb}{0.0, 0.60, 0.65}

\definecolor{mbscolorr}{rgb}{0.6, 0.20, 0.2}

\newcommand{\emb}{{\rule{0cm}{0cm}}}

\newcommand{\cc}[1]{\mathnormal{#1}}

\begin{document}

\title{Skyrme pseudopotentials at next-to-next-to-leading order\\ \normalsize
       Construction of local densities and first symmetry-breaking calculations.}
       
\author{Wouter Ryssens}
\email{wryssens@ulb.ac.be}
\affiliation{Institut d'Astronomie et d'Astrophysique, 
             Universit\'e Libre de Bruxelles, Campus de la Plaine CP 226, BE-1050 Brussels, Belgium}
\affiliation{Center for Theoretical Physics, 
             Sloane Physics Laboratory,
             Yale University, New Haven, USA}
\affiliation{Universit{\'e} de Lyon, 
             Universit{\'e} Claude Bernard Lyon 1, \\
             CNRS, IP2I Lyon / IN2P3,
             UMR 5822, F-69622, Villeurbanne, France}

\author{Michael Bender}
\email{bender@ip2i.in2p3.fr}
\affiliation{Universit{\'e} de Lyon, 
             Universit{\'e} Claude Bernard Lyon 1, \\
             CNRS, IP2I Lyon / IN2P3,
             UMR 5822, F-69622, Villeurbanne, France}
               
\begin{abstract}
There is an ongoing quest to improve on the spectroscopic quality of nuclear energy density 
functionals (EDFs) of the Skyrme type through extensions of its traditional form. One direction 
for such activities is the inclusion of terms of higher order in gradients in the EDF. We
report on exploratory symmetry-breaking calculations performed for an extension of the 
Skyrme EDF that includes central terms with four gradients at next-to-next-to-leading 
order (N2LO) and for which the high-quality parametrization SN2LO1 has been constructed recently
[P.~Becker \textit{et al}, Phys.~Rev.~C 96, 044330 (2017)]. Up to now, the 
investigation of such functionals with higher-order terms was limited to infinite matter
and spherically symmetric configurations of singly- and doubly-magic nuclei. We address
here nuclei and phenomena that require us to consider axial and non-axial deformation, 
both for reflection-symmetric and also reflection-asymmetric shapes, as well as the breaking 
of time-reversal invariance. Achieving these calculations demanded a number of
formal developments. These all resulted from the formulation of the N2LO EDF 
requiring the introduction of new local densities with additional gradients that
are not present in the EDF at NLO. Their choice is not unique, but can differ in the way the gradients are
coupled. While designing a numerical implementation of 
N2LO EDFs in Cartesian 3d coordinate-space representation, we have developed a novel 
definition and a new unifying notation for normal and pair densities that contain 
gradients at arbitrary order. Besides having mnemonic advantages, the new notation allows 
for the easy identification of redundancies and reducibilities in a given set of local 
densities, and the new definition makes it straightforward to construct densities that 
automatically adopt the symmetries of the many-body state they are constructed from. 
The resulting scheme resolves several issues with some of the choices that have been made 
for local densities in the past, in particular when breaking time-reversal symmetry.
Guided by general practical considerations, we propose an alternative form of the N2LO 
contribution to the Skyrme EDF that is built from a different set of densities. It 
has exactly the same physics content, but is much more efficient to handle in formal 
discussions and, compared to the original formulation, leads to a substantial 
reduction of computational cost and memory requirements in deformed codes.
As representative examples for the performance of SN2LO1, we have chosen the ground states 
of even-even Kr and Nd isotopes, the fission barrier of $^{240}$Pu, as well as the 
superdeformed rotational band of $^{194}$Hg.
Overall, for the nuclei and phenomena studied here, the SN2LO1 parametrization does not 
yet present a systematic improvement over standard NLO parametrizations. This finding calls 
for improved fit protocols that better discriminate between NLO and N2LO terms and better
exploit the unique features of the additional degrees of freedom offered by the latter.
\end{abstract}     

\date{30 June 2021}

\maketitle

\section{Introduction}

Methods employing nuclear energy density functionals (EDFs) have found 
wide-spread use in the nuclear physics community, providing the means 
to study both nuclear ground and excited states across the entire nuclear 
chart at a manageable numerical cost~\cite{Bender03}. Functionals based on 
the non-relativistic zero-range Skyrme pseudopotential~\cite{Bender03,Schunck2019} 
are arguably the most widely used, and the description of 
bulk properties, such as the overall trends of nuclear binding energies and 
charge radii across the nuclear chart, can be pushed very far~\cite{Goriely13}.

Despite this, there are inherent limitations to the standard form of the Skyrme 
functional. While generally on par with other types of functionals, 
the spectroscopic properties of Skyrme EDFs still leave much to be 
desired: examples include the description of the ground states of odd-mass 
nuclei~\cite{Bonneau07} and the reproduction of shell gaps in the superheavy 
region~\cite{Dobaczewski15}. A series of systematic studies
\cite{Kortelainen10,Kortelainen12,Kortelainen14} led the UNEDF-Scidac
collaboration to the conclusion that the 
spectroscopic qualities of the traditional Skyrme functional can no longer 
be improved in systematic fashion by changing the optimization procedure. 
For further advancements, one has to look at extensions
of its traditional form instead.

Several possible avenues have been proposed in the literature: the addition of 
tensor terms to the Skyrme pseudopotential has been systematically explored
in Refs.~\cite{Lesinski07,Bender09,Hellemans12,Colo07,Zalewski08}.
Another direction is the introduction of additional density dependencies for the 
coupling constants, both for terms without~\cite{Lesinski06,Erler2010} and with 
gradients~\cite{Krewald1977,Pearson1994,Chamel09}.
We also mention the construction of Skyrme functionals that are 
strictly generated by a density-independent pseudopotential including 
three- and four-body terms with gradients that aim at
spuriosity-free multi-reference calculations~\cite{Sadoudi13a,Sadoudi13b}. 

The approach we will explore here is based on original work by the Jyv{\"a}skyl{\"a}
group~\cite{Carlsson08,Carlsson10a,Carlsson10b,Raimondi11a}. 
They proposed to revisit the possibility of 
adding terms with increasing number of gradients to the Skyrme EDF, 
some of which were in fact already considered in the original 
papers by Skyrme~\cite{Skyrme56,Bell56,Skyrme59a,Skyrme59b}.
This type of extension can be made in a systematic fashion by studying all 
possible terms that can be added up to a given number of gradients: for a 
maximum of $2\ell$ gradients, the resulting EDF is labelled N$\ell$LO. The 
traditional Skyrme pseudopotential, which includes terms up to two gradients, 
is labelled NLO, while Refs.~\cite{Carlsson08,Carlsson10a,Carlsson10b,Raimondi11a} 
explored the possibilities of N3LO EDFs, including up to six gradients.  
Carlsson \emph{et al} \cite{Carlsson08} formulated the most general 
bilinear form of such EDF that is compatible with the usual symmetry requirements, 
whereas Raimondi \emph{et al} constructed the most general contact two-body 
pseudopotential with up to six gradients~\cite{Raimondi11a}. Both approaches 
yield an EDF of the same form, but the one obtained from the latter has a 
smaller number of independent coupling constants, see Appendix~\ref{app:coupling}. 
All of these developments were formulated in terms of gradients coupled to 
spherical tensors, which is well suited for a numerical implementation 
in spherical symmetry as the one documented in Ref.~\cite{Carlsson10a}.
Using this code and a new strategy for density matrix expansion in terms 
of gradients instead of density dependencies, the authors of Ref.~\cite{Carlsson10b} 
 demonstrated that the physics of finite-range interactions, including that of
nonlocal exchange potentials, can be successfully mapped onto a Skyrme 
EDF in a systematically improvable manner when increasing the order in 
gradients. No other parameterization of a higher-order Skyrme EDF has
been published by this group, though.

The further development of N$\ell$LO EDFs was taken up by a collaboration of researchers
from Lyon, Valencia, and 
York~\cite{Davesne13,Davesne14,Davesne15,Davesne15b,Davesne16,Becker15,BeckerPhD,Becker17,Becker19}, 
who opted for a formulation in terms of Cartesian tensors as traditionally 
employed for the standard NLO EDF, and who limited themselves so far to the 
subset of pseudopotential-generated terms that are invariant under local gauge 
transformations. 
These terms are the ones that contribute to the equation of 
   state of homogeneous infinite nuclear matter, and their influence on the 
   latter was 
   studied in detail in 
Refs.~\cite{Davesne13,Davesne14,Davesne15,Davesne15b,Davesne16,Becker15}. 
Again, it was found that the properties of nuclear matter obtained with various other 
approaches and interactions can be successfully mapped onto a Skyrme EDF when 
increasing the order in gradients. Adjusting parameters to infinite matter, 
however, is not sufficient to obtain a predictive parametrization for 
calculations of finite nuclei.

In the next step, this group constructed a spherical mean-field code that
can handle higher-order terms in the Skyrme EDF \cite{BeckerPhD,Becker17}, and
which has been subsequently used for the adjustment of 
the parameters of a Skyrme EDF augmented by central N2LO terms to the properties 
of finite (magic) nuclei~\cite{BeckerPhD,Becker17,Becker19}. Named SN2LO1, this 
parameterization was adjusted with a protocol that is very similar to 
the one used for the adjustment of the SLy5~\cite{ChabanatA,ChabanatB} and 
SLy5* \cite{Pastore13} parametrizations of the NLO EDF. Like SLy5*, SN2LO1 
has been fitted with additional stability constraints imposed via linear-response 
calculations~\cite{Becker15,BeckerPhD,Becker17}. For masses and radii of spherical nuclei, the 
overall performance of these parametrizations is comparable. There is, however,
the possibility that other observables and other nuclei that are usually not 
considered in a fit protocol might be more sensitive to the new terms, and that 
such information can be used to adjust N2LO EDFs that systematically outperform 
standard functionals.

For these reasons, there is an interest to extend calculations with the N2LO EDF
to non-spherical nuclei. On general grounds, it is important to have a possibility to 
calculate observables of more complex nuclei that can inform future fits, and more
specifically it will be instructive to benchmark the performance of
SN2LO1 for deformed and rotating nuclei. For this reason, we study here 
for the first time the properties of deformed nuclei using an N2LO EDF. To this 
end, the N2LO EDF has been implemented into a 3d code in coordinate-space
representation~\cite{RyssensPhD,MOCCa}, that allows for the exploration of 
numerous shape degrees of freedom including non-axial and non-reflection-symmetric
ones, and finite angular momentum. As representative examples, we will
discuss deformation properties of even-even krypton and neodymium isotopes,
the fission barrier of $^{240}$Pu, and the superdeformed rotational band of 
$^{194}$Hg. The latter also serves as a verification that 
time-reversal breaking calculations using the SN2LO1 parametrization are 
stable with respect to unphysical finite-size instabilities that plague many older
Skyrme functionals at NLO~\cite{Hellemans12,Pastore15}.

Designing the implementation of the additional terms into our numerical codes
prompted us to examine in some detail the formulation of the N2LO terms in the Skyrme EDF, 
and in particular the possible definitions of the local densities used in its construction.
There are seven real normal densities and the same number of complex pair 
densities needed to define the most general Skyrme EDF at NLO~\cite{Perlinska04},
which are all well established in the literature. Going to higher-order EDFs
requires one to introduce additional local densities at each 
order~\cite{Carlsson08,Becker15}. There are several possible choices for their
definition that are related to one another through the recoupling of gradient 
operators. These choices are not without consequences. On a formal and practical level, 
it is highly desirable to be able to split the EDF into parts built out of time-even and 
time-odd densities as it is the case for the traditional choice of densities 
entering the Skyrme EDF at NLO. Also, the NLO functional is traditionally formulated
through local densities that also enter the expectation values of frequently-used one-body 
operators. For the densities that are specific to
the EDF at N2LO and higher orders there is no such guidance anymore, and some
of the definitions used in the literature lead to tensor densities whose components do
not all transform in the same way under time-reversal. It is also not automatically
guaranteed that the higher-order densities are linearly independent from one
another. On a practical level, certain choices imply a larger computational burden 
than others. While designing an almost symmetry-unrestricted 3d numerical implementation 
of N2LO EDFs in coordinate-space representation, we have developed 
a unifying notation for mean-field densities that allows for  a much clearer 
discussion of the relevant degrees of freedom and makes it easier to identify 
redundancies in the set of local densities. Guided by general
practical considerations, we propose an alternative version of the form of the 
N2LO EDF of Ref.~\cite{Becker17} that is equivalent, but much more 
efficient to employ both in formal discussions and numerical implementations.

This paper is organized as follows: in Section~\ref{sec:EDF} we present formal 
aspects of the past strategy to
formulate the Skyrme EDF of Ref.~\cite{Becker17} at N2LO just mentioned and point 
to some of its formal and computational inefficiencies, all of which can be 
expected to aggravate when setting up further extensions of the energy functional 
in the future. In Section~\ref{sec:definition} we then propose a novel scheme 
for a universal and systematic notation of local normal and pair
densities and currents at an arbitrary order of gradients, and formulate 
a strategy to define local densities in a systematic manner that allows 
for easy identification of numerically efficient choices for a set of 
non-redundant and irreducible densities and the easy and transparent 
transformation between different such choices.
In Sec.~\ref{sec:N2LO:revisited} we then apply the new definitions and 
notations to reformulate the N2LO Skyrme EDF of Ref.~\cite{Becker17}.
Finally, in Sec.~\ref{sec:calculations} we report on exploratory calculations 
for deformed and rotating nuclei with SN2LO1 and compare with results obtained 
with similarly fitted parametrizations of the traditional NLO EDF.
Section~\ref{sec:conclusion} summarizes the main results of the paper, and
appendices provide further technical and details concerning the properties
of local densities and our implementation of the Skyrme EDF  
as needed for SN2LO1.


\section{The N2LO energy density functional}
\label{sec:EDF}


\subsection{The form of the functional}
\label{sec:formfunc}

The general form of the functional discussed here consists of five terms 
\cite{Bender03},
\begin{equation}
\label{eq:Etot}
E_{\text{tot}}
  =     E_{\text{kin}}
      + E_{\text{Sk}}
      + E_{\text{Coul}}
      + E_{\text{cm}}
      + E_{\text{pair}}
\, ,
\end{equation}
which correspond to the kinetic energy, the Skyrme EDF modeling of 
the strong interaction between the nucleons in the particle-hole channel, 
the Coulomb energy resulting from the electromagnetic repulsion between 
protons, a center-of-mass correction and a pairing term, respectively. In what 
follows, we will mainly address the second term, the Skyrme part, and
we refer the reader to Ref.~\cite{Ryssens15a} for details on the treatment of 
the other terms. 

Following the developments reported in Refs.~\cite{Becker15,BeckerPhD,Becker17}, 
we consider here the Skyrme  EDF as generated by a density-dependent, effective
two-body pseudopotential $\hat{V}_{\rm Sk}$ that consists of a central (C), 
spin-orbit (SO) and a (central) density-dependent (DD) part
\begin{equation}
\hat{V}_{\rm Sk} 
=  \hat{V}^{\rm C}_{\rm LO} 
  +\hat{V}^{\rm C}_{\rm NLO} 
  +\hat{V}^{\rm C}_{\rm N2LO} 
  + \hat{V}^{\rm SO}_{\rm NLO} 
  + \hat{V}^{\rm DD}_{\rm LO} \, .
\label{eq:def:Vsk}
\end{equation}
Like these earlier references, we use a notation where the subscript refers to 
the number of gradients in the generators. Leading-order (LO) terms do not 
contain gradients, next-to-leading-order (NLO) terms contain two gradients, and 
next-to-next-to-leading order (N2LO) terms contain four gradients, and so on. 
The order of these terms refers to their degree of computational complexity and 
numerical cost, but does by no means represent a hierarchy of increasingly 
refined approximations to the nucleon-nucleon interaction such as the orders of 
an effective field theory would do.

The widely-used standard form of the Skyrme EDF combines terms up to 
NLO for central and spin-orbit terms, sometimes augmented also by a tensor 
interaction up to NLO \cite{Lesinski07,Bender09,Hellemans12}, with a 
density-dependent (DD) term at LO. We consider here only its extension to 
N2LO for the central part of the pseudopotential, keeping the standard form
for all other terms, including the other contributions to Eq.~\eqref{eq:Etot}.
Only for this particular extension do high-quality fits 
exist~\cite{Becker17,Becker19}. 
As done in those references, we adopt only the N2LO 
terms that are invariant under local gauge transformations. We refer the 
reader to Ref.~\cite{Becker17} for a detailed discussion of the form of the 
individual terms in Eq.~\eqref{eq:def:Vsk}.

The Skyrme part of the functional generated by the pseudopotential of
Eq.~\eqref{eq:def:Vsk} takes the generic form
\begin{align}
E_{\text{Sk}}^{\rm N2LO}
& =  \int \! d^3 r \,
       \Big[   \mathcal{E}^{(0)}(\vec{r}) 
             + \mathcal{E}^{(2)}(\vec{r})  
             + \mathcal{E}^{(4)}(\vec{r}) \Big] \, , \label{eq:def:N2LOfunc} \\
\mathcal{E}^{(i)}(\vec{r}) 
&=       \sum_{t=0,1}
\Big[  \mathcal{E}_{t,\textrm{e}}^{(i)}(\vec{r})                           
     + \mathcal{E}_{t,\textrm{o}}^{(i)}(\vec{r}) \Big] \quad \text{for $i=0$, 2, 4,} 
\label{eq:def:time}
\end{align}
where $t=0$, 1 labels terms composed of isoscalar and isovector densities,
respectively. Except for the contribution of the density-dependent
interaction, all terms in Eq.~\eqref{eq:def:N2LOfunc} are bilinear in some local
normal one-body densities calculated from an auxiliary many-body state 
that is either a Slater determinant or a Bogoliubov quasiparticle vacuum.
The superscripts (0), (2), and (4) indicate the order of the terms in terms of 
derivatives, whereas the subscripts e and o indicate 
whether the terms are constructed out of time-even or time-odd local 
densities. 

For complete expressions of the LO and NLO energy densities, we refer the 
reader to Ref.~\cite{Ryssens19b}. Leaving the definition of the local
densities to the next section, we specify here the full N2LO
energy density in the notation of Ref.~\cite{Becker17}
\begin{widetext}
\begin{align}
\mathcal{E}_t^{(4)} (\vec{r})
=    \phantom{+} 
   C^{(4)\Delta\rho}_t & \big[ \Delta \rho_t (\vec{r}) \big]^2  
 + C^{(4)\Delta s}_t \big[ \Delta \vec{s}_t (\vec{r}) \big]^2  
  \nonumber \\
   + C^{(4)M\rho}_t & \bigg\{ 
     \rho_t(\vec{r}) \, Q_t(\vec{r})  
 +   \tau_t^2(\vec{r})
 +  2 \sum_{\mu \nu} \tau_{t, \mu \nu}(\vec{r}) \, \tau_{t,\mu \nu} (\vec{r}) 
  - 2 \sum_{\mu \nu} \tau_{t, \mu \nu}(\vec{r}) \, \big[ \nabla_{\mu} \nabla_{\nu} \rho_t(\vec{r}) \big] 
\nonumber \\
&  
 - \big[  \vnabla \cdot \vec{j}_t(\vec{r}) \big]^2
 - 4 \, \vec{j}_t(\vec{r}) \cdot \vec{\Pi}_t(\vec{r})
   \bigg\}
\nonumber \\
%
 + C^{(4)M\vec{s}}_t & \bigg\{
     \vec{s}_t(\vec{r}) \cdot \vec{S}_t(\vec{r})
 +   \vec{T}^2_t(\vec{r})  
 + 2 \sum_{\mu \nu \kappa} K_{t,\mu \nu \kappa}(\vec{r}) \, K_{t,\mu \nu \kappa} (\vec{r})
 - 2 \sum_{\mu \nu \kappa} K_{t,\mu \nu \kappa}(\vec{r}) \, \big[ \nabla_{\mu} \nabla_{\nu} s_{t,\kappa}(\vec{r}) \big] 
   \nonumber \\
& 
 - \sum_{\nu}
   \bigg[ \sum_{\mu} \nabla_{\mu} J_{t,\mu \nu}(\vec{r})  \bigg] 
   \bigg[ \sum_{\kappa} \nabla_{\kappa} J_{t,\kappa \nu}(\vec{r}) \bigg]
   - 4 \sum_{\mu \nu} J_{t,\mu \nu}(\vec{r}) \, V_{t, \mu \nu}  (\vec{r})
   \bigg\}  \, ,
\label{eq:originalN2LO}
\end{align}
\end{widetext}
where the four $C^{(4)}_t$ are coupling constants.
As per usual, densities that are Cartesian tensors of rank~1 are written as
vectors in boldface, and their contractions through inner and outer vector products. For
higher-rank tensors, contractions are explicitly written out as summation over cartesian
components indicated by greek indices.
Besides the well-known densities that the standard NLO Skyrme EDF is constructed from,
the energy density at N2LO additionally depends on the densities 
$Q_t(\vec{r})$,
$\tau_{t,\mu \nu} (\vec{r})$,
$\vec{\Pi}_t(\vec{r})$,
$\vec{S}_t(\vec{r})$,
$K_{t,\mu \nu \kappa} (\vec{r})$, and
$V_{t, \mu \nu}  (\vec{r})$ that will be defined in the next subsection.
For reasons explained in Sec.~\ref{sec:problems}, we do not attempt 
to separate the time-even and time-odd parts of the energy density~\eqref{eq:originalN2LO}
but postpone this to its alternative form given in Sec.~\ref{sec:rewritten}.


\subsection{Traditional representation of densities}
\label{sec:densities}


\subsubsection{The one-body density matrix}
\label{sec:MFdensities:onebodydm}

Depending on the treatment or not of pairing correlations, we will be dealing
with either a single Slater determinant or a single Bogoliubov quasi-particle 
vacuum. We represent the auxiliary state $|\Phi \rangle$ from which the 
densities entering the EDF are constructed using a (for now)
unspecified basis of orthonormal single-particle wave functions $\Psi_k(\vec{r})$
\begin{align} 
\Psi_k(\vec{r}) 
= \, & \sum_{\sigma = \pm} \sum_{q = p,n} \psi_k(\vec{r} , \sigma, q) \, 
       \chi_{\sigma}\,  \xi_{q} 
       \nn \\
= \, & \sum_{\sigma = \pm} \sum_{q = p,n} 
       \langle \vec{r}, \sigma, q | \Psi_k \rangle \, 
       \chi_{\sigma}\,  \xi_{q} 
       \nn \\
= \, & \sum_{\sigma = \pm} \sum_{q = p,n} 
       \langle - | \hat{a}_{\vec{r} \sigma q} \, \hat{a}^\dagger_k | - \rangle \, 
       \chi_{\sigma}\,  \xi_{q} 
\end{align}
which are two-component spinors in both spin and isospin space spanned by the 
unit vectors $\chi_{\sigma}$ and $\xi_{q}$, respectively. Throughout this paper, 
we will assume that protons and neutrons are not mixed on the level of 
single-particle states. This implies that the auxiliary many-body state $|\Phi \rangle$
is the direct product of separate product states for protons and neutrons, 
respectively, $|\Phi \rangle = |\Phi_p \rangle \otimes |\Phi_n \rangle$.
In this case, the full one-body normal and anomalous density matrices for the 
nucleon species $q=n$, $p$ can both be split into a scalar and a vector in 
spin space~\cite{Dobaczewski84,Doba00}. 

Contact pairing interactions as usually used in the context of the Skyrme EDF
naturally lead to a local pairing energy density that can be formulated in terms 
of local pair densities. Because of the anticommutation of fermionic annihilation 
operators, the anomalous density matrix $\kappa_q(\vec{r}\sigma,\vec{r}'\sigma')
= \langle \Phi | \hat{a}_{\vec{r}'\sigma' q} \hat{a}_{\vec{r}\sigma q }| \Phi \rangle$
is skew-symmetric under coordinate exchange. For this reason, one cannot 
construct \emph{local} pair densities out of it in the same manner as is 
done for normal densities. This issue is resolved by using the so-called ``Russian'' 
representation of pair density matrices 
$\tilde{\rho}_q(\vec{r}\sigma,\vec{r}'\sigma')$ 
instead of anomalous density 
matrices $\kappa_q(\vec{r}\sigma,\vec{r}'\sigma')$ \cite{Dobaczewski84,Dobaczewski96,Doba00,RotivalPhD}. 
The full normal and pair one-body density matrix in position and spin space are 
given by
\begin{align}
\label{eq:def:nonlocaldensity}
\rho_q (\vec{r}\sigma, \vec{r}' \sigma')
& \equiv \langle \Phi| \hat{a}^{\dagger}_{\vec{r}' \sigma' q} 
                 \hat{a}_{\vec{r} \sigma q}| \Phi \rangle 
\nonumber \\
&= \sum_{jk} \rho_{kj} \, \psi^*_j(\vec{r}',\sigma') \, \psi_k(\vec{r},\sigma)
 \nonumber \\
&= \tfrac{1}{2} \, \rho_q(\vec{r}, \vec{r}') \, \delta_{\sigma \sigma'} +
  \tfrac{1}{2} \, \vec{s}_q(\vec{r}, \vec{r}') 
  \cdot \langle \sigma |\hat{\boldsymbol{\sigma}} | \sigma' \rangle \, ,  
\end{align}
\begin{align}
\label{eq:def:nonlocalpair}
\tilde{\rho}_q (\vec{r}\sigma, \vec{r}' \sigma')
& \equiv -\sigma' \langle \Phi| \hat{a}_{\vec{r}' -\sigma' q} 
                         \hat{a}_{\vec{r}  \sigma  q} | \Phi \rangle 
\nonumber \\
&= -\sigma' \sum_{jk} \kappa_{kj} \, \psi_j(\vec{r}',-\sigma') \, \psi_k(\vec{r},\sigma)
 \nonumber \\
&= \tfrac{1}{2} \, \tilde{\rho}_q(\vec{r}, \vec{r}') \, \delta_{\sigma \sigma'} +
  \tfrac{1}{2} \, \tilde{\vec{s}}_q(\vec{r}, \vec{r}')
  \cdot  \langle \sigma |\hat{\boldsymbol{\sigma}} | \sigma' \rangle \, ,
\end{align}
where $\hat{\boldsymbol{\sigma}}$ is the Cartesian
vector of Pauli spin matrices. The sums over $j$ and $k$ run over 
the single-particle states of the nucleon species $q$.
The $\rho_{k j}$ and $\kappa_{k j}$ are the elements of the normal
and anomalous density matrices in the single-particle basis 
spanned by the $\Psi_j(\vec{r})$
\begin{align}
\label{eq:rhojk:def}
\rho_{k j}
& \equiv \langle \Phi | \, \hat{a}^\dagger_j \hat{a}_k \, | \Phi \rangle 
    \\
\label{eq:kappajk:def}
\kappa_{k j}
& \equiv \langle \Phi | \, \hat{a}_j \hat{a}_k \,  | \Phi \rangle
\end{align}
The non-local density $\rho_q(\vec{r}, \vec{r}')$, 
the non-local spin density $\vec{s}_q(\vec{r}, \vec{r}')$, the non-local pair 
density $\tilde{\rho}_q(\vec{r},\vec{r}')$ and the non-local spin pair density 
$\tilde{\vec{s}}_q(\vec{r},\vec{r}')$ are defined as
\begin{align}
\label{eq:def:rho:nonlocal_1}
\rho_q(\vec{r}, \vec{r}') 
& \equiv \sum_{\sigma} \rho_q(\vec{r}\sigma, \vec{r}'\sigma) \, , \\
\label{eq:def:s:nonlocal}
s_{q,\kappa}(\vec{r}, \vec{r}') 
& \equiv  \sum_{\sigma \sigma'} \rho(\vec{r}\sigma, \vec{r}'\sigma') \,
          \langle \sigma' | \hat{\sigma}_{\kappa} | \sigma \rangle \, , \\
\label{eq:def:rhot:nonlocal}
\tilde{\rho}_q(\vec{r}, \vec{r}') 
& \equiv \sum_{\sigma} \tilde{\rho}_q(\vec{r}\sigma, \vec{r}'\sigma) \, , \\
\label{eq:def:st:nonlocal}
\tilde{s}_{q, \kappa}(\vec{r}, \vec{r}')  
& \equiv \sum_{\sigma \sigma'} \tilde{\rho}_q(\vec{r}\sigma, \vec{r}'\sigma') \,
              \langle \sigma' | \hat{\sigma}_{\kappa} | \sigma \rangle  \, . 
\end{align}
These non-local densities behave as follows under the exchange of 
$\vec{r} \leftrightarrow \vec{r}'$:
\begin{align}
\rho_q(\vec{r}, \vec{r}')            &= +\rho_q^*(\vec{r}', \vec{r})    \, , && 
\vec{s}_q(\vec{r}, \vec{r}')         &= +\vec{s}_q^*(\vec{r}', \vec{r}) \, , 
\label{eq:nonlocalphsym} \\ 
\tilde{\rho}_q(\vec{r}, \vec{r}')    &= +\tilde{\rho}_q(\vec{r}', \vec{r})  \, , && 
\tilde{\vec{s}}_q(\vec{r}, \vec{r}') &= -\tilde{\vec{s}}_q(\vec{r}',\vec{r})\, .
\label{eq:nonlocalppsym}
\end{align}
The density matrices associated with the time-reversed 
auxiliary many-body state $\check{T} |\Phi \rangle$ are related
to those of the original state $|\Phi \rangle$ through~\cite{Perlinska04}
\begin{align}
\rho_q^T(\vec{r} \sigma, \vec{r}' \sigma') 
        &=   \sigma \sigma' \rho_q^*(\vec{r}-\sigma, \vec{r}'-\sigma') 
\label{eq:Trhononloc} \, ,\\
\tilde{\rho}_q^T(\vec{r} \sigma, \vec{r}' \sigma')
        &=   \sigma \sigma' \tilde{\rho}_q^*(\vec{r}-\sigma, \vec{r}'-\sigma') 
\label{eq:Trhotildenonloc} \, .
\end{align}
These relations then provide the starting point to deduce the symmetry properties
under time-reversal of the local densities that will be defined in what follows.

Inserting the expansion of $\rho_q (\vec{r}\sigma, \vec{r}' \sigma')$
into single-particle wave functions~\eqref{eq:def:nonlocaldensity} into the
definition of the non-local normal densities, one finds
\begin{align}
\label{eq:def:rho:nonlocal}
\rho_q(\vec{r}, \vec{r}') 
& = \sum_{jk} \rho_{kj} \, \Psi^\dagger_j(\vec{r}') \, \Psi_k(\vec{r})
     \, , \\
\label{eq:def:s:nonlocal}
s_{q,\kappa}(\vec{r}, \vec{r}') 
& = \sum_{jk} \rho_{kj} \, \Psi^\dagger_j(\vec{r}') \, \hat{\sigma}_{\kappa} 
    \Psi_k(\vec{r}) \, ,
\end{align}
where the sum over single-particle orbits runs over those of the nucleon species $q$.
Doing the same for the non-local pair densities, Eqs.~\eqref{eq:def:rhot:nonlocal}
and~\eqref{eq:def:st:nonlocal}, the latter can be redefined as
\begin{align}
\label{eq:def:rhot:nonlocal:2}
\tilde{\rho}_q(\vec{r}, \vec{r}') 
& \equiv \sum_{jk} \kappa_{kj} \, \tilde{\varrho}_{jk}(\vec{r},\vec{r}') \, ,
     \\
\label{eq:def:st:nonlocal:2}
\tilde{s}_{q, \mu}(\vec{r}, \vec{r}')  
& \equiv \sum_{jk} \kappa_{kj} \, \tilde{\varsigma}_{jk,\mu}(\vec{r},\vec{r}') \, ,
\end{align}
using the objects
\begin{align}
\tilde{\varrho}_{jk}(\vec{r},\vec{r}') 
& \equiv \sum_{\sigma} \sigma \, \psi_j(\vec{r}',\sigma) \, \psi_k(\vec{r},-\sigma)  \, ,
   \\
\tilde{\varsigma}_{jk,\mu}(\vec{r},\vec{r}')
& \equiv \sum_{\sigma, \sigma'} \sigma' \, \psi_j(\vec{r}',\sigma') \, \psi_k(\vec{r},\sigma) \, 
    \langle -\sigma' | \hat{\sigma}_{\mu} | \sigma \rangle \, .
\end{align}
The first one of these, $\tilde{\varrho}_{jk}(\vec{r},\vec{r}')$ is 
proportional to the two-body wave function of a spin-singlet state, $S=0$,  
with the third component of isospin $T_3 = \pm 1$ and is therefore a 
member of an isospin-triplet, $T = 1$. In contrast, the three Cartesian 
components of $\tilde{\varsigma}_{jk,\mu}(\vec{r},\vec{r}')$ are linear combinations 
of the three possible two-body wave functions of a spin-triplet state,
$S=1$, again with $T = 1$, which gives an indication of the physics 
described by terms that contain these objects. This observation also 
explains the attractiveness of using the ``Russian representation'' 
instead of the traditional one based on $\kappa(\vec{r} \sigma,\vec{r}'\sigma')$
that cannot be further broken down into objects with a physical interpretation. 

These objects have the following symmetries under the simultaneous exchange of 
the single-particle states $j$ and $k$ and their positions
\begin{align}
\label{eq:varrho:exchange}
\tilde{\varrho}_{jk}(\vec{r},\vec{r}')
& = - \tilde{\varrho}_{kj}(\vec{r}',\vec{r}) \, ,
    \\
\label{eq:varsigma:exchange}
\tilde{\varsigma}_{jk,\mu}(\vec{r},\vec{r}') 
& = + \tilde{\varsigma}_{kj,\mu}(\vec{r}',\vec{r}) \, .
\end{align}
Using the phase convention 
$\psi^T_k (\vec{r},\sigma) = -\sigma \, \psi_k(\vec{r},-\sigma)$,
in the time-reversed auxiliary state these objects become
\begin{align}
\label{eq:varrho:T}
\tilde{\varrho}_{jk}^T (\vec{r},\vec{r}')
& = + \tilde{\varrho}_{kj}^*(\vec{r},\vec{r}') \, ,
    \\
\label{eq:varsigma:T}
\tilde{\varsigma}_{jk,\mu}^T (\vec{r},\vec{r}') 
& = - \tilde{\varsigma}_{kj,\mu}^*(\vec{r},\vec{r}') \, .
\end{align}
By the consecutive application of derivatives with respect to either $\vec{r}$ 
or $\vec{r}'$ to the non-local normal and pair densities defined through
Eqs.~\eqref{eq:def:rho:nonlocal_1}--\eqref{eq:def:st:nonlocal},
setting $\vec{r} = \vec{r}'$ afterwards, one can now construct a multitude of 
different local densities, which in turn can be used to build terms of a local 
energy density functional. We will introduce the local densities entering the energy 
densities of Eq.~\eqref{eq:def:N2LOfunc} order-by-order in the next sections.

A final remark is in order: we concern ourselves here chiefly with static 
self-consistent mean-field calculations. Our discussion
will be limited to the normal and pair densities of a single
auxiliary state, and does not extend to the more general calculation of 
transition (also called  mixed) densities between different auxiliary states, 
$\langle \Phi' |\hat{a}^{\dagger}_{\vec{r}' \sigma' q'} \hat{a}_{\vec{r} \sigma q}|\Phi\rangle/\langle \Phi' | \Phi\rangle$,  
$\langle \Phi' |\hat{a}^{\dagger}_{\vec{r}' \sigma' q'} \hat{a}^{\dagger}_{\vec{r} \sigma q}|\Phi\rangle/\langle \Phi' | \Phi\rangle$ and 
$\langle \Phi' |\hat{a}_{\vec{r}' \sigma' q'} \hat{a}_{\vec{r} \sigma q}|\Phi\rangle/\langle \Phi' | \Phi\rangle$.
The calculation of these objects is required, for example, when restoring broken 
symmetries or when employing the generator coordinate method~\cite{Bender03}. 
As these objects are in general complex-valued functions, the following discussion 
regarding the reality of specific densities has to be replaced by an analysis of the 
hermiticity of the corresponding densities, that is to say their symmetry under exchange of $|\Phi\rangle$ 
and $|\Phi' \rangle$, which would lead to the same conclusions for preferable
choices for the definition of local densities.


\subsubsection{LO: Local densities appearing in $\mathcal{E}^{(0)}$}
\label{sec:LO}

When discarding the possibility of proton-neutron mixing as done here, it
is only possible to construct three different gradientless local densities 
at leading order. These are the normal local density $\rho_q(\vec{r})$, 
the normal spin density $\vec{s}_q(\vec{r})$,  and the local 
pair density $\tilde{\rho}_q(\vec{r})$,
\begin{align}
\rho_q(\vec{r})
& \equiv \rho_q(\vec{r}, \vec{r}') \big|_{\vec{r} = \vec{r}'} \, , 
\label{eq:def:rho} \\ 
\vec{s}_q(\vec{r})
& \equiv \vec{s}_q(\vec{r}, \vec{r}') \big|_{\vec{r} = \vec{r}'} \, ,
\label{eq:def:vecs} \\
\tilde{\rho}_q(\vec{r})
& \equiv \tilde{\rho}_q(\vec{r}, \vec{r}') \big|_{\vec{r} = \vec{r}'} \, .
\label{eq:def:rhotilde}
\end{align}
The local part of the spin pair density $\tilde{\vec{s}}_q$(\vec{r}) is 
identical to zero, which can be easily deduced from 
Eq.~\eqref{eq:nonlocalppsym}, and which explains the absence of a
fourth local density.\footnote{We note, however, that a local isoscalar spin 
pair density appears naturally when also considering proton-neutron-mixing $T=0$ 
pairing correlations \cite{Perlinska04,Rohozinski10a}.
} 
From Eq.~\eqref{eq:nonlocalphsym} follows that both 
$\rho_q(\vec{r})$ and $\vec{s}_q(\vec{r})$ are real-valued functions, 
even if the matrix elements $\rho_{kj}$ and the single-particle wave functions 
are not. By contrast, $\tilde{\rho}_q(\vec{r})$ is in general a complex-valued 
spatial function.

Using Eq.~\eqref{eq:Trhononloc} and Eq.~\eqref{eq:Trhotildenonloc}, we
can determine the local densities of the time-reversed auxiliary state
\begin{align}
\rho^T_q(\vec{r})          & = + \rho_q(\vec{r})     \, , \nn \\
\vec{s}^T_q(\vec{r})       & = - \vec{s}_q(\vec{r})  \, , \nn \\
\tilde{\rho}^T_q(\vec{r})  & = + \tilde{\rho}^*_q(\vec{r})  \, .
\end{align}
Hence $\rho_q(\vec{r})$ is time-even, whereas $\vec{s}_q(\vec{r})$ is time-odd. 
The pair density $\tilde{\rho}_q(\vec{r})$ in its entirety is neither time-even 
nor time-odd: its real part is time-even, while its imaginary part is time-odd.


\subsubsection{NLO: Local densities additionally appearing in $\mathcal{E}^{(2)}$}
\label{sec:NLO}

By acting with either one or two derivatives on the non-local densities,
we obtain five different normal and two different pair densities 
that enter the EDF at NLO.  These normal local densities are given by
\begin{align}
\tau_q(\vec{r}) 
& \equiv \vnabla \cdot \vnabla' \, \rho_q(\vec{r}, \vec{r}')\Big|_{\vec{r} = \vec{r}'} \, ,
\label{eq:def:tau}
    \\
\vec{j}_q(\vec{r}) 
& \equiv -\tfrac{\iunit}{2} \big(  \vnabla - \vnabla'    \big)
   \rho_q(\vec{r}, \vec{r}') \Big|_{\vec{r} = \vec{r}'} \, , 
\label{eq:def:vecj}
\\
T_{q,\mu}(\vec{r})
& \equiv  \vnabla \cdot \vnabla'  \,
   s_{q, \mu}(\vec{r}, \vec{r}') \Big|_{\vec{r} = \vec{r}'} \, ,
\label{eq:def:vect} \\
F_{q,\mu}(\vec{r})
& \equiv \tfrac{1}{2} \sum_{\nu} \left[ \nabla_{\mu}\nabla'_{\nu} + \nabla'_{\mu}\nabla_{\nu}  \right] \,
   s_{q, \nu}(\vec{r}, \vec{r}') \Big|_{\vec{r} = \vec{r}'} \, ,
\label{eq:def:vect}
\\
J_{q, \mu \nu}(\vec{r})
& \equiv -\tfrac{\iunit}{2} \big(\nabla_{\mu}-\nabla'_{\mu} \big) \,
   s_{q, \nu} (\vec{r}, \vec{r}') \Big|_{\vec{r} = \vec{r}'} \, ,
\label{eq:def:Jmunu}
\end{align}
where $\mu$, $\nu$, $\kappa$ label the three cartesian directions of space.
Three of these densities, which are $\tau_q(\vec{r})$, $\vec{T}_q(\vec{r})$, and $\vec{F}_q(\vec{r})$,
are kinetic-type densities involving the successive application of two gradients 
acting on different coordinates, while the two others, that is
$\vec{j}_q(\vec{r})$ and $J_{q,\mu\nu}(\vec{r})$, are of current type, involving 
the difference of two gradient operators acting on different coordinates. 
The density $\vec{F}_q(\vec{r})$ in general only appears for parametrizations
that include explicit tensor forces, such as considered in 
Refs.~\cite{Lesinski07,Bender09,Hellemans12}, and does not play a role for the 
presently available parametrizations considering N2LO terms; we have included 
it here for completeness' sake.
There are a few additional possible densities with
two gradients, but they cannot be used to construct suitable scalar contributions 
to the energy density at this order
\cite{Dobaczewski96b,Doba00}.

For the pair densities that can contribute to the pair EDF at 
NLO \cite{Dobaczewski84,Sadoudi13a}, we write
\begin{align}
\tilde{\tau}_{q}(\vec{r})
& \equiv
   \vnabla \cdot \vnabla' \, \tilde{\rho}_q(\vec{r}, \vec{r}')\Big|_{\vec{r} = \vec{r}'} \, ,
\label{eq:def:tautilde}
\\
\tilde{J}_{q, \mu \nu}(\vec{r})
& \equiv -\tfrac{\iunit}{2} \big(\nabla_{\mu}-\nabla'_{\mu} \big) \,
   \tilde{s}_{q, \nu} (\vec{r}, \vec{r}') \Big|_{\vec{r} = \vec{r}'} \, .
\label{eq:def:Jmunutilde}
\end{align}
where we have one kinetic-type ($\tilde{\tau}_{q}$) and one current-type
pair density ($\tilde{J}_{q, \mu \nu}$). 

Together with the three gradientless densities, the seven densities presented
in this section form the backbone of the many different forms of the Skyrme 
functional up to NLO found in the literature.\footnote{The pair
densities contributing to the NLO pair functional are only rarely incorporated, 
notable exceptions being Refs.~\cite{Dobaczewski84,Sadoudi13a} where 
the same coupling constants determine the particle-hole and pairing parts
of the EDF. Most applications of the Skyrme EDF, however, use an independent
pairing EDF that is constructed only out of the LO pair densities such as the
one of Eq.~\eqref{eq:pair:EDF:ULB} that will be used here.
} 
In fact, it can be shown that, in
the absence of proton-neutron mixing, together with some of their derivatives 
these local densities are sufficient to express \textit{any} local energy 
density with up to two gradients that is compatible with the usual symmetries of
the nuclear Hamiltonian~\cite{Perlinska04}.

These traditional definitions have the advantage 
that all normal NLO densities are real-valued functions and are either 
time-even or time-odd:
\begin{align}
\tau^T_{q}(\vec{r})       &= + \tau_{q}   (\vec{r})  \, , 
&  
\vec{j}^T_q(\vec{r})      &= - \vec{j}_q(\vec{r})   \, , \nn  \\
\vec{T}^T_{q}(\vec{r})    &= - \vec{T}_{q}(\vec{r})  \, ,
&  
J^T_{q, \mu \nu}(\vec{r}) &= + J_{q, \mu \nu}(\vec{r}) \, , \nn \\
\vec{F}^T_{q}(\vec{r})    &= - \vec{F}_{q}(\vec{r})  \, .
\end{align}
As at LO, the pair densities are neither time-even nor time-odd
\begin{align}
\tilde{\tau}^T_{q}(\vec{r})      = + \tilde{\tau}^{*}_{q}(\vec{r}) \, , \quad & \,  
\tilde{J}^T_{q,\mu \nu}(\vec{r}) = + \tilde{J}^{*}_{q,\mu \nu}(\vec{r}) \, .
\end{align}
Again, their real parts are time-even, while the imaginary parts are  time-odd. 


\subsubsection{N2LO: Local densities additionally appearing in $\mathcal{E}^{(4)}$}
\label{sec:N2LO}

We are aware of two existing conventions in the literature for densities at 
N2LO: a scheme based on spherical tensors 
introduced in Ref.~\cite{Carlsson08}, and a scheme based on cartesian tensors, 
more analogous to the traditional densities of the NLO Skyrme functional, 
introduced first in Ref.~\cite{Becker15} and amended in 
Ref.~\cite{Becker17}. As we will discuss in what follows, these two schemes are 
but two out of many different, but equivalent, possibilities to define such 
densities. Of the two approaches, we will stay closer to the choice of 
Ref.~\cite{Becker17}, as the spherical tensors used in Ref.~\cite{Carlsson08}
are complicated to use when considering deformed nuclei.

The choice made in Ref.~\cite{Becker17} consists of the following four local 
normal densities
\begin{align}
\label{eq:def:Q}
Q_{q}(\vec{r})
& \equiv \Delta \, \Delta' \, \rho_{q} (\vec{r}, \vec{r}')
                    \big|_{\vec{r} = \vec{r}'} \, ,
\\
\label{eq:def:sn2lo}
\vec{S}_q(\vec{r})   
& \equiv \Delta \, \Delta' \, \vec{s}_q(\vec{r}, \vec{r}') \big|_{\vec{r} = \vec{r}'} \, ,
\\ 
\label{eq:def:pin2lo} 
\Pi_{q, \mu} (\vec{r}) 
& \equiv -\tfrac{\iunit}{2} \big(\nabla_{\mu}-\nabla'_{\mu} \big) \,
        \big( \vnabla \cdot \vnabla' \big)  
     \rho_q(\vec{r}, \vec{r}') \big|_{\vec{r} = \vec{r}'} \, ,
\\
V_{q, \mu \nu} (\vec{r})
& \equiv  - \tfrac{\iunit}{2} \big(\nabla_{\mu}-\nabla'_{\mu} \big) \,
      ( \vnabla \cdot \vnabla'  ) \,
                    s_{q, \nu} (\vec{r}, \vec{r}')
                    \big|_{\vec{r} = \vec{r}'} \, ,
\label{eq:def:Vmunu}
\end{align}
which are all real spatial functions by construction. 
In addition to these four new normal densities, the authors of Ref.~\cite{Becker17} define
also the following kinetic-type normal tensor densities
\begin{align}
\tau_{q, \mu\nu} (\vec{r})
& \equiv \nabla_{\mu} \, \nabla'_{\nu} \, \rho_{q}(\vec{r}, \vec{r}')
          \big|_{\vec{r} = \vec{r}'} \, ,
\label{eq:def:taun2}
\\
K_{q, \mu\nu \kappa} (\vec{r})
& \equiv \nabla_{\mu} \, \nabla'_{\nu} \, s_{q,\kappa}(\vec{r}, \vec{r}')
    \big|_{\vec{r} = \vec{r}'} \, ,
\label{eq:def:Kn2}
\end{align}
which are in general complex spatial functions. Note that not every component
of either density is a complex number with non-zero real and imaginary parts: 
$\tau_{xx}(\vec{r})$, $\tau_{yy}(\vec{r})$ and $\tau_{zz}(\vec{r})$ are for 
instance all real functions as they make up the scalar part of the tensor.
The densities $\tau_q(\vec{r})$, $\vec{T}_q(\vec{r})$ and $\vec{F}_q(\vec{r})$
that already entered $\mathcal{E}^{(2)}(\vec{r})$ are different tensor 
contractions of the full tensors $\tau_{q, \mu \nu}(\vec{r})$ and $K_{q, \mu\nu \kappa}(\vec{r})$
with the Kronecker symbol $\delta_{\lambda \chi}$. Although these densities 
only contain two gradients (not three or four), they only become 
relevant at the N2LO level since there are no other densities at the NLO level that 
they can be combined with to a form a scalar term in the energy density \cite{Dobaczewski96b}.

Even though they do not play a role in the functional of Ref.~\cite{Becker17}, 
we can construct the following pair densities by analogy
\begin{align}
\label{eq:def:Q}
\tilde{Q}_{q}(\vec{r})
&=  \Delta \, \Delta' \, \tilde{\rho}_{q} (\vec{r}, \vec{r}')
                    \Big|_{\vec{r} = \vec{r}'} \, ,
\\
\label{eq:def:Vmunutilde} 
\tilde{V}_{q, \mu \nu} (\vec{r})
&=  - \tfrac{\iunit}{2}  \left( \nabla_{\mu}-\nabla'_{\mu} \right)
         (\vnabla \cdot \vnabla') \, 
                    \tilde{s}_{q, \nu} (\vec{r}, \vec{r}')
                    \Big|_{\vec{r} = \vec{r}'} \, , \\
\tilde{\tau}_{q,\mu \nu} &= 
    \nabla_{\mu} \, \nabla'_{\nu}  \,
    \tilde{\rho}_q(\vec{r}, \vec{r}') \Big|_{\vec{r} = \vec{r}'} \, .
\end{align}
Again, in the absence of proton-neutron mixing, it follows from 
Eq.~\eqref{eq:nonlocalppsym} that the pair densities
$\tilde{\vec{S}}_{q}(\vec{r})$ and $\tilde{K}_{q, \mu \nu \kappa}(\vec{r})$ 
that are the analogues of the normal densities defined in Eqs.~\eqref{eq:def:sn2lo}
and~\eqref{eq:def:Kn2} are zero, as is the pair current $\tilde{\Pi}_{q, \mu} (\vec{r})$
that is the analogue of Eq.~\eqref{eq:def:pin2lo}.

Using Eq.~\eqref{eq:Trhononloc}, one can easily
establish the following relations for the new normal densities
\begin{alignat}{4}
Q^T_{q}(\vec{r})           &= + Q_{q}(\vec{r})          \, ,  & \quad 
\vec{S}^T_q(\vec{r})       &= - \vec{S}_q(\vec{r})      \, ,\\
V^T_{q, \mu \nu} (\vec{r}) &= + V_{q, \mu \nu} (\vec{r})\, ,  & \quad 
\vec{\Pi}^T_{q} (\vec{r})  &= - \vec{\Pi}_{q} (\vec{r}) \, , \\
\tau^T_{q, \mu \nu}        &= + \tau^*_{q, \mu \nu}     \, ,  & \quad 
K_{q,\mu\nu \kappa}^T      &= - K^*_{q,\mu\nu \kappa}   \, .
\label{eq:def:Vmunu}
\end{alignat} 
The new real densities $Q_q(\vec{r})$ and $V_{q,\mu\nu}(\vec{r})$ are both 
time-even while $\vec{S}_q(\vec{r})$ and $\vec{\Pi}_{q}(\vec{r})$ are time-odd. 
By contrast, the complex densities 
$\tau_{q,\mu \nu}(\vec{r})$ and $K_{q, \mu \nu \kappa}(\vec{r})$ do not have a
definite behavior under time-reversal:
$\Re \{ \tau_{q, \mu \nu}(\vec{r}) \}$ 
and $\Im \{ K_{q, \mu \nu \kappa}(\vec{r})\}$ are time-even,
whereas $\Im \{\tau_{q, \mu \nu}(\vec{r}) \}$ and 
$\Re \{K_{q, \mu \nu \kappa}(\vec{r})\} $ are time-odd. Ultimately,
this is a consequence of the operator $\nabla_{\mu} \, \nabla'_{\nu}$
that generates these densities in Eqs.~\eqref{eq:def:taun2} and~\eqref{eq:def:Kn2}
not being hermitian.

In similar fashion, we can use Eq.~\eqref{eq:Trhotildenonloc} for the pair
densities and deduce
\begin{align}
\tilde{Q}^T_q(\vec{r})              &= + \tilde{Q}^*(\vec{r}) \, ,  \\
\tilde{V}^T_{q,\mu\nu}(\vec{r})     &= + \tilde{V}_{q, \mu\nu}^*(\vec{r}) \, ,  \\
\tilde{\tau}^T_{q,\mu \nu}(\vec{r}) &= + \tilde{\tau}^*_{q,\mu\nu}(\vec{r}) \, .
\end{align}
As is the case for the pair densities entering the LO and NLO functional,
all of the new pair densities are complex functions, with real
parts that are time even while the imaginary parts are time odd.


\subsubsection{Problems with this scheme}
\label{sec:problems}

There are three main reasons to devise a better accounting of the mean-field
densities. The first one is that one cannot keep up in a reasonable way the 
letter-naming scheme of Ref.~\cite{Becker17}: at N2LO level one has to introduce
four new normal densities (and two pair densities) in addition 
to higher-rank tensor versions of those already present at NLO. Combined with those needed 
for the leading order Skyrme functional, the full N2LO functional is formulated 
in terms of eleven distinct local densities. Each of these leads to a 
distinct potential for which also a symbol has to be chosen.
At higher orders this number will be
further inflated such that at some point one will run out of letters of the roman and
greek alphabets that can be unambiguously used to represent these densities and the
corresponding potentials, not to mention the increasing mnemonic difficulties to associate
these symbols with the object that they represent.
A more extensible notation is desired, which preferably links
the underlying operator structure to the symbol used for a given density. 

The second is a problem of efficiency: at any given order in gradients there are
many ways to recouple gradient operators that act on the non-local densities. 
For example: the imaginary part of $\tau_{q,\mu \nu}(\vec{r})$ can be rewritten 
as a linear combination of external gradients acting on the current density
$\vec{j}_q(\vec{r})$ \cite{KBJMnotes}:
\begin{align}
\Im \{ \tau_{q, \mu \nu} (\vec{r}) \} 
=    \tfrac{1}{2} \big[    \nabla_{\mu} \, j_{q, \nu}(\vec{r})
                         - \nabla_{\nu} \, j_{q, \mu}(\vec{r})  \big].
\label{eq:reducible_example}
\end{align}
This relation can be easily shown using Eq.~\eqref{eq:recoupleC}, which will
explained in the next subsection. We will call densities that can be rewritten 
as linear combinations of gradients of lower order densities \emph{reducible}. 
Equation~\eqref{eq:reducible_example} demonstrates that 
$\Im \{ \tau_{q\mu \nu}(\vec{r}) \}$ is reducible: it can be completely 
eliminated from any formulation of the EDF in favor of terms involving 
gradients acting on the components of $\vec{j}_q(\vec{r})$.

Another example is the definition of an alternative density touched upon briefly
in Ref.~\cite{BeckerPhD}, which involves four gradients,
\begin{align}
\mathcal{T}_q(\vec{r}) \equiv
 (\vnabla' \cdot \vnabla) \, (\vnabla' \cdot \vnabla) \, \left.\rho_q(\vec{r},\vec{r}')\right|_{\vec{r} = \vec{r}'} \, .
\end{align}
While $\mathcal{T}_q(\vec{r})$ is a valid local density that can be used to 
define an EDF, with the help of relations that will be explained in the next
subsection one can show that it can be rewritten as a linear combination 
of the N2LO density $Q_q(\vec{r})$ defined in Eq.~\eqref{eq:def:Q} that also 
contains four gradients and second derivatives of the kinetic tensor density 
$\tau_{q, \mu \nu}(\vec{r})$ that contains just two gradients
\begin{align}
\mathcal{T}_q(\vec{r}) = Q_q(\vec{r}) + \Delta \tau_q(\vec{r})
            - \sum_{\mu \nu} \nabla_{\mu} \nabla_{\nu} \tau_{q,\mu\nu}(\vec{r}) \, .
\label{eq:redundancy_example}
\end{align}
It is clear that when choosing the local densities $Q_q(\vec{r})$, 
$\tau_{q,\mu \nu}(\vec{r})$, and $\tau_q(\vec{r})$ to express an EDF,
one does not need to additionally consider $\mathcal{T}_q(\vec{r})$.
\footnote{Note that $Q_q(\vec{r})$ and $\mathcal{T}_q(\vec{r})$ become equal
in infinite homogeneous nuclear matter. Also, the integral over 
$\mathcal{T}_q(\vec{r})$ 
is always equal to the integral over $Q_q(\vec{r})$, although for systems 
with a surface they will look unalike when plotted. What will be different for such 
systems are integrals where either $\mathcal{T}_q(\vec{r})$ or $Q_q(\vec{r})$
multiplies another function or density, such as bilinear or trilinear
terms in the EDF.}

Because of the possibility of redundancies such as 
Eq.~\eqref{eq:redundancy_example} and the existence of reducible densities, it 
is not trivial to choose a set of mean-field densities at arbitrary order in 
gradients. The problem can arrive in two different ways: either when generating
the EDF from a pseudopotential~\eqref{eq:def:Vsk}, which necessitates to 
recouple the gradients in such a way that they can be expressed through suitable 
local densities~\cite{Sadoudi13a,Raimondi11a,Perlinska04,Engel75}, or when setting 
up the EDF directly as a combination of densities that respects the usual 
symmetry requirements \cite{Zalewski08,Carlsson08,Dobaczewski95,Dobaczewski96b}. 
In either case one would like to allow for the most general EDF but without 
introducing extra overhead for the densities, neither formally in the multitude of 
symbols to define and memorize, nor numerically in the number of objects to be calculated. 
As will be sketched in Appendix~\ref{app:funfact}, at the NLO level this 
only concerns the kinetic densities. For these, however, it is known for long that
using one or the other of the possible forms can lead to subtle differences when 
setting up semi-classical approximations.

While this problem is still somewhat limited at the N2LO level
detailed here, the amount of possible redundancies rises  
markedly when going to higher orders. From general arguments about the
symmetries of a local bilinear EDF whose terms contain in total six gradients, 
one can deduce that the construction of the additional central terms that 
emerge at N3LO will require two kinetic-type densities involving six 
derivatives, a scalar one and a pseudovector spin density, as well as two 
current-type densities involving five derivatives, one being a vector density, 
the other a rank-2 spin-current-type density. In addition, four higher-rank 
tensor densities will appear. One of the possible definitions of 
the additional tensor densities
is such that their contraction with a Kronecker symbol gives
$Q_s(\vec{r})$,  $\vec{S}_q(\vec{r})$, $\vec{\Pi}_q(\vec{r})$ and 
$V_{q, \mu \nu}(\vec{r})$, but there are many other possible definitions that 
are all related through the recoupling of derivatives similar in spirit 
to Eq.~\eqref{eq:redundancy_example}. 
We feel it is important that we are able to make an informed choice 
of the local  densities used to construct an energy density. In our view, 
an ideal choice would be to keep only irreducible and non-redundant 
densities, and preferably those that have some practical advantages 
such as requiring less memory storage compared to other possible choices.

As a third problem, the components of the two normal tensor densities 
$\tau_{q, \mu \nu}(\vec{r})$ and $K_{q, \mu \nu \kappa}(\vec{r})$ introduced
in Sec.~\ref{sec:N2LO} do not all behave in the same way under time-reversal, such
that one cannot attribute a definite global behavior under time-reversal
to these densities. This can ultimately be traced back to the 
fact that these local densities are generated by a non-hermitian combination
of gradient operators ($\nabla_\mu \nabla_\nu'$)
and therefore are complex-valued. This makes separating 
the densities into time-even and time-odd categories impossible, complicates
the separation of terms of the EDF as either time-even or time-odd,\footnote{This 
issue was overlooked in the earlier Ref.~\cite{Becker15}, where a term 
bilinear in $\tau_{q,\mu\nu}(\vec{r})$ was wrongly assigned as time-even and a term 
bilinear in $K_{q, \mu \nu \kappa}(\vec{r})$ was wrongly assigned as time-odd.
} 
and explains why this has not been done in Eq.~\eqref{eq:originalN2LO}.

For the pair densities, this issue already arises at leading order, but the 
literature has so far simply accepted this inconvenient feature, see for 
example Refs.~\cite{Hellemans12,Sadoudi13a}. We would prefer a prescription 
that guarantees  the reality of all densities, and therefore make a clear 
delineation of time-even and time-odd possible for both types of
densities and terms of the EDF.

In the next section, we will introduce a definition and a notation for 
normal and pair densities that is extensible to any order of derivatives and that 
guarantees real densities that have well-defined signs under time-reversal. In addition, 
it will help us identify redundancies and relations between different densities
by making their operator structure explicit.


\section{A new notation and definition for local densities}
\label{sec:definition}


\subsection{Systematic counting notation}
\label{sec:notation}


\subsubsection{Normal densities}
\label{subsec:notation:normal}

In order to approach the problem of defining local normal and pair densities in 
a more systematic way, we propose the following definition for normal
densities obtained by applying derivatives to the non-local density 
$\rho(\vec{r}, \vec{r}')$ and then setting the positions $\vec{r}$ and 
$\vec{r}'$ to be equal
\begin{align}
\label{eq:def:Dden}
D_q^{A,B} (\vec{r}) 
& \equiv  \Re \big\{ \hat{A}' \hat{B} \, \rho_q(\vec{r}, \vec{r}') \big\}  \Big|_{\vec{r} = \vec{r}'}
      \, , \\
\label{eq:def:Cden}
C_q^{A, B} (\vec{r}) 
& \equiv  \Im \big\{ \hat{A}' \hat{B}\, \rho_q(\vec{r}, \vec{r}') \big\}  \Big|_{\vec{r} = \vec{r}'}
     \, ,
\end{align}
where the ``left'' operator $\hat{A}'$ is a combination of derivatives acting on 
the primed spatial coordinates in the non-local density, whereas the  ``right'' operator
$\hat{B}$ is a combination of derivatives acting on the spatial coordinates of
the unprimed coordinates.

Similarly, we also define $D$ and $C$ objects starting from the non-local 
spin density $\vec{s}_q(\vec{r}, \vec{r}')$
\begin{align}
\label{eq:def:Dsvec}
D_q^{A, B\sigma} (\vec{r})
& \equiv  \Re \big\{ \hat{A}' \hat{B} \, \vec{s}_q(\vec{r}, \vec{r}') \big\}  \Big|_{\vec{r} = \vec{r}'}
\, , \\
\label{eq:def:Csvec}
C_q^{A, B\sigma} (\vec{r})
& \equiv \Im \big\{ \hat{A}' \hat{B} \, \vec{s}_q(\vec{r}, \vec{r}') \big\}  \Big|_{\vec{r} = \vec{r}'}\, .
\end{align}
The generic definitions made in Eqs.~\eqref{eq:def:Dden}--\eqref{eq:def:Csvec}
omit the tensor structure of the densities and currents on the left-hand side as they are 
specific to each choice of operators $\hat{A}$ and $\hat{B}$. The indices 
labeling the Cartesian components of the densities and currents can easily 
be accommodated in this kind of notation through subscripts: as an example we write
\begin{align}
D_{q, \mu \nu \kappa}^{\nabla, \nabla \sigma} (\vec{r}) 
&=   \Re \big\{  \nabla'_{\mu} \nabla_{\nu} \, {s}_{q, \kappa}(\vec{r}, \vec{r}') \big\}  \Big|_{\vec{r} = \vec{r}'} \, .
\end{align}
Note that the order of subscripts follows the order of operators in the 
superscripts.
Tensor contractions with Kronecker symbols, i.e.\ scalar products between the
operators $\hat{A}$ and $\hat{B}$ in the definition of the local densities,
can be accommodated with the use of brackets; we define for example
\begin{align}
D_{q, \kappa}^{(\nabla, \nabla) \sigma} (\vec{r}) 
& = \sum_{\mu} D_{q, \mu \mu \kappa}^{\nabla, \nabla \sigma} (\vec{r})
    \nn \\
& = \sum_{\mu}  \Re \left\{ \nabla'_{\mu} \nabla_{\mu}  \, 
    {s}_{q, \kappa}(\vec{r}, \vec{r}') \right\}  \Big|_{\vec{r} = \vec{r}'}
   \nn \\
& = \Re \left\{ \vnabla' \cdot \vnabla  \, 
   {s}_{q, \kappa}(\vec{r}, \vec{r}') \right\}  \Big|_{\vec{r} = \vec{r}'}  \, .
\end{align}
Similarly, as a shorthand for vector products, we introduce 
\begin{align}
C_{q, \mu}^{1, \nabla \times \sigma} (\vec{r}) 
&= \sum_{\nu \kappa} \Im \left\{  \epsilon_{\mu \nu \kappa} \nabla_{\nu} \, 
   {s}_{q, \kappa}(\vec{r}, \vec{r}') \right\}  \Big|_{\vec{r} = \vec{r}'}
   \nn \\
\Leftrightarrow \qquad
\vec{C}_{q}^{1, \nabla \times \sigma} (\vec{r})
& = \Im \left\{  \vnabla \times \vec{s}_{q}(\vec{r}, \vec{r}') \right\}  \Big|_{\vec{r} = \vec{r}'}
\, , 
\end{align}
where $\epsilon_{\mu\nu\kappa}$ is the skew-symmetric rank-3 Levi-Civita tensor.

The definitions~\eqref{eq:def:Dden}--\eqref{eq:def:Csvec} automatically 
lead to local densities that are a real function of the position $\vec{r}$
and that are either even or odd under time-reversal. In addition, in this form
they can be efficiently implemented in numerical codes. For formal manipulations, 
such as the derivation of the single-particle Hamiltonian from the EDF that will
be sketched in Sec.~\ref{sec:singleh}, or the derivation of the residual interaction as
used in QRPA from the EDF, or the construction of the MR~extension of the EDF, it 
is necessary to use an alternative definition that does not necessitate to take 
the real or imaginary part of some expression. 

Using that the real and imaginary parts of a complex number can be obtained as 
$\Re \{ z \} = \tfrac{1}{2} ( z + z^*)$ and $\Im \{ z \} = -\tfrac{\iunit}{2} ( z - z^*)$,
respectively, and exploiting the symmetry relation 
$\rho_q(\vec{r}, \vec{r}') = \rho_q^*(\vec{r}', \vec{r})$  
of the non-local density~\eqref{eq:nonlocalphsym}, the expressions~\eqref{eq:def:Dden}
and~\eqref{eq:def:Cden} for the local densities $D_q^{A, B} (\vec{r})$ and 
$C_q^{A, B} (\vec{r})$ can be rewritten in terms of an hermitian or anti-hermitian 
combination of the operators $\hat{A}$ and $\hat{B}$
\begin{align}
\label{eq:def:Dden:2}
D_q^{A, B} (\vec{r}) 
& = \Re \big\{ \hat{A}' \hat{B} \, \rho_q(\vec{r}, \vec{r}') \big\}  \Big|_{\vec{r} = \vec{r}'}
          \nonumber \\
& = \hat{A}' \hat{B} \, \tfrac{1}{2} \big[ \rho_q(\vec{r}, \vec{r}') + \rho^*_q(\vec{r}, \vec{r}') \big]
    \Big|_{\vec{r} = \vec{r}'}
          \nonumber \\
& = \tfrac{1}{2} \big[   \hat{A}' \hat{B} \, \rho_q(\vec{r}, \vec{r}') 
                      + \hat{A} \hat{B}' \, \rho^*_q(\vec{r}', \vec{r}) \big]
    \Big|_{\vec{r} = \vec{r}'}
          \nonumber \\
& = \tfrac{1}{2} \big(  \hat{A}' \hat{B} + \hat{A} \hat{B}' \big) \, \rho_q(\vec{r}, \vec{r}')  
    \Big|_{\vec{r} = \vec{r}'}
\, , \\
\label{eq:def:Cden:2}
C_q^{A, B} (\vec{r}) 
& = \Im \big\{  \hat{A}' \hat{B} \, \rho_q(\vec{r}, \vec{r}') \big\}  \Big|_{\vec{r} = \vec{r}'}
           \nonumber \\
& = \hat{A}' \hat{B} \, \big( -\tfrac{\iunit}{2} \big) \, 
    \big[  \rho_q(\vec{r}, \vec{r}') - \rho^*_q(\vec{r}, \vec{r}') \big]
    \Big|_{\vec{r} = \vec{r}'} \, ,
           \nonumber \\
& = -\tfrac{\iunit}{2} \big[  \hat{A}' \hat{B} \, \rho_q(\vec{r}, \vec{r}') 
                      - \hat{A} \hat{B}' \, \rho^*_q(\vec{r}', \vec{r}) \big]
    \Big|_{\vec{r} = \vec{r}'}
               \nonumber \\
& = -\tfrac{\iunit}{2} \big(  \hat{A}' \hat{B} - \hat{A} \hat{B}' \big) \, \rho_q(\vec{r}, \vec{r}') 
    \Big|_{\vec{r} = \vec{r}'} \, ,
\end{align}
When going from the second to the third line in each of these manipulations, 
the coordinates $\vec{r}$ and $\vec{r}'$ are exchanged in the second term, 
which can be done when both coordinates are to be set equal afterwards.

The non-local spin density exhibits the same symmetry 
$\vec{s}_q(\vec{r}, \vec{r}') = \vec{s}_q^*(\vec{r}', \vec{r})$ under exchange of 
$\vec{r}$ and $\vec{r}'$ as $\rho_q(\vec{r}, \vec{r}')$, Eq.~\eqref{eq:nonlocalphsym}, 
therefore the symmetrized expressions for $D^{A,B\sigma}_{q,\kappa}(\vec{r})$ and 
$C^{A,B\sigma}_{q,\kappa}(\vec{r})$ are obtained with the same manipulations and take 
the same structure as their homologues without spin
\begin{align}
\label{eq:def:Dsvec:2}
D^{A,B\sigma}_{q,\kappa}(\vec{r})
& =      \Re \big\{ \hat{A}' \hat{B} s_{q,\kappa}(\vec{r}, \vec{r}') \big\}
         \Big|_{\vec{r} = \vec{r}'}
         \nn \\
& =      \tfrac{1}{2} \, \big( \hat{A}' \hat{B} + \hat{A} \hat{B}' \big) \, s_{q,\kappa}(\vec{r}, \vec{r}')  
         \Big|_{\vec{r} = \vec{r}'} \, ,
         \\
\label{eq:def:Csvec:2}
C^{A,B\sigma}_{q,\kappa}(\vec{r})
& =      \Im \big\{ \hat{A}' \hat{B} s_{q,\kappa}(\vec{r}, \vec{r}')   \big\} \,
         \Big|_{\vec{r} = \vec{r}'} 
         \nn \\
& =    -\tfrac{\iunit}{2} \, 
         \big( \hat{A}' \hat{B} - \hat{A} \hat{B}' \big) \, s_{q,\kappa}(\vec{r}, \vec{r}')  
         \Big|_{\vec{r} = \vec{r}'} \, .
\end{align}
From the hermiticity of the density matrix $\rho_{jk} = \rho_{kj}^*$ and of 
the kernels $\Psi^\dagger_j(\vec{r}') \, \Psi_k(\vec{r}) = \big[ \Psi^\dagger_k(\vec{r}) \, \Psi_j(\vec{r}') \big]^*$ 
and $\Psi^\dagger_j(\vec{r}') \, \hat{\sigma}_\mu \, \Psi_k(\vec{r})
= \big[ \Psi^\dagger_k(\vec{r}) \, \hat{\sigma}_\mu \, \Psi_j(\vec{r}')\big]^*$, 
follows that the combined contributions of the states $j$
and $k$ to the local densities~\eqref{eq:def:Dden}--\eqref{eq:def:Csvec}
through the expansion of the non-local densities into single-particle 
states, Eqs.~\eqref{eq:def:rho:nonlocal} and~\eqref{eq:def:s:nonlocal}, that is 
$\big\{ \rho_{jk} \, \Psi^\dagger_j(\vec{r}) \, \Psi_k(\vec{r})
+\rho_{kj} \, \Psi^\dagger_k(\vec{r}) \, \Psi_j(\vec{r}) \big\}$ and 
$\big\{ \rho_{jk} \, \Psi^\dagger_j(\vec{r}) \, \hat{\sigma}_\mu \, \Psi_k(\vec{r})
+\rho_{kj} \, \Psi^\dagger_k(\vec{r}) \, \hat{\sigma}_\mu \, \Psi_j(\vec{r}) \big\}$, 
are both real. This has been implicitly used when attributing the real and imaginary 
part of operators $\hat{A}$ and $\hat{B}$ acting on some part of these objects to the sum 
and difference of the operators as done in Eqs.~\eqref{eq:def:Dden:2}--\eqref{eq:def:Csvec:2}.
The possibility to do so is limited to normal densities, which will lead to a
fundamental difference between normal densities and the pair densities that are 
defined in the next subsection. In the canonical single-particle basis, 
$\rho$ is diagonal~\cite{Schunck2019} and all contributions to a local, normal 
density from any individual single-particle state are even automatically real. In 
fact, in this basis, the densities can be efficiently calculated as
\begin{align}
\label{eq:def:Dden:3}
D_q^{A, B} (\vec{r}) 
& = \sum_{k} \rho_{kk} \, 
    \Re \Big\{ \big[ \hat{A} \Psi^\dagger_k(\vec{r}) \big] \, \big[ \hat{B} \, \Psi_k(\vec{r}) \big] \Big\} \, ,
     \\
\label{eq:def:Cden:3}
C_q^{A, B} (\vec{r}) 
& = \sum_{k} \rho_{kk} \, 
    \Im \Big\{ \big[ \hat{A} \Psi^\dagger_k(\vec{r}) \big] \, \big[ \hat{B} \, \Psi_k(\vec{r}) \big] \Big\} \, ,
    \\
\label{eq:def:Dsvec:2}
D^{A,B\sigma}_{q,\kappa}(\vec{r})
& =  \sum_{k} \rho_{kk} \, 
    \Re \Big\{ \big[ \hat{A} \Psi^\dagger_k(\vec{r}) \big] \, \hat{\sigma}_{\kappa} \, \big[ \hat{B} \, \Psi_k(\vec{r}) \big] \Big\} \, ,
    \\
\label{eq:def:Csvec:3}
C^{A,B\sigma}_{q,\kappa}(\vec{r})
& = \sum_{k} \rho_{kk} \, 
    \Im \Big\{ \big[ \hat{A} \Psi^\dagger_k(\vec{r}) \big] \, \hat{\sigma}_{\kappa} \, \big[ \hat{B} \, \Psi_k(\vec{r}) \big] \Big\} \, ,
\end{align}
where the sum runs over the canonical single-particle states $k$ of the nucleon species $q$. 
The adaptation of these expressions to arbitrary single-particle bases is straightforward, 
but less efficient for numerical implementations because of the double summation over 
single-particle states that is avoided in the canonical basis.


\subsubsection{Pair densities}
\label{subsec:notation:pair}

The definition of local pair and spin-pair densities along the same lines requires a few
additional considerations. These are necessary because of conceptual differences between 
normal and pair densities. Most importantly, all ingredients of the non-local pair densities 
$\tilde{\rho}_q(\vec{r}, \vec{r}')$ and $\tilde{\vec{s}}_q(\vec{r}, \vec{r}')$ as
defined in Eqs.~\eqref{eq:def:rhot:nonlocal:2} and~\eqref{eq:def:st:nonlocal:2}, that 
is the anomalous density matrix $\kappa_{kj}$ and the two-body wave functions 
$\tilde{\varrho}_{jk}(\vec{r},\vec{r}')$ and $\tilde{\varsigma}_{jk,\mu}(\vec{r},\vec{r}')$, 
transform as rank-two tensors under basis changes, and not as hermitian matrices. 
Related to this is that the two non-local pair densities $\tilde{\rho}_q(\vec{r}, \vec{r}')$ 
and $\tilde{\vec{s}}_q(\vec{r}, \vec{r}')$ not only transform differently from the normal 
non-local densities under time-reversal and exchange of $\vec{r} \leftrightarrow \vec{r}'$, 
but also differently from each other, see Sec.~\ref{sec:densities}. There are also systematic 
differences between the behavior of normal and pair densities under other similarity 
transformations. In addition, the anomalous density matrix $\kappa_{kj}$ cannot always be
chosen to be real, even in the canonical basis. All of this has consequences for the 
efficient definition of pair densities. More detailed explanations of these properties 
can be found in Appendix~\ref{app:pairdens}. Taking them into account, we propose 
to define local pair densities as follows
\begin{align}
\label{eq:def:Ddenpair} 
\tilde{D}_q^{A, B} (\vec{r})
& \equiv \sum_{j < k} \kappa_{kj}
         \big( \hat{A}' \hat{B} + \hat{A} \hat{B}' \big) 
         \Re \big\{ \tilde{\varrho}_{jk}(\vec{r},\vec{r}') \big\} \Big|_{\vec{r} = \vec{r}'} \, , 
   \\
\label{eq:def:Cdenpair} 
\tilde{C}_q^{A, B} (\vec{r}) 
& \equiv \sum_{j < k} \kappa_{kj}
         \big( \hat{A}' \hat{B} + \hat{A} \hat{B}' \big) 
         \Im \big\{ \tilde{\varrho}_{jk}(\vec{r},\vec{r}') \big\} \Big|_{\vec{r} = \vec{r}'} \, ,  
   \\
\label{eq:def:Dsvecpair}
\tilde{D}^{A, B\sigma}_{q,\mu} (\vec{r}) 
& \equiv \sum_{j < k} \kappa_{kj} 
         \big( \hat{A}' \hat{B} - \hat{A} \hat{B}' \big) 
         \Re \big\{ \tilde{\varsigma}_{jk,\mu}(\vec{r},\vec{r}') \big\}  \Big|_{\vec{r} = \vec{r}'} \, , 
   \\
\label{eq:def:Csvecpair}
\tilde{C}^{A, B\sigma}_{q,\mu} (\vec{r}) 
& \equiv \sum_{j < k} \kappa_{kj}
         \big( \hat{A}' \hat{B} - \hat{A} \hat{B}' \big) 
         \Im \big\{ \tilde{\varsigma}_{jk,\mu}(\vec{r},\vec{r}') \big\}  \Big|_{\vec{r} = \vec{r}'} \, .
\end{align}
As done above for normal densities, the possible Cartesian tensor components of the operators
$\hat{A}$ and $\hat{B}$ are suppressed for sake of a compact generic notation. 
Features of these  definitions are
\begin{itemize}
\item
Unlike the local normal densities defined through Eqs.~\eqref{eq:def:Dden}--\eqref{eq:def:Csvec},
the local pair densities cannot be considered a priori to be real under all circumstances. 
They are, however, defined such that only the elements of $\kappa_{kj}$ are complex, not the 
spatial functions they multiply. 
\item
For many situations of practical interest, such as static calculations that conserve 
some anti-linear anti-hermitian symmetry \cite{RyssensPhD}, there exists a 
specific gauge in which the anomalous density matrix $\kappa_{ij}$ can be chosen
to be real in the center-of-mass frame. This is the choice that is made in the 
majority of numerical codes for static self-consistent mean-field calculations.
In such case, all local pair densities defined through 
Eqs.~\eqref{eq:def:Ddenpair}--\eqref{eq:def:Csvecpair} are real and have well-defined
behavior under time-reversal, see Eqs.~\eqref{eq:def:Ddenpair:T}--\eqref{eq:def:Csvecpair:T}
in what follows. 
\item
In one way or another, all conserved spatial symmetries of the single-particle 
states in the canonical basis of an HFB calculation are transferred to the local pair 
densities as defined through Eqs.~\eqref{eq:def:Ddenpair}--\eqref{eq:def:Csvecpair}, 
such that the local pair densities respect the same set of spatial symmetries as the
local normal densities in a symmetry-restricted HFB code, see Appendix~\ref{app:symmetries}.
\item
Contrarily to the normal densities, the distinction between $\tilde{D}$ and $\tilde{C}$ 
objects has to be made through the real and imaginary parts of $\tilde{\varrho}_{jk}(\vec{r},\vec{r}')$ 
and $\tilde{\varsigma}_{jk,\mu}(\vec{r},\vec{r}')$, instead of the sum or difference
of the gradient operators. Because of the symmetries of $\kappa_{jk}$, $\tilde{\varrho}_{jk}(\vec{r},\vec{r}')$ 
and $\tilde{\varsigma}_{jk,\mu}(\vec{r},\vec{r}')$ under exchange of single-particle labels and 
positions, summing up a pair density with $\big( \hat{A}' \hat{B} - \hat{A} \hat{B}' \big)$ 
acting on $\tilde{\varrho}_{jk}(\vec{r},\vec{r}')$ gives zero when setting $\vec{r} = \vec{r}'$, 
as does doing the same with $\big( \hat{A}' \hat{B} + \hat{A} \hat{B}' \big)$ acting on
$\tilde{\varsigma}_{jk,\mu}(\vec{r},\vec{r}')$.
\item
The skew-symmetry of $\kappa_{jk}$, 
$\big( \hat{A}' \hat{B} + \hat{A} \hat{B}' \big) \tilde{\varrho}_{jk}(\vec{r},\vec{r}')$, 
and $\big( \hat{A}' \hat{B} - \hat{A} \hat{B}' \big) \, \tilde{\varsigma}_{jk,\mu}(\vec{r},\vec{r}')$ 
under exchange of $j$ and $k$ has been used to reduce the summations in 
Eqs.~\eqref{eq:def:Ddenpair}--\eqref{eq:def:Csvecpair}
to just half of the possible combinations of single-particle levels $j$ and $k$, 
which cancels an implicit factor factor $1/2$ from the symmetrization or anti-symmetrization 
of the operators $\hat{A}$ and $\hat{B}$.
\item
Unless $\hat{A} = \hat{B}$, the sums and differences of these operators in 
Eqs.~\eqref{eq:def:Ddenpair}--\eqref{eq:def:Csvecpair} cannot be further simplified. With that,
the definitions of the pair densities are much closer to Eqs.~\eqref{eq:def:Dden:2}---\eqref{eq:def:Csvec:2}
than to the more efficient Eqs.~\eqref{eq:def:Dden}---\eqref{eq:def:Csvec} for normal densities.
\end{itemize}
%


\subsection{Useful relations}

In static calculations, all local normal densities $D$ and $C$ are real by 
construction, which implies that they all have definite behavior under time-reversal. 
One can easily show that, for arbitrary combinations of derivative operators 
$\hat{A}$ and $\hat{B}$,  
\begin{alignat}{2}
\big[D_q^{A, B}\big]^T(\vec{r})       &= + D_q^{A, B}(\vec{r})       \, ,  \\
\big[C_q^{A, B}\big]^T(\vec{r})       &= - C_q^{A, B}(\vec{r})       \, ,   \\
\big[D_q^{A, B\sigma}\big]^T(\vec{r}) &= - D_q^{A, B\sigma}(\vec{r}) \, ,   \\
\big[C_q^{A, B\sigma}\big]^T(\vec{r}) &= + C_q^{A, B\sigma}(\vec{r}) \, .
\end{alignat}
Provided the anomalous density matrix $\kappa_{ij}$ can be chosen to be real, 
analogue equations also hold for the local pair densities
\begin{alignat}{2}
\label{eq:def:Ddenpair:T} 
\big[\tilde{D}_q^{A, B}\big]^T(\vec{r})       &= + \tilde{D}_q^{A, B}(\vec{r})       \, ,  \\
\label{eq:def:Cdenpair:T} 
\big[\tilde{C}_q^{A, B}\big]^T(\vec{r})       &= - \tilde{C}_q^{A, B}(\vec{r})       \, , \\
\label{eq:def:Dsvecpair:T}
\big[\tilde{D}_q^{A, B\sigma}\big]^T(\vec{r}) &= - \tilde{D}_q^{A, B\sigma}(\vec{r}) \, , \\
\label{eq:def:Csvecpair:T}
\big[\tilde{C}_q^{A, B\sigma}\big]^T(\vec{r}) &= + \tilde{C}_q^{A, B\sigma}(\vec{r}) \, .
\end{alignat}
The local densities defined here have a few more practical properties. The 
first is that they exhibit definite signs under the exchange of left and right operators
\begin{alignat}{2}
D^{A, B}_q(\vec{r}) &= + D^{B, A}_q(\vec{r}) \, , &
C^{A, B}_q(\vec{r}) &= - C^{B, A}_q(\vec{r}) \, ,  
\nonumber \\
D^{A, B\sigma}_q(\vec{r}) &= + D^{B, A\sigma}_q(\vec{r}) \, , & \, 
C^{A, B\sigma}_q(\vec{r}) &= - C^{B, A\sigma}_q(\vec{r}) \, , \nonumber \\
\tilde{D}^{A, B}_q(\vec{r}) &= + \tilde{D}^{B, A}_q(\vec{r}) \, , \quad  &
\tilde{C}^{A, B}_q(\vec{r}) &= + \tilde{C}^{B, A}_q(\vec{r}) \, ,  
\nonumber \\
\tilde{D}^{A, B\sigma}_q(\vec{r}) &= - \tilde{D}^{B, A\sigma}_q(\vec{r}) \, ,  &
\tilde{C}^{A, B\sigma}_q(\vec{r}) &= - \tilde{C}^{B, A\sigma}_q(\vec{r}) \, , 
\nonumber \\
\label{eq:symmetry}
\end{alignat}
where it is implied that the tensor indices in the subscripts need to be 
exchanged accordingly.
These relations are consequences of the symmetries of the non-local densities under
exchange of the positions $\vec{r}$ and $\vec{r}'$, Eqs.~\eqref{eq:nonlocalphsym} 
and \eqref{eq:nonlocalppsym}. They imply that some densities and currents 
are automatically zero when the generating operators $\hat{A}$ and $\hat{B}$ are equal,
examples being $C^{A, A}_q(\vec{r}) = C^{A, A\sigma}_q(\vec{r}) = 0$ and
$\tilde{D}^{A, A\sigma}_q(\vec{r}) = \tilde{C}^{A, A\sigma}_q(\vec{r}) = 0$. The latter,
however is only found when not mixing protons and neutrons~\cite{Perlinska04,Rohozinski10a}.

Another set of relations that will be useful for the identification of 
reducible densities in Sec.~\ref{sec:reducibility},
are the following equalities, 
\begin{align}
\label{eq:recoupleD}
D^{\nabla A, B}_q(\vec{r}) 
& = \nabla D^{A, B}_q(\vec{r}) - D^{A,\nabla B}_q(\vec{r}) \, , \\
\label{eq:recoupleC}
C^{\nabla A, B}_q(\vec{r}) 
& = \nabla C^{A, B}_q(\vec{r}) - C^{A,\nabla B}_q(\vec{r}) \, .
\end{align}
Equations~\eqref{eq:recoupleD} and \eqref{eq:recoupleC} follow directly from the 
chain rule for derivatives and 
analogous relations hold for the spin, pair and spin pair 
densities and currents.


\subsection{Dictionary}
\label{sec:dictionary}
Having explored the consequences of this new notation and the sometimes
new definitions that it implies, we are now in a position to rewrite the 
densities employed in Ref.~\cite{Becker17} using the new notation.
We have chosen the letters ``D'' and ``C'' for these objects as shorthands for 
\emph{density} and \emph{current} respectively. For normal densities, 
the $C$ objects, Eqs.~\eqref{eq:def:Cden} and \eqref{eq:def:Csvec} do indeed 
resemble currents. For example, $C^{1, \nabla}_{\mu}(\vec{r})$ can be 
identified with
\begin{align}
C^{1, \nabla}_{\mu} (\vec{r})
& = \Im  \big\{ \nabla_{\mu} \rho_q(\vec{r}, \vec{r}')  \big\}  
    \Big|_{\vec{r} = \vec{r}'}
    \nonumber\\
& = -\tfrac{\iunit}{2} \big[   \nabla_{\mu} \rho_q(\vec{r}, \vec{r}') 
                     - \nabla_{\mu} \rho^*_q(\vec{r}, \vec{r}') \big]
    \Big|_{\vec{r} = \vec{r}'} 
    \nonumber\\
& = -\tfrac{\iunit}{2} \big( \nabla_{\mu} - \nabla_{\mu}' \big) \, \rho_q(\vec{r}',\vec{r})
    \Big|_{\vec{r} = \vec{r}'} 
    \nonumber \\
& = j_{\mu}(\vec{r})  \, .
\end{align}
Similarly, all other local densities used in Eq.~\eqref{eq:originalN2LO}
and elsewhere in the literature can be unambiguously expressed in the
new notation.
At leading and next-to-leading order, every normal density can be identified 
with exactly one $D$ or $C$ object, matching their generating operator structure
\begin{align}
\label{eq:dictionaryNLO:ph}
\rho_q(\vec{r})        &  \rightarrow D^{1,1}_q(\vec{r})            \, ,    
                       &
s_{q,\mu}(\vec{r})     &  \rightarrow D^{1,\sigma}_{q,\mu}(\vec{r}) \, ,  \nonumber \\
\tau_q(\vec{r})        &  \rightarrow D_q^{(\nabla,\nabla)}(\vec{r}) \, ,
                       & 
T_{q,\mu}(\vec{r})     &  \rightarrow D^{(\nabla,\nabla) \sigma}_{q,\mu}(\vec{r}) \, , \nonumber\\
                    &&
F_{q,\mu}(\vec{r})     &  \rightarrow D_{q,\mu}^{\nabla,(\nabla \sigma)}(\vec{r}) \, ,  \nonumber\\
j_{q,\mu}(\vec{r})     &  \rightarrow C^{1,\nabla}_{q,\mu}(\vec{r})  \, ,
                       & 
J_{q,\mu \nu}(\vec{r}) &  \rightarrow C^{1,\nabla \sigma }_{q,\mu \nu}(\vec{r}) \, . 
\end{align}
As $\tilde{\rho}_q(\vec{r})$, $\tilde{J}_{q,\mu\nu}(\vec{r})$, and 
$\tilde{\tau}_q(\vec{r})$ are all split in two separate objects, 
there is no one-to-one correspondence between the traditional and the 
new notation. Instead, one finds
\begin{align}
\tilde{\rho}_q(\vec{r}) 
& \rightarrow \tilde{D}^{1,1}_q(\vec{r})   + \iunit \, \tilde{C}^{1,1}_q(\vec{r})    \, ,  \nonumber  \\
\tilde{\tau}_q(\vec{r})
& \rightarrow \tilde{D}^{(\nabla,\nabla)}_q(\vec{r})  + \iunit \, \tilde{C}^{(\nabla,\nabla)}_q(\vec{r})    \, ,  \nonumber \\
\tilde{J}_{q, \mu \nu}(\vec{r}) 
&  \rightarrow \tilde{C}^{1,\nabla \sigma}_{q, \mu \nu}(\vec{r}) - \iunit \, \tilde{D}^{1,\nabla \sigma}_{q, \mu \nu}(\vec{r})   \, .
\label{eq:dictionaryNLO:pp}
\end{align}
The additional sign that accompanies $\tilde{D}^{1,\nabla \sigma}_{q, \mu \nu}(\vec{r})$, 
follows from the symmetry~\eqref{eq:varsigma:exchange} of the $S=1$ two-body wave function
$\tilde{\varsigma}_{jk,\mu}(\vec{r},\vec{r}')$.
The separation of the traditional pair densities into a $\tilde{D}$ and $\tilde{C}$ object 
separates the real and imaginary parts of the operators acting on the two-body wave
functions $\tilde{\varrho}_{jk}(\vec{r},\vec{r}')$ and $\tilde{\varsigma}_{jk,\mu}(\vec{r},\vec{r}')$
in Eqs.~\eqref{eq:def:Ddenpair}--\eqref{eq:def:Csvecpair}. But the $\tilde{D}$ and $\tilde{C}$ objects
still can be complex when $\kappa_{jk}$ entering the same equations is complex.

For four of the normal densities that appear in the N2LO functional
of \cite{Becker17}, Eq.~\eqref{eq:originalN2LO}, that is
$Q_q(\vec{r}) $, $V_{q, \mu\nu}(\vec{r}) $, $\vec{S}_q(\vec{r}) $ and 
$\vec{\Pi}_q(\vec{r}) $, the mapping is one-to-one as well:
\begin{align}
\label{eq:dictionaryN2LO:ph}
Q_q(\vec{r})          & \rightarrow D_q^{\Delta,\Delta}(\vec{r}) \, ,                                          
                      & 
S_{q,\mu}(\vec{r})    & \rightarrow D^{\Delta,\Delta \sigma}_{q,\mu}(\vec{r}) 
\nonumber \, ,\\
\Pi_{q,\mu}(\vec{r})    & \rightarrow C^{(\nabla, \nabla) \nabla}_{q,\mu}(\vec{r}) \, ,
                        &
V_{q, \mu \nu}(\vec{r}) & \rightarrow C^{(\nabla, \nabla) \nabla \sigma}_{q,\mu \nu}(\vec{r}) \, .
\end{align}
The densities $\tau_{q, \mu \nu}(\vec{r}) $ and $K_{q, \mu \nu \kappa}(\vec{r})$
that also appear in the N2LO functional of Ref.~\cite{Becker17}, Eq.~\eqref{eq:originalN2LO},
can only be written as a sum of a $D$ and a $C$ object each, reflecting the fact 
that they are neither time-even nor time-odd. We have for these
\begin{align}
\tau_{q,\mu \nu}(\vec{r})      & \rightarrow 
    D^{\nabla,\nabla}_{q,\nu \mu} (\vec{r}) 
+ \iunit \, C^{\nabla,\nabla}_{q,\nu \mu} (\vec{r}) \, ,  \label{eq:dicttau}\\
K_{q, \mu \nu \kappa}(\vec{r}) & \rightarrow 
    D^{\nabla,\nabla \sigma}_{q,\nu \mu \kappa}(\vec{r})  
+ \iunit \, C^{\nabla,\nabla \sigma}_{q,\nu \mu \kappa}(\vec{r})\label{eq:dictK}  \, ,
\end{align}
where Eq.~\eqref{eq:recoupleC} can be used to further reduce 
the currents
\begin{align}
C^{\nabla,\nabla}_{q,\nu \mu} (\vec{r}) 
&= \tfrac{1}{2} \, \big[ \nabla_\nu C^{1,\nabla}_{q,\mu} (\vec{r}) 
 - \nabla_\mu C^{1,\nabla}_{q,\nu} (\vec{r}) \big]   \, , \label{eq:reducetau}\\
C^{\nabla,\nabla \sigma}_{q,\nu \mu \kappa}(\vec{r}) 
&=  \tfrac{1}{2} \, \big[ \nabla_\nu C^{1,\nabla \sigma}_{q,\mu} (\vec{r}) 
 - \nabla_\mu C^{1,\nabla \sigma}_{q,\nu} (\vec{r}) \big] \, . \label{eq:reduceK}
\end{align}
Eq.~\eqref{eq:reducetau} is nothing but Eq.~\eqref{eq:reducible_example}
in the new notation. This type of formal transformation will be discussed from a
more general point of view in Sec.~\ref{sec:reducibility}.
Note that the indices $\mu \nu$ in Eqs.~\eqref{eq:dicttau} and \eqref{eq:dictK}
are inverted on the
right hand side as compared to the left hand side: the ordering of the
indices in the traditional notation for $\tau_{q, \mu \nu}(\vec{r})$ and
$K_{q, \mu \nu \kappa}(\vec{r})$ as defined through Eqs.~\eqref{eq:def:taun2} 
and~\eqref{eq:def:Kn2} is based on the ordering of the spatial and 
spin arguments of the non-local density matrices on the left-hand side of
Eqs.~\eqref{eq:def:nonlocaldensity} and~\eqref{eq:def:nonlocalpair}, such that 
one starts with the indices of the components of the operators acting on
non-primed coordinates followed by the indices of the components of the 
operators acting on primed coordinates. Our notation is based on the ordering 
of the spatial and spin arguments of creation and annihilation operators in 
the matrix elements on the right-hand side of Eqs.~\eqref{eq:def:nonlocaldensity} 
and~\eqref{eq:def:nonlocalpair}, such that it starts with the indices labeling
the Cartesian tensor components of the operator $\hat{A}$ followed by the indices 
labeling the tensor components of the operator $\hat{B}$.

The N2LO pair densities are complex in general as well, and also need to be
replaced by a combination of $\tilde{D}$ and $\tilde{C}$ objects:
\begin{align}
\tilde{\tau}_{q, \mu \nu}(\vec{r}) 
& \rightarrow   \tilde{D}^{\nabla, \nabla}_{q,\nu \mu}(\vec{r}) 
              +\iunit \, \tilde{C}^{\nabla, \nabla}_{q,\nu \mu}(\vec{r}) \, , \nonumber \\
\tilde{V}_{q, \mu \nu}(\vec{r}) 
& \rightarrow \tilde{C}^{(\nabla, \nabla) \nabla \sigma}_{q,\mu \nu}(\vec{r}) 
  - \iunit \, \tilde{D}^{(\nabla, \nabla) \nabla \sigma}_{q,\mu \nu}(\vec{r}) \, , \nonumber \\
\tilde{Q}_q(\vec{r})          & \rightarrow 
    \tilde{D}_q^{\Delta,\Delta}(\vec{r}) 
+  \iunit \, \tilde{C}_q^{\Delta,\Delta}(\vec{r}) \, ,  
\label{eq:dictionaryN2LOpairing}
\end{align}
where we emphasize again the reordering of indices for 
$\tilde{\tau}_{q, \mu \nu}(\vec{r})$ as for $\tau_{q,\mu \nu}(\vec{r})$.
The various $\tilde{D}$ or $\tilde{C}$ objects used to express pair densities have 
the same operator structure as the homologous normal densities, but they do 
not have the same interpretation in terms of what is usually called density 
and current. The reason is that the definition of pair densities sums
up the local parts of two-body wave functions that provide an efficient 
factorization of specific traces of the two-body density matrix of a paired 
quasiparticle vacuum and that enter the pairing energy as obtained from 
contact interactions \cite{Dobaczewski96}. While the local normal densities 
are truly one-body objects that do not only factorize the particle-hole part of 
the N2LO EDF (and some other many-body operators), but also can be used 
to directly calculate numerous one-body observables, the pair densities 
only appear in matrix elements that are of two-body (or higher) nature.

As a conclusion to this section we see four main advantages of the $D(\vec{r})$, 
$C(\vec{r})$ notation for normal and pair densities as compared to the 
extension of the historical notations to higher-order in gradients. These advantages are:
\begin{enumerate}
\item \textbf{Extensibility}: 
      the notation can be extended indefinitely, up to any order in derivatives.
\item \textbf{Clarity}: the operator structure of any given density is 
      incorporated directly into the notation.
\item \textbf{Reality}: By construction, all normal $D(\vec{r})$ and $C(\vec{r})$
      objects are real spatial functions when doing single-reference EDF
      calculations. Similarly, all pair densities $\tilde{D}(\vec{r})$ and 
      $\tilde{C}(\vec{r})$ are real whenever $\kappa_{ij}$ can 
      be chosen to be real, which is the case for the vast majority of use cases
      of interest for static mean-field calculations.
\item \textbf{Time-reversal}: $D(\vec{r})$ and $C(\vec{r})$ objects have definite 
      behavior under time-reversal: they are either time-even or time-odd, never mixed.
\end{enumerate}
Taken together, these aspects make the $D(\vec{r})$ and $C(\vec{r})$ notation significantly 
easier to work with for functionals at N2LO and higher order. The new 
notation eliminates the need for ever-expanding lists of letters for densities, 
whose precise definition and symmetry properties can easily be confused both in writing 
and in practical implementations. The same advantages also carry over to the 
definition and notation of local potentials entering the single-particle Hamiltonian
and the pair potentials, which will be addressed in Sec.~\ref{sec:singleh}
and Sec.~\ref{sec:singleDelta}, respectively.


\subsection{Advantageous choices of densities}
\label{sec:choice}


\subsubsection{Redundancy and reducibility}
\label{sec:reducibility}

The recoupling relations, Eqs.~\eqref{eq:recoupleD} and \eqref{eq:recoupleC} 
have striking implications on the redundancy and reducibility of the local 
densities. To start, we consider a spinless normal density $D$, with 
$a+b=N$, $N>0$, total gradient operators, 
which are split into $a$ gradients acting on the primed coordinate and $b$ gradients
acting on the unprimed coordinate. If $a$ is non-zero, through repeated 
application of Eq.~\eqref{eq:recoupleD} we get
\begin{widetext}
\begin{align}
\label{eq:redundancy_a}
D^{\nabla^a, \nabla^b}_{q,\mu_1 \mu_2 \ldots \mu_{a+b}}(\vec{r}) = 
  (-1)^{a}  D^{1, \nabla^{a+b}}_{q,\mu_1 \mu_2 \ldots \mu_{a+b}}(\vec{r}) 
 + \sum_{n=1}^{a} (-1)^{n+1} \left( \prod_{k=1}^{n} \nabla_{\mu_k} \right)
D^{\nabla^{a-n}, \nabla^b}_{q, \mu_{n+1} \ldots \mu_{a+b}}(\vec{r}) 
\, . 
\end{align}
If $b$ is non-zero, we have that
\begin{align}
\label{eq:redundancy_b}
D^{\nabla^a, \nabla^b}_{q,\mu_1 \mu_2 \ldots \mu_{a+b}}(\vec{r})  &= 
 (-1)^{b}  D^{\nabla^{a+b}, 1}_{q,\mu_1 \mu_2 \ldots \mu_{a+b}}(\vec{r})
 + \sum_{n=1}^{b} (-1)^{n+1} \left( \prod_{k=0}^{n-1} \nabla_{\mu_{a+b-k}} \right) 
D^{\nabla^{a}, \nabla^{b-n}}_{q, \mu_1 \mu_2 \ldots \mu_{a+b-n}}(\vec{r})  \, . 
\end{align}
\end{widetext}
If both $a$ and $b$ are non-zero, we can sum Eqs.~\eqref{eq:redundancy_a}
and~\eqref{eq:redundancy_b}, and use Eq.~\eqref{eq:nonlocalphsym} to obtain
\begin{align}
\label{eq:eqredundant}
2 & D^{\nabla^a, \nabla^b}_{q,\mu_1 \mu_2 \ldots \mu_{a+b}}(\vec{r}) 
   \nn \\
& = \big[ (-1)^{a} + (-1)^{b} \big] D^{1, \nabla^{a+b}}_{q,\mu_1 \mu_2 \ldots \mu_{a+b}}(\vec{r})  
    + \cdots \, , 
\end{align}
where we have dropped terms involving derivatives of lower order densities.
When $a+b$ is an odd number, the first term on the rhs in this equation 
vanishes, meaning that $D^{\nabla^a, \nabla^b}(\vec{r})$ can be rewritten 
in terms of gradients of lower-order densities. Hence, if the total 
number of derivatives in $D^{A, B}(\vec{r})$ is odd, the density is 
reducible. This result is valid as well when either $a$ or $b$ (but not both) 
are zero: in that case either Eq.~\eqref{eq:redundancy_a} or 
Eq.~\eqref{eq:redundancy_b} implies this result by itself.

Arguments along the same line hold for currents, spin densities and
spin currents, as well as all types of pair densities and currents,
provided (i) one uses the correct symmetry relations for
the exchange of $\vec{r}$ and $\vec{r}'$ for the underlying non-local densities,
Eqs.~\eqref{eq:nonlocalphsym} and \eqref{eq:nonlocalppsym}, and (ii) one does
not use any index corresponding to a Pauli $\sigma$ matrix in the recoupling. 
The following statements summarize the reducibility of densities in terms of the
number of derivatives in the operator structure: 
\begin{align}
\text{If } a+b \text{ is \textbf{odd}, then}
\left\{
\begin{tabular}{l}
 $D_q^{\nabla^a, \nabla^b}(\vec{r})$ \\
 $D_q^{\nabla^a, \nabla^b\sigma}(\vec{r})$ \\
 $\tilde{D}_q^{\nabla^a, \nabla^b}(\vec{r})$ \\
 $\tilde{C}_q^{\nabla^a, \nabla^b}(\vec{r})$ \\
\end{tabular}
\right\} \text{is reducible.}  \nonumber \\
\text{If }a + b\text{ is \textbf{even}, then}
\left\{
\begin{tabular}{l}
 $C_q^{\nabla^a, \nabla^b}(\vec{r})$ \\
 $C_q^{\nabla^a, \nabla^b\sigma}(\vec{r})$ \\
 $\tilde{D}_q^{\nabla^a, \nabla^b \sigma}(\vec{r})$ \\
 $\tilde{C}_q^{\nabla^a, \nabla^b \sigma}(\vec{r})$ \\
\end{tabular}
\right\} \text{is reducible.}  \nonumber
\end{align}
It is thus possible to see immediately from the notation whether a density 
is reducible, and whether it can be eliminated from the definition of the 
functional. Eq.~\eqref{eq:reducible_example}, which we used as an example of 
reducibility, is a particular case of these more general considerations.

Eq.~\eqref{eq:eqredundant} also implies a strong result with respect to the 
redundancy of any set of local densities: one out of any pair of densities 
$D^{\nabla^a, \nabla^b}$, $D^{\nabla^{a'}, \nabla^{b'}}$ of equal tensor rank 
is redundant, 
if the total number of gradients is equal, $a+b = a'+b'$. In other words, we 
need only pick at most a single spinless $D$ object at any given number of 
derivatives to allow for the most general EDF. As the analogue of Eq.~\eqref{eq:eqredundant} 
holds for the $C$ objects as well, we need at most one 
spinless $C$ object at that order; all others will automatically be redundant. 
The same reasoning holds for densities
and currents with Pauli matrices. From this argument follows that going from 
N$(\ell-1)$LO to N$\ell$LO in the Skyrme EDF one has to introduce at most
four completely new normal densities: a scalar $D$ and a pseudoscalar 
$D^{\sigma}_{\mu}$ with 
$2\ell$ gradients, as well as a vector $C_{\mu}$ and a pseudotensor
$C^{\sigma}_{\mu\nu}$ with $2\ell-1$ gradients. Similarly, to extend an  
N$(\ell-1)$LO EDF to N$\ell$LO, one needs at most four different pair 
densities: a $\tilde{D}$ and a $\tilde{C}$ with $2\ell$ gradients, together
with a $\tilde{D}_{\mu}^{\sigma}$ and a $\tilde{C}_{\mu}^{\sigma}$ with 
$2\ell-1$ gradients, provided protons and neutrons are not mixed.

Combining the reducibility and the redundancy argument, we can simply enumerate
the local densities we need, as we will do below.


\subsubsection{Computational considerations}
\label{sec:computational}

A final consideration in the selection of densities should be their efficiency
of computation and storage. At LO and NLO, there is 
no particular reason for thinking about computational efficiency: the largest 
objects in terms of storage space are $J_{q, \mu \nu}(\vec{r})$ and its pair
analogue $\tilde{J}_{q, \mu \nu}(\vec{r})$ which respectively represent 9 real
and 9 complex (18 real) spatial functions. At N2LO however, storing the 
density $K_{q, \mu \nu \kappa}(\vec{r})$ completely requires 27 complex 
functions (54 real spatial functions), which can be reduced to 18 complex 
functions (36 real functions) when exploiting its symmetries with respect to
permutations of the indices. The storage required for the local densities
inevitably grows with increasing order in derivatives; while not necessarily being
prohibitive at N2LO level yet, it is of practical interest to limit this growth
where possible.

More problematic than the storage of the densities themselves is the storage of
the derivatives of the single-particle wave functions. For the NLO Skyrme 
functional, one needs only to calculate and store the single-particle wave functions 
themselves, their gradient and finally their Laplacian, but not the full set of 
their second derivatives. For the N2LO functional as formulated in 
Ref.~\cite{Becker17} however, one needs to calculate and store all six linearly
independent combinations of second derivatives of the single-particle wave functions 
when summing up the densities.
For example, when written out in full as a sum over single-particle wave functions,
the density $V_{q, \mu \nu}(\vec{r})$ as defined in Ref.~\cite{Becker17} 
and used in Eq.~\eqref{eq:originalN2LO} is given by
\begin{align}
\label{eq:laplacianexample}
V_{q, \mu \nu}(\vec{r}) 
& = \Im \Big\{ \sum_{jk} \rho_{kj} \sum_\kappa 
\big[ \nabla_\kappa \Psi_j^\dagger (\vec{r}) \big]
\big[ \nabla_\kappa \nabla_{\mu} \hat{\sigma}_{\nu} \Psi_k (\vec{r}) \big] \Big\} \, .
\end{align}
Implementing the calculation of 
$V_{q, \mu\nu}(\vec{r}) = C^{(\nabla, \nabla) \nabla \sigma}_{q,\mu\nu}(\vec{r})$ 
through Eq.~\eqref{eq:laplacianexample} requires the calculation and storage
of the second-order derivatives of the single-particle wave functions. They 
are required as well for the calculation of the current
$\bold{\Pi}_{q}(\vec{r}) = \vec{C}^{(\nabla, \nabla) \nabla}_{q}(\vec{r})$, 
and (if included in the pair functional) for the pair density
$\tilde{V}_{q, \mu \nu}(\vec{r})$. This is a substantial amount of storage, as 
the single-particle wave functions are two-component complex-valued spinors and 
there are six independent second order derivatives to be calculated. As will
be discussed in Sec.~\ref{sec:singleh}, this matter can become even more critical when
evaluating the action of the single-particle Hamiltonian on the single-particle states,
as the terms originating from the variation of the EDF with respect to the density 
$V_{q, \mu \nu}(\vec{r})$ even require many combinations of third derivatives.
This computational issue is particularly troubling for approaches that deal with 
large single-particle model spaces. In coordinate space approaches such as ours,
calculating the derivatives on the mesh is the most time-consuming 
task~\cite{Ryssens15a} and the storage of the single-particle wave functions 
and their derivatives dwarfs all other memory requirements. 

For the N2LO functional, however, one can deduce that storage of all
second-order derivatives is not necessary when making a suitable choice
of densities. Indeed, using the recoupling relations we can 
rewrite Eq.~\eqref{eq:laplacianexample} as
\begin{align}
C^{(\nabla, \nabla) \nabla \sigma}_{q,\mu\nu}(\vec{r})
= \, & - C^{\Delta, \nabla \sigma}_{q,\mu\nu}(\vec{r})
     + \tfrac{1}{2} \big[ \Delta C^{1, \nabla \sigma}_{q, \mu\nu}(\vec{r}) \big] \nonumber \\
  &  - \tfrac{1}{2} \sum_{\kappa} \big[ \nabla_{\kappa} \nabla_{\mu} C^{1, \nabla\sigma}_{q, \kappa \nu} (\vec{r}) \big] \, .
\label{eq:laplacianexample_final}
\end{align}
A direct implementation of Eq.~\eqref{eq:laplacianexample_final} only requires 
the gradient and Laplacian of the single-particle wave functions for 
the calculation of the first term on the right hand side, while the two other 
terms are only external derivatives of densities of lower order that are 
already present in the NLO functional. Similar relations that avoid the storage 
of all second-order derivatives (other than the Laplacian) of the single-particle 
wave functions can easily be derived for 
$\bold{\Pi}_{q}(\vec{r}) = \vec{C}^{(\nabla, \nabla) \nabla}_{q}(\vec{r})$
and $\tilde{V}_{q, \mu \nu}(\vec{r})$. An implementation that uses the original 
choice of densities of Ref.~\cite{Becker17} requires the storage of ten sets
of spinors (the wave functions, three first-order derivatives and six independent
second-order derivatives) while an implementation that avoids 
$V_{q}(\vec{r}), \bold{\Pi}_q$ and $\tilde{V}_{q}(\vec{r})$ can make do with 
only half the memory: it requires only five sets of spinors. There
is of course a corresponding gain in computational time: the first implementation
requires the execution of nine derivative operations (three first-order ones and 
six second-order ones) while the second requires only six. If derivatives 
dominate the computational cost, as they do in our case, the CPU time required
by the first implementation is roughly 50\% larger than that required for
the second implementation. Our experience is that this naive estimate is 
qualitatively correct for typical mesh sizes employed for heavy nuclei. While 
the differences in requirements might seem modest, we emphasize that the 
additional computational burden can be completely eliminated at no cost, 
i.e.\ both implementations perform an \emph{equivalent} calculation.

The more general lesson to be learned here is that contractions of derivatives
should be recoupled to Laplacians wherever possible. Further balancing the number
of Laplacian operators on the ``left" (acting on $\vec{r}'$) and on the 
``right" (acting on $\vec{r}$) further helps reduce storage and computing costs.
For example, the density $D_q^{1, \Delta \Delta}(\vec{r})$ requires the 
computation and storage of the double Laplacian of the single-particle 
wave functions, but is redundant with the density $D_q^{\Delta, \Delta}(\vec{r})$
which requires storing only the single Laplacian.

While it is inevitable that the computational requirements (both CPU time and 
storage) will augment with increasing order in gradients of the Skyrme interaction,
we believe that it should not be increased more than necessary, and informed
choices of local densities can help with that, especially where the derivatives of 
single-particle wave functions are concerned.


\subsubsection{Comparison to other schemes}

The definitions and notations proposed above in terms of Cartesian tensor densities
are mnemonic, systematic and extensible, and therefore allow for expressing any local 
EDF of arbitrary order in gradients. To the best of our knowledge, the only other 
framework that offers the same features is the formulation of the local 
EDF in terms of spherical tensor densities proposed and employed in 
Refs.~\cite{Carlsson08,Carlsson10a,Raimondi11a}.
While both schemes can be equivalently used to express the same physics, using one
or the other can make an enormous practical difference for formal and numerical applications. 
In a framework that is based on spherical tensors, the coupling and recoupling of gradients 
follows the rules of angular-momentum coupling. This makes formal manipulations 
less transparent than a Cartesian framework, and leads to numerous straightforward, 
but cumbersome, angular-momentum coupling coefficients that appear in the final 
expressions.\footnote{The relation between the definition of the coupling constants
of the Cartesian formulation of the generating N2LO pseudopotential $\hat{V}^{\rm C}_{\rm N2LO}$
of Eq.~\eqref{eq:def:Vsk} that is used here and its representation in the spherical tensor 
framework of Refs.~\cite{Carlsson08,Carlsson10a,Raimondi11a} can be found in 
Ref.~\cite{Davesne13}.} 
When working with spherical tensors, there is a natural preference for 
a stretched coupling of gradient operators, as this automatically ensures that 
the resulting 
densities to be non-redundant and irreducible~\cite{Carlsson08}. These densities, 
however, might not be the computationally most advantageous ones.

While a framework using spherical tensors can be naturally
applied to systems that exhibit spherical symmetry, it can become cumbersome to 
use in numerical codes for deformed nuclei. One reason is that the higher-rank 
spherical tensors inevitably combine gradients into different Cartesian
directions, because of the inherent definition of spherical vector components~\cite{Carlsson08}. 
This can lead to the computational inconveniences sketched in Sect.~\ref{sec:application}
when several gradients are present in a given term. As can be deduced from 
Table~XXI of Ref.~\cite{Carlsson08}, there is not always a one-to-one correspondence 
between the spherical tensor densities as defined there and advantageous choices for
cartesian tensor densities. The spherical tensor framework of 
Refs.~\cite{Carlsson08,Carlsson10a,Raimondi11a} also automatically treats the 
various irreducible representations of high-rank tensors as different objects, which 
increases the number of densities and potentials that have to be tracked in 
calculations for deformed nuclei where all irreducible representations are non-zero.
In such situation it is more efficient to use the full tensor instead of its decomposition,
as done in the Cartesian scheme presented above.


\section{The N2LO energy density functional revisited}
\label{sec:N2LO:revisited}


\subsection{Towards a more efficient form of the functional}
\label{sec:application}

It is now rather straightforward to enumerate a set of non-redundant, irreducible
normal densities to form the particle-hole part of the EDF at N2LO and, that 
have definite behavior under time-reversal and are guaranteed to be real. 
From the considerations above, it follows that a non-redundant set of densities
that contain a given number of derivatives contains at most 
eight different ones. Among those, however, two normal densities are always 
reducible, and two pair densities vanish by choice when assuming that protons 
and neutrons are not mixed at the level of single-particle wave functions. 

There are four densities without gradients
\begin{align}
D_q^{1,1}(\vec{r}), \quad
D^{1, \sigma}_{q,\mu}(\vec{r}), \quad
\tilde{D}^{1,1}_{q}(\vec{r}),  \quad
\tilde{C}^{1,1}_{q}(\vec{r}) \, ,
\end{align}
which are obviously unique as they do not contain gradient operators that
could be recoupled in different ways. The 
current densities $C^{1,1}_{q}(\vec{r})$ and $C_{q,\mu}^{1,\sigma}(\vec{r})$ 
always vanish for symmetry reasons, whereas $\tilde{D}_{q,\mu}^{1, \sigma}(\vec{r})$ 
and $\tilde{C}_{q,\mu}^{1, \sigma}(\vec{r})$ only have to considered when mixing
protons and neutrons.

There are four densities containing a single gradient operator that have to be
considered
\begin{align}
C^{1,\nabla}_{q,\mu}(\vec{r}) ,  \quad
C^{1, \nabla\sigma}_{q,\mu\nu}(\vec{r}),  \quad
\tilde{C}^{1,\nabla\sigma}_{q,\mu\nu}(\vec{r}),  \quad
\tilde{D}^{1,\nabla\sigma}_{q,\mu \nu}(\vec{r}) \, .
\end{align}
At this order, the densities $D^{1,\nabla}_q(\vec{r})$ and 
$D^{1,\nabla\sigma}_{q}(\vec{r})$ are reducible, whereas 
$\tilde{D}_{q}^{1, \nabla}(\vec{r})$ and $\tilde{C}_{q}^{1, \nabla}(\vec{r})$
only have to considered when mixing protons and neutrons.

With two gradients, we can again choose four irreducible densities
\begin{align}
D^{\nabla,\nabla}_{q,\mu\nu}(\vec{r}), \quad
D^{\nabla, \nabla\sigma}_{q,\mu \nu \kappa}(\vec{r}), \quad
\tilde{D}^{\nabla,\nabla}_{q,\mu \nu}(\vec{r}), \quad
\tilde{C}^{\nabla,\nabla}_{q,\mu \nu}(\vec{r}) \, .
\label{eq:choice_LO}
\end{align}

It is implied by Eq.~\eqref{eq:choice_LO} that we construct the lower-rank 
densities with two gradients as contractions of these densities with a 
Kronecker $\delta_{\mu \nu}$, such that we altogether remain very close to the 
traditional choice using the densities using $\tau_{q,\mu \nu}(\vec{r})$
and $K_{q, \mu \nu \kappa} (\vec{r})$. The main difference is that the new
definitions automatically eliminate the reducible contributions of the 
full tensor densities, see Eq.~\eqref{eq:reducible_example} for an example, 
thereby leading to real densities (gauge permitting for the pair densities).

At the level of three derivatives, we can write down four irreducible 
densities with one free gradient index
\begin{align}
C^{\Delta,\nabla}_{q,\mu}(\vec{r}), \quad
C^{\Delta, \nabla\sigma}_{q,\mu}(\vec{r}),  \quad
\tilde{D}^{\Delta,\nabla\sigma}_{q,\mu\nu}(\vec{r}), \quad
\tilde{C}_{q,\mu \nu}^{\Delta,\nabla\sigma}(\vec{r}) \, .
\end{align}
At this order our choice of normal densities differs 
entirely from Ref.~\cite{Becker17}. Note how we have systematically 
chosen the Laplacian and non-contracted gradient to act on different 
coordinates. Not doing so would significantly increase the number of 
combinations of derivatives to be calculated when constructing local 
densities and, even more, the corresponding term in the single-particle 
Hamiltonian.

Finally, with four gradients we need only scalar objects. We choose
\begin{align}
D^{\Delta,\Delta}_{q}(\vec{r}),  \quad
D^{\Delta, \Delta \sigma}_{q,\mu}(\vec{r}),  \quad
\tilde{D}^{\Delta,\Delta}_{q}(\vec{r}),  \quad
\tilde{C}^{\Delta,\Delta}_{q}(\vec{r}) \, ,
\end{align}
where we have opted to recouple all contracted gradients into Laplacians, 
balanced between primed and unprimed coordinates.

The adjustment of the only available N2LO parametrization, SN2LO1, 
did not include pairing degrees of freedom~\cite{Becker17}, so we will limit 
ourselves in what follows to the normal densities. Our choice for these can be 
summarized as
\begin{eqnarray}
\mathsf{R}_q &= \Big( D^{1,1}_q, \,
                       D_{q, \mu}^{1,\sigma}, \,
                       D^{\nabla, \nabla}_{q, \mu \nu}, \, 
                       C^{1, \nabla \sigma}_{q, \mu \nu}, \,
                       C^{1, \nabla}_{q,\mu}, \, 
                       D^{\nabla, \nabla \sigma}_{q, \mu \nu \kappa},
\nonumber \\
             & 
                       D^{\Delta, \Delta}_q,  \,
                       C^{\Delta, \nabla \sigma}_{q, \mu \nu}, \,
                       D^{\Delta, \Delta \sigma}_{q, \mu}, \,
                       C^{\Delta, \nabla}_{q, \mu}
               \Big) \, .
\label{eq:densityvector}
\end{eqnarray}
As in Ref.~\cite{Ryssens19b}, we employ the shorthand notation of a density
vector when talking about all normal local densities simultaneously. 
We index this object with latin indices, as in $\mathsf{R}_{q,a}$, to indicate 
a particular density as an element of the vector.
Of the different elements of Eq.~\eqref{eq:densityvector}, only 
$C^{\Delta, \nabla}_{q, \mu}(\vec{r})$ has not been discussed before. It is 
related to the higher-order current density 
$\bold{\Pi}_{q}(\vec{r}) = \bold{C}^{(\nabla, \nabla) \nabla }_{q}(\vec{r})$ 
as used in Ref.~\cite{Becker17} as follows
\begin{align}
C^{\Delta, \nabla}_{q, \mu}(\vec{r}) 
& =  
    -                     C^{(\nabla, \nabla)\nabla }_{q, \mu}(\vec{r})
    + \tfrac{1}{2} \big[ \Delta C^{1,\nabla}_{q, \mu} (\vec{r}) \big]
\nonumber \\
&   - \tfrac{1}{2} \sum_{\nu} \big[ \nabla_{\mu} \nabla_{\nu} C^{1,\nabla}_{q, \nu} (\vec{r}) \big] \, ,
\label{eq:rewriting_pi}
\end{align}
which is the analogue of Eq.~\eqref{eq:laplacianexample_final} without spin.

We reiterate that our choice of local densities, 
Eq.~\eqref{eq:densityvector}, is not unique, but we believe it to be one of the 
choices that both considerably simplifies the formal expressions and 
also enormously reduces the complexity of the numerical implementation of the
N2LO functional form of Ref.~\cite{Becker17}. While we do not consider 
here the expressions for N2LO pairing, tensor or spin-orbit terms, 
we expect that the choices made here will also be optimal for these
extensions of the Skyrme EDF. For terms at even higher orders in gradients, the 
guiding principles we discussed can easily be used to construct optimal sets of 
densities in such cases as well and we believe that these will be compatible
with the choices made here for densities up to N2LO.

Despite these simplifications, the practical implementation of the 
calculation of all densities discussed here is tedious. One particular aspect 
is especially error-prone: the properties of all components of all densities 
under the symmetries conserved by the implementation. For reference, we present 
such symmetry relations for the case of a 3D coordinate-space representation 
such as ours in Appendix~\ref{app:symmetries}.

We lack one more definition in order to write down the form of the functional
as used for the SN2LO1 parametrization of Ref.~\cite{Becker17} in 
the new notation and with the new choice of densities. As with the 
historical notation, we define the isoscalar and isovector densities
\begin{alignat}{1}
D_{0}^{A,B}(\vec{r}) &\equiv  
    D_{n}^{A,B}(\vec{r}) 
  + D_{p}^{A,B}(\vec{r}) \, ,  \\
D_{1}^{A,B}(\vec{r}) &\equiv  
    D_{n}^{A,B}(\vec{r}) 
  - D_{p}^{A,B}(\vec{r}) \, ,  \\
C_{0}^{A,B}(\vec{r}) &\equiv  
    C_{n}^{A,B}(\vec{r}) 
  + C_{p}^{A,B}(\vec{r}) \, ,  \\
C_{1}^{A,B}(\vec{r}) &\equiv 
    C_{n}^{A,B}(\vec{r}) 
  - C_{p}^{A,B}(\vec{r}) \, ,  
\end{alignat}
and similar for the spin densities. Note,
however, that the construction of isoscalar and isovector
pair and spin-pair densities requires additional 
considerations~\cite{Perlinska04,Rohozinski10a,Sadoudi13a}. 
In calculations that do not mix protons and neutrons as
assumed here, using isovector pair densities instead of proton and 
neutron pair densities unnecessarily complicates the formulation 
of the pairing EDF and the HFB equations and will therefore not
be addressed here.


\subsection{An alternative functional form of SN2LO1}
\label{sec:rewritten}

With these definitions at hand we now have all the tools 
to write down the LO, NLO and N2LO energy densities from 
Section~\ref{sec:formfunc} as a function of the coupling constants 
$\cc{A}^{(i,j)}_{t, \textrm{e/o}}$ and the densities:
\begin{widetext}
\begin{align}
\label{eq:SkTeven:0}
\mathcal{E}^{(0)}_{\rm Sk, e} (\vec{r})
  = & \sum_{t=0,1}
      \Big[ 
      \cc{A}_{t,\textrm{e}}^{(0,1)}    \big( D^{1,1}_t \big)^2
    + \cc{A}^{(0,2)}_{t,\textrm{e}}    \big( D^{1,1}_0 \big)^\alpha 
                                       \big( D^{1,1}_t \big)^2
      \Big] \, , 
\\
\label{eq:SkTeven:2}
\mathcal{E}^{(2)}_{\rm Sk,e} (\vec{r})
  = & \sum_{t=0,1}
      \Big[
      \cc{A}^{(2,1)}_{t,\textrm{e}} D_t^{1,1} \big( \Delta D_t^{1,1} \big)    
    + \cc{A}^{(2,2)}_{t,\textrm{e}} D_t^{1,1} D_t^{(\nabla,\nabla)} 
    + \cc{A}^{(2,3)}_{t,\textrm{e}} \sum_{\mu \nu} C^{1, \nabla \sigma}_{t, \mu \nu}
                                                   C^{1, \nabla \sigma}_{t, \mu \nu} 
    + \cc{A}^{(2,4)}_{t,\textrm{e}} D_t^{1,1} 
           \big(  \nabla \cdot \vec{C}^{1,\nabla \times \sigma}_t \big)
      \Big] \, , 
\\
\label{eq:SkTeven:4}
  \mathcal{E}^{(4)}_{\text{Sk,e}} (\vec{r})
  = & \sum_{t=0,1}
     \Big[
     \cc{A}^{(4,1)}_{t,\textrm{e}} \big(\Delta D^{1,1}_t \big) \big( \Delta D^{1,1}_t \big)
   + \cc{A}^{(4,2)}_{t,\textrm{e}} D^{1,1}_t \, D^{\Delta, \Delta}_t
   + \cc{A}^{(4,3)}_{t,\textrm{e}} D^{(\nabla, \nabla)}_t D^{(\nabla, \nabla)}_t
\nonumber \\
   & \quad
   + \cc{A}^{(4,4)}_{t,\textrm{e}} \sum_{\mu\nu} D^{\nabla, \nabla}_{t, \mu\nu}
                                                 D^{\nabla, \nabla}_{t, \mu\nu}       
   + \cc{A}^{(4,5)}_{t,\textrm{e}} \sum_{\mu\nu} D^{\nabla, \nabla}_{t,\mu\nu}
                                              \big( \nabla_{\mu} \nabla_{\nu}  D^{1,1}_t \big)
\nonumber \\
   & \quad
   +  \cc{A}^{(4,6)}_{t,\textrm{e}} \sum_{\mu\nu} C^{1,\nabla\sigma}_{t,\mu \nu} 
      \big( \Delta C^{1,\nabla\sigma}_{t,\mu \nu} \big)
   +  \cc{A}^{(4,7)}_{t,\textrm{e}}  \sum_{\mu\nu\kappa} 
            \big(       \nabla_{\mu} C^{1,\nabla\sigma}_{t,\mu \kappa} \big)
            \big(       \nabla_{\nu} C^{1,\nabla\sigma}_{t,\nu \kappa} \big)
   +  \cc{A}^{(4,8)}_{t,\textrm{e}} \sum_{\mu\nu} 
       C^{1,\nabla\sigma}_{t,\mu\nu} C^{\Delta,\nabla \sigma}_{t,\mu\nu}
\Big] \, , \\
\label{eq:SkTodd:0}
\mathcal{E}^{(0)}_{\text{Sk,o}}(\vec{r})
  = & \sum_{t=0,1}
      \Big[ 
      \cc{A}^{(0,1)}_{t,\textrm{o}}         
                 \vec{D}^{1,\sigma}_t \cdot  \vec{D}^{1,\sigma}_t
    + \cc{A}^{(0,2)}_{t,\textrm{o}} (D^{1,1}_0 )^\alpha \, 
                 \vec{D}^{1,\sigma}_t \cdot  \vec{D}^{1,\sigma}_t
      \Big] \, ,
      \\
\label{eq:SkTodd:2}
\mathcal{E}^{(2)}_{\text{Sk,o}}(\vec{r})
  = &  \sum_{t=0,1}
      \Big[
     \cc{A}^{(2,1)}_{t,\textrm{o}}   
           \vec{D}^{1, \sigma}_{t} \cdot \big(  \Delta \vec{D}^{1, \sigma}_{t} \big) 
    + \cc{A}^{(2,2)}_{t,\textrm{o}}  
            \vec{D}^{1,\sigma}_{t} \cdot         \vec{D}^{(\nabla, \nabla)\sigma}_{t} 
    + \cc{A}^{(2,3)}_{t,\textrm{o}}   
           \vec{C}^{1,\nabla}_{t}  \cdot         \vec{C}^{1,\nabla}_{t}  
    + \cc{A}^{(2,4)}_{t,\textrm{o}}    
           \vec{D}_{t}^{1,\sigma}  \cdot  \big( \vnabla \times \vec{C}^{1, \nabla}_t\big)
     \Big]
\, , 
\\
\label{eq:SkTodd:4}
\mathcal{E}^{(4)}_{\text{Sk,o}}(\vec{r})
  =  & \sum_{t=0,1}
      \Big[ 
       \cc{A}^{(4,1)}_{t,\textrm{o}}  
           \big(   \Delta \vec{D}^{1,\sigma}_t \big) \cdot \big( \Delta\vec{D}^{1, \sigma}_t \big)
  +    \cc{A}^{(4,2)}_{t,\textrm{o}}  
                      \vec{D}^{1,\sigma}_t \cdot \vec{D}^{\Delta, \Delta\sigma}_t
  +    \cc{A}^{(4,3)}_{t,\textrm{o}}  
             \vec{D}^{(\nabla, \nabla) \sigma}_t  \cdot 
             \vec{D}^{(\nabla, \nabla) \sigma}_t
\nonumber \\
&   \quad
  +    \cc{A}^{(4,4)}_{t,\textrm{o}} \sum_{\mu \nu \kappa}
             D_{\mu \nu \kappa}^{ \nabla, \nabla \sigma}
             D_{\mu \nu \kappa}^{ \nabla, \nabla \sigma}
  +    \cc{A}^{(4,5)}_{t,\textrm{o}}   \sum_{\mu \nu \kappa}
             D_{\mu \nu \kappa}^{ \nabla, \nabla \sigma} 
         \big( \nabla_{\mu} \nabla_{\nu} D_{\kappa}^{1, \sigma} \big)
 \, ,  \nonumber 
\\
&  \quad
  +    \cc{A}^{(4,6)}_{t,\textrm{o}} 
                      \vec{C}_t^{1, \nabla} \cdot \big( \Delta \vec{C}_t^{1, \nabla} \big)
  +    \cc{A}^{(4,7)}_{t,\textrm{o}} 
                 \big(\vnabla \cdot \vec{C}_t^{1, \nabla}\big)
                 \big(\vnabla \cdot \vec{C}_t^{1, \nabla}\big)
  +    \cc{A}^{(4,8)}_{t,\textrm{o}}  
              \vec{C}^{1, \nabla}_t \cdot \vec{C}^{\Delta,\nabla}_t
     \Big]
\, ,
\end{align}
\end{widetext}
where we have dropped the explicit position dependence of the densities for 
brevity. Other than being formulated in terms of $D$ and $C$ objects, 
the LO and NLO energy densities are identical to those in a traditional NLO functional. 
For the N2LO energy density, some additional work is required to rewrite
all terms using only our choice of densities. These additional steps are sketched
in Appendix~\ref{app:BeckerCD}, together with the 
relation between the four coupling constants $C^{(4)}$ in 
Eqs.~\eqref{eq:originalN2LO} and the $A^{(4,j)}_{t, \rm e/o}$. If one wants to 
link the EDF to an effective interaction, as we do 
here, the coupling constants $\cc{A}^{(i,j)}_{t/e}$ are related to each other 
and the parameters of the pseudopotential. In this case, the N2LO energy
density can be parameterized in terms of four numbers, 
$t^{(4)}_1,t^{(4)}_2,x^{(4)}_1$ and $x^{(4)}_2$. The relevant expressions for 
the coupling constants in terms of these parameters
are given in Appendix~\ref{app:coupling}. 

If the connection to an effective interaction is maintained, then this 
form of the EDF is locally gauge invariant. This in turn implies that a
continuity equation relates the divergence of the current 
$\vec{j}_q(\vec{r}) = \bold{C}_q^{1,\nabla}(\vec{r})$ to the time derivative of 
the density $\rho_q(\vec{r}) = D^{1,1}_q(\vec{r})$~\cite{Raimondi11b}
\begin{equation}
\label{eq:continuity:normal}
\frac{\partial}{\partial t} D^{1,1}_q(\vec{r},t)
= - \frac{\hbar}{m} \vnabla \cdot \vec{C}^{1,\nabla}_q(\vec{r},t) \, .
\end{equation}
Since we consider here only static calculations, Eq.~\eqref{eq:continuity:normal} 
implies that the divergence of the current density vanishes identically, as does the
penultimate term in the N2LO time-odd energy density, Eq.~\eqref{eq:SkTodd:4}. 
For locally gauge invariant EDFs, another continuity equation relates 
$C^{1, \nabla \sigma}_{q,\mu\nu}(\vec{r})$ to the time derivative of 
$D^{1,\sigma}_{q,\kappa}(\vec{r})$~\cite{Raimondi11b}
\begin{equation}
\label{eq:continuity:spin}
\frac{\partial}{\partial t} D^{1,\sigma}_{q,\kappa} (\vec{r},t)
= - \frac{\hbar}{m} \sum_\mu \nabla_\mu C^{1,\nabla \sigma}_{q,\mu \kappa} (\vec{r},t) \, ,
\end{equation}
such that the penultimate term of the 
time-even N2LO energy density, Eq.~\eqref{eq:SkTeven:4}, vanishes as well
in static calculations. We have not dropped either of those two terms 
from the expressions in this section and the following ones, as they will 
always contribute to the total energy for time-dependent 
approaches as well as for static calculations that use possible extensions
of the functional that do not conserve local gauge invariance.

If the EDF is generated as the expectation value of a \emph{density-independent} 
effective interaction without any further manipulation of coupling constants, then the 
functional is automatically self-interaction free \cite{Bender09b,Stringari78a}. 
This means that the total energy of a system composed of just one particle is its 
kinetic energy. We have checked numerically that, when the relations between 
the coupling constants of Appendix~\ref{app:coupling} hold and the density 
dependence is omitted for the purpose of such test, the contribution of the Skyrme EDF to the 
single-particle energy and the total energy is zero up to numerical noise.
Self-interaction freedom can only be tested in a non-self-consistent 
calculation, where for an arbitrarily generated localized single-particle state 
the expectation value of the single-particle Hamiltonian and the total energy are
evaluated by taking only this states' contributions into account.
Performing such test requires the breaking of rotational symmetry as 
well as time-reversal symmetry, implying that all local densities
are non-vanishing. Such calculation constitutes a powerful check that both 
Eqs.~\eqref{eq:SkTeven:0}--\eqref{eq:SkTodd:4} and
the expressions in Ref.~\cite{Becker17} have correctly been derived from the 
generating effective interaction~\eqref{eq:def:Vsk} and correctly 
been implemented in our code. We have also verified that, for such symmetry 
broken cases, the total energy calculated from the single-particle energies 
at self-consistency (see Ref.~\cite{Ryssens15a} for details) is equal 
to a direct integration of the energy density at the keV level, as is the
case for standard NLO functionals.


\subsection{Mean-field potentials and single-particle Hamiltonian}
\label{sec:singleh}

Using the density vector $\mathsf{R}$ as defined in Eq.~\eqref{eq:densityvector} 
for compact notation, the expression for the energy can be rewritten
as $E_{\rm tot}(\mathsf{R})$. The individual terms in the single-particle
Hamiltonian are then obtained by rewriting the variation of the energy
with respect to the full density matrix $\rho(\vec{r}'\sigma', \vec{r}\sigma)$ 
as the sum of variations with respect to the local densities

\begin{align}
\label{eq:variation}
h_q (\vec{r}\sigma, \vec{r}' \sigma') 
&\equiv \frac{\delta E_{\rm tot}}{\delta \rho(\vec{r}'\sigma,\vec{r} \sigma)}
\nonumber \\
&= \sum_{a} \int \! d^3r'' \,
  \frac{\delta E_{\rm tot}(\mathsf{R})}{\delta \mathsf{R}_{q,a}(\vec{r}'')}
  \frac{\delta \mathsf{R}_{q,a}(\vec{r}'')}
       {\delta \rho_q(\vec{r}' \sigma', \vec{r} \sigma)}
 \, .
\end{align}
The single-particle Hamiltonian is a $2 \times 2$ matrix in the space 
of spin $1/2$ spinors, whose matrix elements are given by
\begin{widetext}
\begin{align}
\label{eq:sphamiltonian:matrix}
h_{jk}\emb\
  = \langle \Psi_j | \hat{h}_q | \Psi_k \rangle
& = \iint \! d^3r \, d^3r' \, \sum_{\sigma \sigma'}
    \psi^*_j(\vec{r} \sigma) \, 
    h_q(\vec{r} \sigma, \vec{r}' \sigma') \,
    \psi_k(\vec{r}' \sigma') 
     \nn \\
& = \iint \! d^3r \, d^3r' \,
    \big( \psi^*_j(\vec{r} +), \psi^*_j(\vec{r} -) \big)
    \left( \begin{array}{cc} 
           h_q(\vec{r} +, \vec{r}' +) & h_q(\vec{r} +, \vec{r}' -)  \\
           h_q(\vec{r} -, \vec{r}' +) & h_q(\vec{r} -, \vec{r}' -) 
     \end{array} \right)
     \left( \begin{array}{c}
     \psi_k(\vec{r}' +) \\ 
     \psi_k(\vec{r}' -)
     \end{array} \right) \, ,
\end{align}
\end{widetext}
where we assume that the single-particle states $j$ and $k$ are of
the same nucleon species $q$.
The derivative of the energy with respect to the density 
$\mathsf{R}_{q,a}(\vec{r})$ can be identified as an associated mean-field 
potential $\mathsf{F}_{q,a}(\vec{r})$, defined as
\begin{align}
\label{eq:def:potentials:F}
\mathsf{F}_{q,a}(\vec{r})
\equiv \frac{\delta E_{\rm tot}(\mathsf{R})}{\delta \mathsf{R}_{q,a}(\vec{r})}\, .
\end{align}
Each potential $\mathsf{F}_{q,a}(\vec{r})$ depends in general on the density 
vectors $\mathsf{R}_{p}(\vec{r})$ and $\mathsf{R}_{n}(\vec{r})$ of both 
nucleon species. 

We use the letters $F$ and $G$ to distinguish between potentials
corresponding to $D$ and $C$ densities, respectively. As there are four generic 
types of local densities and currents, there are four different types of generic
potentials that are given by 
\begin{align}
\label{eq:def:Fden}
F^{AB}_q(\vec{r}'')
\equiv \, & \frac{\delta E_{\rm tot}}{\delta D^{AB}_q(\vec{r}'')} 
  =         \int \! d^3r'''' \, \frac{\delta \mathcal{E}(\vec{r''''}) }{\delta D^{AB}_q(\vec{r}'')}  
            \, ,
         \\
\label{eq:def:Gden}
G^{AB}_q(\vec{r}'')
\equiv \, & \frac{\delta E_{\rm tot}}{\delta C^{AB}_q(\vec{r}'')}  
  =         \int \! d^3r'''' \, \frac{\delta \mathcal{E}(\vec{r''''}) }{\delta C^{AB}_q(\vec{r}'')}  
            \, ,
         \\
\label{eq:def:Fsvec}
F^{AB\sigma}_q(\vec{r}'')
\equiv \, & \frac{\delta E_{\rm tot}}{\delta D^{AB\sigma}_q(\vec{r}'')} 
  =         \int \! d^3r'''' \, \frac{\delta \mathcal{E}(\vec{r''''}) }{\delta D^{AB\sigma}_q(\vec{r}'')}  
            \, ,
         \\
\label{eq:def:Gsvec}
G^{AB\sigma}_q(\vec{r}'')
\equiv \, & \frac{\delta E_{\rm tot}}{\delta C^{AB\sigma}_q(\vec{r}'')}  
  =         \int \! d^3r'''' \, \frac{\delta \mathcal{E}(\vec{r''''})}{\delta C^{AB\sigma}_q(\vec{r}'')}  
            \, .
\end{align}
The cartesian tensor structure, behavior under index exchange and symmetry properties 
carry over directly from a local density to its associated potential. The identification of 
potentials~\eqref{eq:def:Fden}--\eqref{eq:def:Gsvec} with the 
existing literature is analogous to the mapping of the densities outlined
in Sec.~\ref{sec:application}; for example $F^{1,1}_q(\vec{r})$ corresponds to 
the central potential that is commonly called $U_q(\vec{r})$.

Similar to the density vector $\mathsf{R}_q(\vec{r})$,
we define the potential vector $\mathsf{F}_q(\vec{r})$ that is composed of 
all the mean-field potentials
\begin{align}
\mathsf{F}_q &= \Big\{ F^{1,1}_q, 
                       F_{q, \mu}^{1,\sigma}, 
                       F^{\nabla, \nabla}_{q, \mu \nu}, 
                       G^{1, \nabla \sigma}_{q, \mu \nu},
                       G^{1, \nabla}_q,
                       F^{\nabla, \nabla \sigma}_{q, \mu \nu \kappa},
\nn \\
             & 
                       F^{\Delta, \Delta}_q, 
                       G^{\Delta, \nabla \sigma}_{q, \mu \nu},
                       F^{\Delta, \Delta \sigma}_{q, \mu},
                       G^{\Delta, \nabla}_{q, \mu}
               \Big\} \, .
\label{eq:potentialvector}
\end{align}
The contribution to the single-particle Hamiltonian from the term containing 
derivatives with respect to $D^{AB}_q(\vec{r}'')$ in the chain rule when deriving the energy 
with respect to $\rho_q(\vec{r}' \sigma', \vec{r} \sigma)$ is
\begin{align}
\int \! d^3r'' \,  &
\frac{\delta D^{AB}_q(\vec{r}'')}{\delta \rho_q(\vec{r}' \sigma', \vec{r} \sigma)} \, 
\frac{\delta E_{\rm tot}}{\delta D^{AB}_q(\vec{r}'')}
    \nn \\
& = \int \! d^3r'' \,
    \tfrac{1}{2} \, 
    \Big\{ \big[ \hat{A}'' \delta_{\vec{r}'' \vec{r} } \big] \,
           F^{AB}_q(\vec{r}'') \,            
           \big[ \hat{B}'' \delta_{\vec{r}'' \vec{r}'} \big]
    \nn \\
&   \quad
          +\big[ \hat{B}'' \delta_{\vec{r}'' \vec{r} } \big] \,
           F^{AB}_q(\vec{r}'') \, 
           \big[ \hat{A}'' \delta_{\vec{r}'' \vec{r}'} \big]
   \Big\} \, \delta_{\sigma \sigma'} \, .
\end{align}
In the matrix element~\eqref{eq:sphamiltonian:matrix}, 
$\big[ \hat{A}'' \delta_{\vec{r}'' \vec{r} } \big]$ and 
$\big[ \hat{B}'' \delta_{\vec{r}'' \vec{r} } \big]$ become 
``gradient operators acting to the left'' on the single-particle state 
with coordinate $\vec{r}$, whereas $\big[ \hat{B}'' \delta_{\vec{r}'' \vec{r}'} \big]$
and $\big[ \hat{A}'' \delta_{\vec{r}'' \vec{r}'} \big]$ are ``gradient 
operators acting to the right'' on the single-particle state 
with coordinate $\vec{r}'$. Through integration by parts, however, the 
gradients acting on the wave function on one side can be transferred 
to act on the potential $F^{AB}_q(\vec{r}'')$ times the wave function 
on the other side.

The latter is a necessity when one is interested in the action of
the single-particle Hamiltonian on a state that is not its eigenstate,
which is an ingredient of many schemes for the iterative diagonalization 
of the single-particle Hamiltonian, such
as the heavy-ball method used by us \cite{Ryssens19b}. This is the form
into which we will bring the expression for $h_q(\vec{r} \sigma, \vec{r}'\sigma')$.
In general, one finds
\begin{widetext}
\begin{align}
\label{eq:def:Fden:right}
\int \! d^3r''
\frac{\delta D^{AB}_q(\vec{r}'')}{\delta \rho_q(\vec{r}' \sigma', \vec{r} \sigma')} \, 
\frac{\delta E_{\rm tot}}{\delta D^{AB}_q(\vec{r}'')}
& = + \tfrac{1}{2} 
    \Big(   (-)^{n_A} \, \hat{A} \big\{ F^{AB}_q(\vec{r}) \, \big[ \hat{B} \delta_{\vec{r} \vec{r}' } \big] \big\}
          + (-)^{n_B} \, \hat{B} \big\{ F^{AB}_q(\vec{r}) \, \big[ \hat{A} \delta_{\vec{r} \vec{r}' } \big] \big\}
    \Big) \, \delta_{\sigma \sigma'}  \, ,
    \\
\label{eq:def:Gden:right}
\int \! d^3r'' \,  
\frac{\delta C^{AB}_q(\vec{r}'')}{\delta \rho_q(\vec{r}' \sigma', \vec{r} \sigma)} \, 
\frac{\delta E_{\rm tot}}{\delta C^{AB}_q(\vec{r}'')}
& = - \tfrac{\iunit}{2} \, 
    \Big( (-)^{n_A} \, \hat{A} \big\{ G^{AB}_q(\vec{r}) \, \big[ \hat{B} \delta_{\vec{r} \vec{r}' } \big] \big\}
        - (-)^{n_B} \, \hat{B} \big\{ G^{AB}_q(\vec{r}) \, \big[ \hat{A} \delta_{\vec{r} \vec{r}' } \big] \big\}
   \Big) \, \delta_{\sigma \sigma'} \, ,
   \\
\label{eq:def:Fvec:right}
\int \! d^3r'' \,  
\frac{\delta D^{AB\sigma}_q(\vec{r}'')}{\delta \rho_q(\vec{r}' \sigma', \vec{r} \sigma)} \,
\frac{\delta E_{\rm tot}}{\delta D^{AB\sigma}_q(\vec{r}'')}
& = + \tfrac{1}{2} 
    \Big(   (-)^{n_A} \, \hat{A} \big\{ F^{AB\sigma}_q(\vec{r}) \, \big[ \hat{B} \delta_{\vec{r} \vec{r}' } \big] \big\}
          + (-)^{n_B} \, \hat{B} \big\{ F^{AB\sigma}_q(\vec{r}) \, \big[ \hat{A} \delta_{\vec{r} \vec{r}' } \big] \big\}
    \Big) \, \langle \sigma | \hat{\sigma} | \sigma' \rangle \, ,
    \\
\label{eq:def:Gvec:right}
\int \! d^3r'' \,  
\frac{\delta C^{AB\sigma}_q(\vec{r}'')}{\delta \rho_q(\vec{r} \sigma, \vec{r}' \sigma')} \, 
\frac{\delta E_{\rm tot}}{\delta C^{AB\sigma}_q(\vec{r}'')}
& =  - \tfrac{\iunit}{2} \, 
    \Big( (-)^{n_A} \, \hat{A} \big\{ G^{AB\sigma}_q(\vec{r}) \, \big[ \hat{B} \delta_{\vec{r} \vec{r}' } \big] \big\}
        - (-)^{n_B} \, \hat{B} \big\{ G^{AB\sigma}_q(\vec{r}) \, \big[ \hat{A} \delta_{\vec{r} \vec{r}' } \big] \big\}
    \Big) \, \langle \sigma | \hat{\sigma} | \sigma '\rangle  \, ,
\end{align}
\end{widetext}
where $n_A$ and $n_B$ are the order in gradient operators of $\hat{A}$ and $\hat{B}$, respectively. 
Using these generic expressions, the calculation of the single-particle Hamiltonian can be automatized, 
both for its matrix elements as well as for its action on a given single-particle state.

The considerations concerning the often significant differences in computational 
cost of alternative choices for local densities sketched in Sec.~\ref{sec:computational} 
also apply to the numerical application of the 
contributions~\eqref{eq:def:Fden:right}--\eqref{eq:def:Gvec:right} to the single-particle Hamiltonian: 
for reasons of storage and computational time, it is advantageous to couple higher-order 
derivatives to Laplacians whenever possible, and to use a symmetric form where $\hat{A} = \hat{B}$ 
whenever possible for $D^{AB}_q(\vec{r})$ and $D^{AB\sigma}_q(\vec{r})$, such that the two terms in 
Eqs.~\eqref{eq:def:Fden:right} and~\eqref{eq:def:Fvec:right} can be combined into one. For 
non-redundant $C^{AB}_q(\vec{r})$ and $C^{AB\sigma}_q(\vec{r})$ this cannot be done, but at 
least for many of the widely-used locally gauge-invariant EDFs there are choices for which 
$\hat{A}$ and/or $\hat{B}$ acting on either $G^{AB}_q(\vec{r})$ and/or $G^{AB\sigma}_q(\vec{r})$ 
gives zero when the expressions for these potentials are proportional to
the right-hand side of a continuity equation such as Eqs.~\eqref{eq:continuity:normal} or
\eqref{eq:continuity:spin}, see for example Ref.~\cite{Hellemans12}. 
There also is a natural preference to express the EDF through local densities and currents 
that enter the expectation values of one-body operators that are frequently used to formulate 
constraints on shape degrees of freedom, angular momentum, etc, such that these additional
terms do not introduce additional operator structures in the resulting single-particle Hamiltonian.

The single-particle Hamiltonian obtained from the N2LO EDF 
\eqref{eq:SkTeven:0}--\eqref{eq:SkTodd:4} naturally breaks down into ten parts
\begin{align}
h_q(\vec{r} \sigma, \vec{r}'\sigma') & = 
    h^{(0)}_{q,\text{e}}(\vec{r}\sigma, \vec{r}'\sigma') 
  + h^{(0)}_{q,\text{o}}(\vec{r}\sigma, \vec{r}'\sigma') \nonumber  \\
& + h^{(1)}_{q,\text{e}}(\vec{r}\sigma, \vec{r}'\sigma') 
  + h^{(1)}_{q,\text{o}}(\vec{r}\sigma, \vec{r}'\sigma') \nonumber  \\
& + h^{(2)}_{q,\text{e}}(\vec{r}\sigma, \vec{r}'\sigma') 
  + h^{(2)}_{q,\text{o}}(\vec{r}\sigma, \vec{r}'\sigma') \nonumber  \\
& + h^{(3)}_{q,\text{e}}(\vec{r}\sigma, \vec{r}'\sigma') 
  + h^{(3)}_{q,\text{o}}(\vec{r}\sigma, \vec{r}'\sigma') \nonumber  \\
& + h^{(4)}_{q,\text{e}}(\vec{r}\sigma, \vec{r}'\sigma') 
  + h^{(4)}_{q,\text{o}}(\vec{r}\sigma, \vec{r}'\sigma') \, ,
\label{eq:hbreakdown}
\end{align}
which are time-even and time-odd structures with up to four gradients
acting on the single-particle states
\begin{widetext}
\begin{align}
\label{eq:h0e}
h^{(0)}_{q, \text{e}} (\vec{r}\sigma, \vec{r}' \sigma')
& = \delta_{\sigma \sigma'} \,  F_{q}^{1,1} (\vec{r})  \, \delta_{\vec{r} \vec{r}'}
    \phantom{\sum_\mu}
   \, , \\
\label{eq:h0o}
h^{(0)}_{q, \text{o}} (\vec{r}\sigma, \vec{r}'\sigma')  
& = \sum_{\mu}  \langle \sigma | \hat{\sigma}_{\mu} | \sigma' \rangle \, 
   F_{q, \mu}^{1,\sigma} (\vec{r}) \, \delta_{\vec{r} \vec{r}'}
   \, , \\
\label{eq:h1e}
h^{(1)}_{q, \text{e}}(\vec{r}\sigma, \vec{r}'\sigma')
& = - \tfrac{\iunit}{2} \sum_{\mu \nu} \langle \sigma | \hat{\sigma}_{\nu} | \sigma' \rangle
      \big[ \nabla_{\mu} \, G^{1, \nabla\sigma}_{q, \mu \nu}(\vec{r})
           + G^{1, \nabla\sigma}_{q,\mu \nu} (\vec{r}) \, \nabla_{\mu}
      \big] \,  \delta_{\vec{r} \vec{r}'} 
    \, ,  \\
\label{eq:h1o}
h^{(1)}_{q, \text{o}} (\vec{r}\sigma, \vec{r}'\sigma') 
& = - \tfrac{\iunit}{2}  \, \delta_{\sigma \sigma'}  \sum_{\mu} 
      \big[    \nabla_{\mu} \, G^{1, \nabla}_{q,\mu}(\vec{r}) 
            +  G^{1, \nabla}_{q, \mu} (\vec{r}) \, \nabla_{\mu} 
     \big] \, \delta_{\vec{r} \vec{r}'}
    \, , \\
\label{eq:h2e}
h^{(2)}_{q, \text{e}}(\vec{r}\sigma, \vec{r}'\sigma')
& = - \delta_{\sigma \sigma'}  
    \sum_{\mu \nu} \nabla_{\mu} \, F^{\nabla,\nabla}_{q,\mu \nu}(\vec{r})   \, \nabla_{\nu} \, \delta_{\vec{r}' \vec{r}}
    \, ,  \\
\label{eq:h2o}
h^{(2)}_{q, \text{o}} (\vec{r}\sigma, \vec{r}'\sigma') 
& = - \sum_{\mu \nu \kappa} \langle \sigma | \hat{\sigma}_{\kappa} | \sigma' \rangle
        \nabla_{\mu} \, F^{\nabla, \nabla \sigma}_{q,\mu \nu \kappa}(\vec{r}) \, \nabla_{\nu} \, 
        \, \delta_{\vec{r} \vec{r}'}
    \, , \\
\label{eq:h3e}
h^{(3)}_{q, \text{e}} (\vec{r}\sigma, \vec{r}'\sigma')
& = - \tfrac{\iunit}{2} \sum_{\mu \nu} \langle \sigma | \hat{\sigma}_{\nu} | \sigma' \rangle \,
     \left[   \nabla_{\mu}  \, G^{\Delta, \nabla \sigma}_{q, \mu \nu} (\vec{r}) \, \Delta
             + \Delta \, G^{\Delta, \nabla \sigma}_{q, \mu \nu} (\vec{r})         \, \nabla_{\mu} 
    \right] \, \delta_{\vec{r} \vec{r}'}
\, , \\
\label{eq:h3o}
h^{(3)}_{q, \text{o}}(\vec{r}\sigma, \vec{r}'\sigma')
& = - \tfrac{\iunit}{2} \, \delta_{\sigma \sigma'} 
     \sum_{\mu} \left[ 
          \nabla_{\mu}  \, G^{\Delta, \nabla}_{q, \mu} (\vec{r}) \, \Delta
          + \Delta \, G^{\Delta, \nabla}_{q, \mu} (\vec{r})        \, \nabla_{\mu}
    \right]  \, \delta_{\vec{r} \vec{r}'}
    \, , \\
\label{eq:h4e}
h^{(4)}_{q, \text{e}} (\vec{r}\sigma, \vec{r}'\sigma')
& = \delta_{\sigma \sigma'}  \, 
    \Delta \, F^{\Delta, \Delta}_q(\vec{r}) \, \Delta \, \delta_{\vec{r} \vec{r}'}
    \phantom{\sum_\mu} \, , \\
\label{eq:h4o}
h^{(4)}_{q, \text{o}}(\vec{r}\sigma, \vec{r}'\sigma')
& = \sum_{\mu} \langle \sigma | \hat{\sigma}_{\mu} | \sigma' \rangle
    \Delta \, F^{\Delta, \Delta\sigma}_{q, \mu}(\vec{r}) \, \Delta \, 
    \delta_{\vec{r} \vec{r}'} \, . 
\end{align}
\end{widetext}
The reader is advised that there is an implicit product rule in 
Eqs.~\eqref{eq:h1e}--\eqref{eq:h4o}: the gradients act on everything
to their right. However, $\hat{h}_{q}(\vec{r}\sigma, \vec{r}'\sigma')$
is usually used such that there is no other function depending on the position
$\vec{r}$ other than the potentials and the delta function $\delta_{\vec{r} \vec{r}'}$. 
In that sense, the presence of $\delta_{\vec{r} \vec{r}'}$ keeps the derivatives 
``inside'' of $\hat{h}_{q}(\vec{r}\sigma, \vec{r}'\sigma')$. The integration
over $d^3r'$ in a matrix element of $\hat{h}_{q}$ such as~\eqref{eq:sphamiltonian:matrix} 
then transfers the derivatives to whatever function of the position $\vec{r}$ that 
is under the integral. With this, all contributions from the local Skyrme EDF 
(meaning that it only depends on densities that are all evaluated at the same 
position $\vec{r}$) lead to local terms in the single-particle Hamiltonian
as expected. While the direct term of the Coulomb energy is a non-local 
functional of local densities, it nevertheless also leads to a local 
contribution to the single-particle Hamiltonian. The Coulomb exchange energy 
in Slater approximation is again a local functional of local densities like 
the Skyrme interaction~\cite{Bender03,Ryssens15a}, such that it also leads to a local contribution to 
the single-particle Hamiltonian. Further details on these standard terms
that we do not address here can be found for example in Ref.~\cite{Ryssens15a}. 
However, when using the exact Coulomb exchange energy instead of the Slater 
approximation, one would have to deal with a non-local functional of non-local 
densities that inevitably leads to a manifestly non-local contribution to 
$h_{q}(\vec{r} \sigma, \vec{r}' \sigma')$, both resulting in a substantial 
increase of computational time. 

For completeness' sake, we include the expressions for 
all potentials $\mathsf{F}_{q,a}(\vec{r})$ obtained from the N2LO
functional defined through~\eqref{eq:SkTeven:0}--\eqref{eq:SkTodd:4}
in Appendix~\ref{app:potentials}. 

The single-particle Hamiltonian is a $2 \times 2$ matrix in the space of 
single-particle spinors; hence, it can be decomposed into the complete 
set of four linearly independent complex $2 \times 2$ matrices, which can be 
chosen to be the union of the unit matrix $\hat{1}_{\sigma}$ and the Pauli spin matrices
\begin{equation}
\hat{h}_q(\vec{r},\vec{r}')
= \hat{h}_{q,1}(\vec{r},\vec{r}') \, \hat{1}_{\sigma} 
  + \hat{\vec{h}}_{q,\sigma}(\vec{r},\vec{r}') \cdot \hat{\vsigma} \, ,
\end{equation}
The terms in Eqs.~\eqref{eq:h0e}--\eqref{eq:h4o} containing $\delta_{\sigma \sigma'}$
are the coordinate-space representation of $\hat{h}_{q,1} \hat{1}_{\sigma}$, whereas those
containing $\langle \sigma | \hat{\sigma}_{\mu} | \sigma' \rangle$ are the coordinate-space
representation of $\hat{\vec{h}}_{q,\sigma} \cdot \hat{\vsigma}$ in this decomposition.
The former are related to variation with respect to $D^{AB}_q(\vec{r})$ and
$C^{AB}_q(\vec{r})$, whereas the latter are obtained from variation with respect 
to $D^{AB\sigma}_q(\vec{r})$ and $C^{AB\sigma}_q(\vec{r})$. When working with 
irreducible non-redundant local densities, time-even (time-odd) terms in 
$\hat{h}_{q,1}$ have an even (odd) number of gradients, whereas for  $\hat{h}_{q,\sigma}$
it is the opposite.

It has to be stressed that there is no unequivocal relation between the time-even 
and time-odd LO, NLO, and N2LO parts of the EDF and the terms with a given number of 
gradients in the single-particle Hamiltonian~\eqref{eq:hbreakdown}. The reason is that 
the order $\ell$ of terms in $h^{(\ell)}_{q,\eta}(\vec{r} \sigma, \vec{r}'\sigma')$,
$\eta = \text{e}$, $\text{o}$, is determined by the number of internal gradients 
in the definition of the local density whose variation generates the respective 
term in Eq.~\eqref{eq:variation}. While the terms with three and four gradients 
in $h_q(\vec{r} \sigma, \vec{r}'\sigma')$ are necessarily generated by the N2LO 
terms $\mathcal{E}^{(4)}_{\textrm{Sk},\eta}(\vec{r})$, in the EDF, the N2LO 
terms also contribute to all lower-order parts of the single-particle 
Hamiltonian. Similarly, the NLO terms $\mathcal{E}^{(2)}_{\textrm{Sk},\eta}$ 
contribute to all terms in the single-particle Hamiltonian with up to two
gradients. Only the LO terms are necessarily limited to the gradientless terms
in $\hat{h}_q$.
In addition, the density-dependent time-odd terms from Eq.~\eqref{eq:SkTodd:0} 
also contribute to the time-even part $h^{(0)}_{q,\rm e}(\vec{r}\sigma, \vec{r}'\sigma')$
of the single-particle Hamiltonian, but of course only for time-reversal-invariance
breaking many-body states. 

For some contributions to the single-particle Hamiltonian, considering N2LO terms
makes the tensor structure of these terms more complex. For NLO functionals, only
the scalar contribution to $F^{\nabla, \nabla}_{q,\mu \nu}(\vec{r})$ is needed, 
leading to an effective mass that is a scalar. It is only in the presence of N2LO terms, 
that $F^{\nabla, \nabla}_{q,\mu \nu}(\vec{r})$ becomes a rank-2 tensor. Similarly,
for standard NLO terms without explicit tensor interaction, only the components of
$F^{\nabla, \nabla \sigma}_{q,\mu \nu \kappa}(\vec{r})$ with $\mu = \nu$ are non-zero,
leading to a spin effective mass that is a vector. In the presence of genuine
NLO tensor forces, there is an additional contribution for which the elements with
$\mu = \kappa$ are also non-zero, and which is traditionally treated as a separate 
term in the single-particle Hamiltonian~\cite{Hellemans12}. For EDFs including the
N2LO terms of  Ref.~\cite{Becker17}, however, the potential 
$F^{\nabla, \nabla \sigma}_{q,\mu \nu \kappa}(\vec{r})$ necessarily becomes a full 
rank-3 tensor.


\subsection{The pairing energy density functional}
\label{Sec:pairing:EDF}

Like the majority of studies based on the Skyrme interaction, we use
a simple pairing energy functional with parameters that are independent
from those that define the particle-hole part of the pairing EDF. Its 
actual form is the one of a gradientless density-dependent contact pairing 
interaction of surface type \cite{Terasaki95,Rigollet,Hellemans12}. 
Its gauge-invariant representation takes the form
\begin{align} 
\label{eq:pair:EDF:ULB}
E_{\text{pair}}
& = \int \! d^3r 
\sum_{q=p,n} \frac{V_q}{4}
\Bigg[ 1 - \frac{\rho_0(\vec{r})}{\rho_c} \Bigg] 
\tilde\rho_q(\vec{r}) \, \tilde\rho_q^*(\vec{r}) 
   \nn \\
& = \int \! d^3r 
   \sum_{q=p,n} \frac{V_q}{4} 
  \Bigg[ 1 - \frac{\rho_0(\vec{r})}{\rho_c} \Bigg] \,
  \Big\{ \tilde{D}^{1,1}_q(\vec{r}) \, \tilde{D}^{1,1*}_q(\vec{r}) 
  \nn \\
& \qquad \qquad \qquad 
       + \tilde{C}^{1,1}_q(\vec{r}) \, \tilde{C}^{1,1*}_q(\vec{r}) \Big\} \, ,
\end{align}
where the $V_q$ and $\rho_c$ are the parameters of the pairing interaction.
As the single-reference EDF is real by construction, its pairing part necessarily 
consists of symmetrized products of pair densities and the complex conjugates of 
pair densities, such that only products of either two real parts or products 
of two imaginary parts contribute, but no cross terms between the real and imaginary
parts of the various densities. Even when working in a gauge where these densities
are complex, expressing the pairing EDF through $\tilde{D}^{1,1}_q(\vec{r})$ and 
$\tilde{C}^{1,1}_q(\vec{r})$ clearly separates terms that are constructed
from densities with different spatial symmetries, cf.\ the discussion of 
Tables~\ref{tab:densymmetries:pairte} and~\ref{tab:densymmetries:pairto} 
in Appendix~\ref{app:symmetries}.

The adaptation to the new notation of possible higher-order terms, 
either in gradients \cite{Perlinska04} or in the number of 
densities \cite{Sadoudi13a,Sadoudi13b}, in the pairing EDF follows the same 
principles but leads to lengthy expressions that will be detailed elsewhere. 

Local pairing EDFs such as Eq.~\eqref{eq:pair:EDF:ULB}, which are generated by 
contact pairing interactions, have to be regularized in one way or another in order to 
suppress the divergence of the pairing energy with increasing basis size. For the
calculations presented in Sec.~\ref{sec:calculations} this has been done
by introducing a state-dependent cutoff as done earlier in 
Refs.~\cite{Gall1994,Ryssens19b,Terasaki95,Rigollet}. In practice, this is achieved 
by multiplying the two-body wave functions $\tilde{\varrho}_{jk}(\vec{r},\vec{r}')$
and $\tilde{\varsigma}_{jk,\mu}(\vec{r},\vec{r}')$ with state-dependent cutoff 
factors, that is by making the substitutions
\begin{align}
\tilde{\varrho}_{jk}(\vec{r},\vec{r}') 
& \to f_j \, f_k \, \tilde{\varrho}_{jk}(\vec{r},\vec{r}') \, ,
  \nn \\
\tilde{\varsigma}_{jk,\mu}(\vec{r},\vec{r}')
& \to f_j \, f_k \, \tilde{\varsigma}_{jk,\mu}(\vec{r},\vec{r}') \, .
\label{eq:cutoff_def}
\end{align}
when summing up the pair densities~\eqref{eq:def:Ddenpair}--\eqref{eq:def:Csvecpair}. 
For the HFB calculations reported in Sec.~\ref{sec:calculations}, we use cutoffs 
$f_m$ that (i) are defined in the ``Hartree-Fock basis'' 
that diagonalizes the single-particle Hamiltonian 
and (ii) depend on the distance of the given single-particle level from the 
Fermi energy, see Refs.~\cite{Gall1994,Terasaki95,Rigollet,Ryssens15a,Ryssens19b} 
for details. Note that the cutoff should only be introduced in the pair densities
used to calculate the pairing EDF and the associated pair fields. All
non-energetic observables should still be calculated with unmodified
pair densities instead.


\subsection{Pairing fields}
\label{sec:singleDelta}

The HFB equation is given by
\begin{equation}
\label{eq:HFB:equation}
\left( \begin{array}{cc}
h - \lambda & \Delta \\
-\Delta^* & -h^*  + \lambda
\end{array} \right)
\left( \begin{array}{c}
U_k \\ 
V_k
\end{array} \right)
= E_k \, 
\left( \begin{array}{c}
U_k \\ 
V_k
\end{array} \right) \, ,
\end{equation}
where $\lambda$ is a Lagrange multiplier for the
adjustment of the average particle number \cite{Schunck2019}, 
and
\begin{align}
\Delta_{jk} 
& \equiv \frac{\delta E_{\text{pair}}}{\delta \kappa^{*}_{jk}} \, ,
           \\
\Delta^{*}_{jk} 
& \equiv \frac{\delta E_{\text{pair}}}{\delta \kappa_{jk}} \, .
\end{align}
When deriving the HFB equation, it is assumed that $\kappa_{jk} = -\kappa_{kj}$ and 
$\kappa^{*}_{jk} = - \kappa^{*}_{kj}$ are independent degrees of freedom that are 
varied separately \cite{Schunck2019}.
For this reason, the distinction between pair 
densities and their complex conjugate should be formally kept even when they can be 
chosen to be real. In such case all pair densities without complex conjugation entering
the EDF contain only $\kappa$, while all densities with explicit complex conjugation 
contain only $\kappa^*$.

Defining, analogously to Eq.~\eqref{eq:densityvector}, $\tilde{\mathsf{R}}_q(\vec{r})$ 
as the vector of all pair densities entering a given pairing EDF 
and $\tilde{\mathsf{R}}^*_q(\vec{r})$ as the vector of their complex conjugates, 
the matrix elements $\Delta_{jk}$ and $\Delta_{jk}^*$ can be calculated as
\begin{align}
\label{eq:Delta:def}
\Delta_{jk} 
  = \frac{\delta E_{\text{{tot}}}}{\delta \kappa^{*}_{jk}}
& = \int \! d^3r \, \sum_a
    \frac{\delta E_{\text{{tot}}}}{\delta \tilde{\mathsf{R}}^{*}_{q,a}(\vec{r}) } 
          \frac{\delta \tilde{\mathsf{R}}^{*}_{q,a}(\vec{r}) }{\delta \kappa^{*}_{jk}} \, ,
    \\
\label{eq:DeltaS:def}
\Delta_{jk}^{*} 
  = \frac{\delta E_{\text{{tot}}}}{\delta \kappa_{jk}}
& = \int \! d^3r \, \sum_a
    \frac{\delta E_{\text{{tot}}}}{\delta \tilde{\mathsf{R}}_{q,a}(\vec{r}) } 
          \frac{\delta \tilde{\mathsf{R}}_{q,a}(\vec{r}) }{\delta \kappa_{jk}} \, .
\end{align}
Analogously to the mean-field potentials defined for the 
single-particle Hamiltonian, we define pair fields as
\begin{align}
\tilde{\mathsf{F}}_{q,a}(\vec{r}) \equiv \frac{\delta E_{\rm {tot}}}{\delta \mathsf{R}_{q,a}^{*}(\vec{r})} \, , \\
\tilde{\mathsf{F}}^*_{q,a}(\vec{r}) \equiv \frac{\delta E_{\rm {tot}}}{\delta \mathsf{R}_{q,a}(\vec{r})} \, .
\end{align}
As before, we use $\tilde{F}, \tilde{G}, \tilde{F}^*$ and $\tilde{G}^*$ to 
distinguish between potentials corresponding to $\tilde{D}^*, \tilde{C}^*, \tilde{D}$
and $\tilde{C}$ densities, leading to eight different types of generic pairing 
potentials. For pair densities without spin, these are
\begin{align}
\tilde{F}^{A,B}_q(\vec{r})      &\equiv \frac{\delta E_{\rm {tot}}}{\delta D^{A,B*}_{q}(\vec{r})} \, ,
\\
\tilde{F}^{A,B*}_q(\vec{r})    &\equiv \frac{\delta E_{\rm {tot}}}{\delta D^{A,B}_{q}(\vec{r})} \, ,
\\
\tilde{G}^{A,B}_q(\vec{r})      &\equiv \frac{\delta E_{\rm {tot}}}{\delta C^{A,B*}_{q}(\vec{r})} \, ,
\\
\tilde{G}^{A,B*}_q(\vec{r})    &\equiv \frac{\delta E_{\rm {tot}}}{\delta C^{A,B}_{q}(\vec{r})} \, .
\end{align}
For local pair densities with spin, we define  $\tilde{F}_q^{A,B\sigma}(\vec{r})$,
$\tilde{G}_q^{A,B\sigma}(\vec{r}), \tilde{F}_q^{A,B\sigma*}(\vec{r})$
and $\tilde{G}_q^{A,B\sigma*}(\vec{r})$ analogously. For the simple 
pairing functional used here, Eq.~\eqref{eq:pair:EDF:ULB}, only 
{four} such potentials {
($\tilde{F}^{1,1}_q(\vec{r}),\tilde{F}^{1,1*}_q(\vec{r}),\tilde{G}^{1,1}_q(\vec{r})$ and $\tilde{G}_q^{1,1*}(\vec{r})$)}
are relevant.

In the two-basis-method~\cite{Gall1994, Ryssens19b} we employ to solve the
HFB equation, only the matrix elements $\Delta_{jk}$ and 
$\Delta^{*}_{jk}$ in a basis of limited size
 are needed at each iteration instead of the action of the full pair Hamiltonian 
on some quasiparticle wave function. In this case, the calculation
of contributions from densities at arbitrary order in gradients from
Eqs.~\eqref{eq:def:Ddenpair}--\eqref{eq:def:Csvecpair} is 
straightforward and requires the construction of the spatial functions
\begin{align}
   \frac{\delta \tilde{D}^{A,B}_q(\vec{r}) }{\delta \kappa_{jk}}
  & = \left.
      \tfrac{1}{2} \big( \hat{A}' \hat{B} + \hat{A} \hat{B}' \big)
      \Re \big\{ \tilde{\varrho}_{kj}(\vec{r},\vec{r}') 
                -\tilde{\varrho}_{jk}(\vec{r},\vec{r}') \big\} \right|_{\vec{r} = \vec{r}'}
      \nn \\
  & = \big( \hat{A}' \hat{B} + \hat{A} \hat{B}' \big) 
      \Re \big\{ \tilde{\varrho}_{kj}(\vec{r},\vec{r}') \big\} 
      \Big|_{\vec{r} = \vec{r}'} \, , \label{eq:temp1}
      \\
\frac{\delta \tilde{C}^{A,B}_q(\vec{r}) }{\delta \kappa_{jk}}
& = \big( \hat{A}' \hat{B} + \hat{A} \hat{B}' \big) 
    \Im \big\{ \tilde{\varrho}_{jk}(\vec{r},\vec{r}') \big\} 
    \Big|_{\vec{r} = \vec{r}'} \, , 
    \\
\frac{\delta \tilde{D}^{A,B \sigma}_q(\vec{r}) }{\delta \kappa_{jk}}
& = \big( \hat{A}' \hat{B} - \hat{A} \hat{B}' \big)
    \Re \big\{ \tilde{\varsigma}_{kj}(\vec{r},\vec{r}') \big\} 
    \Big|_{\vec{r} = \vec{r}'} \, ,
    \\
\frac{\delta \tilde{C}^{A,B \sigma}_q(\vec{r}) }{\delta \kappa_{jk}}
& = \big( \hat{A}' \hat{B} - \hat{A} \hat{B}' \big)
    \Im \big\{ \tilde{\varsigma}_{kj}(\vec{r},\vec{r}') \big\}
    \Big|_{\vec{r} = \vec{r}'}  \, ,      
\end{align}

here the factor $1/2$ on the right-hand side of Eq.~\eqref{eq:temp1} originates from
extending the summation over $j$ and $k$ in Eq. (75) to all combinations.
Because of the skew symmetry of $\kappa_{ij} = -\kappa_{ji}$,
the variation with respect to $\kappa_{ij}$ yields two contributions.
Using the skew symmetry of $\tilde\varrho_{jk}(\vec{r},\vec{r}')$, these
can be combined again into a single term. This intermediate step is omitted
in the other three relations. Analogously, one finds for their
complex conjugates
\begin{align}
\frac{\delta \tilde{D}^{A,B*}_q(\vec{r}) }{\delta \kappa_{jk}^{*}}
& = \big( \hat{A}' \hat{B} + \hat{A} \hat{B}' \big) 
    \Re \big\{ \tilde{\varrho}_{kj}^{*}(\vec{r},\vec{r}') \big\} 
    \Big|_{\vec{r} = \vec{r}'} \, ,
    \\
\frac{\delta \tilde{C}^{A,B*}_q(\vec{r}) }{\delta \kappa_{jk}^{*}}
& = \big( \hat{A}' \hat{B} + \hat{A} \hat{B}' \big) 
    \Im \big\{ \tilde{\varrho}_{kj}^{*}(\vec{r},\vec{r}') \big\} 
    \Big|_{\vec{r} = \vec{r}'} \, , 
    \\
\frac{\delta \tilde{D}^{A,B \sigma*}_q(\vec{r}) }{\delta \kappa_{jk}^{*}}
& = \big( \hat{A}' \hat{B} - \hat{A} \hat{B}' \big) 
    \Re \big\{ \tilde{\varsigma}_{kj}^{*}(\vec{r},\vec{r}') \big\} 
    \Big|_{\vec{r} = \vec{r}'} \, ,
    \\
\frac{\delta \tilde{C}^{A,B \sigma*}_q(\vec{r}) }{\delta \kappa_{jk}^{*}}
& = \big( \hat{A}' \hat{B} - \hat{A} \hat{B}' \big)
    \Im \big\{ \tilde{\varsigma}_{kj}^{*}(\vec{r},\vec{r}') \big\}
    \Big|_{\vec{r} = \vec{r}'}  \, .      
\end{align}
For sake of compact notation we omitted again the possible indices 
of the spatial cartesian tensor components of these objects.

For the pairing EDF~\eqref{eq:pair:EDF:ULB} used for the calculations
discussed in what follows, one obtains in new notation 
\begin{widetext}
\begin{align}
\label{eq:Delta:ULB:def}
\Delta_{jk} 
  = \frac{\delta E_{\text{pair}}}{\delta \kappa^{*}_{jk}}
& = \int \! d^3r \, 
    \bigg[ \frac{\delta E_{\text{pair}}}{\delta \tilde{D}^{1,1*}_q(\vec{r}) } 
           \frac{\delta \tilde{D}^{1,1*}_q(\vec{r}) }{\delta \kappa^{*}_{jk}} 
          +\frac{\delta E_{\text{pair}}}{\delta \tilde{C}^{1,1*}_q(\vec{r}) } 
           \frac{\delta \tilde{C}^{1,1*}_q(\vec{r}) }{\delta \kappa^{*}_{jk}} 
    \bigg]
    \nn \\
& = f_j f_k
    \int \! d^3r \, \frac{V_q}{2} 
    \bigg[ 1 - \frac{D^{1,1}_0(\vec{r})}{\rho_c}\bigg] \, 
    \Big[ \tilde{D}^{1,1}_q(\vec{r}) \,          
          \Re \big\{ \tilde{\varrho}_{kj}^{*}(\vec{r},\vec{r}) \big\}
         +\tilde{C}^{1,1}_q(\vec{r}) \, 
          \Im \big\{ \tilde{\varrho}_{kj}^{*}(\vec{r},\vec{r}) \big\}
    \Big] \, , \nonumber \\
& = f_j f_k
    \int \! d^3r \, 
    \Big[
          \tilde{F}^{1,1}_q(\vec{r}) \,          
          \Re \big\{ \tilde{\varrho}_{kj}^{*}(\vec{r},\vec{r}) \big\}
         +\tilde{G}^{1,1}_q(\vec{r}) \, 
          \Im \big\{ \tilde{\varrho}_{kj}^{*}(\vec{r},\vec{r}) \big\}
    \Big] \, ,
           \\
\label{eq:DeltaS:ULB:def}
\Delta^{*}_{jk} 
  = \frac{\delta E_{\text{pair}}}{\delta \kappa_{jk}} 
& = \int \! d^3r \, 
    \bigg[ \frac{\delta E_{\text{pair}}}{\delta \tilde{D}^{1,1}_q(\vec{r}) } 
           \frac{\delta \tilde{D}^{1,1}_q(\vec{r}) }{\delta \kappa_{jk}} 
          +\frac{\delta E_{\text{pair}}}{\delta \tilde{C}^{1,1}_q(\vec{r}) } 
           \frac{\delta \tilde{C}^{1,1}_q(\vec{r}) }{\delta \kappa_{jk}} 
    \bigg]
    \nn \\
& = f_j f_k
    \int \! d^3r \, \frac{V_q}{2} \, 
    \bigg[ 1 - \frac{D^{1,1}_0(\vec{r})}{\rho_c}\bigg]  \, 
    \Big[ \tilde{D}^{1,1*}_q(\vec{r}) \,          
          \Re \big\{ \tilde{\varrho}_{kj}(\vec{r},\vec{r}) \big\}
         +\tilde{C}^{1,1*}_q(\vec{r}) \, 
          \Im \big\{ \tilde{\varrho}_{kj}(\vec{r},\vec{r}) 
    \Big] \, , \nonumber \\
& = f_j f_k
    \int \! d^3r \, 
    \Big[
          \tilde{F}^{1,1,*}_q(\vec{r}) \,          
          \Re \big\{ \tilde{\varrho}_{kj}(\vec{r},\vec{r}) \big\}
         +\tilde{G}^{1,1,*}_q(\vec{r}) \, 
          \Im \big\{ \tilde{\varrho}_{kj}(\vec{r},\vec{r}) \big\}
    \Big] \, ,
\end{align}
\end{widetext}
where $f_j$ and $f_k$ are the cutoff factors of Eq.~\eqref{eq:cutoff_def} and
we have used that by definition
$\delta \tilde{D}^{1,1}_q(\vec{r})   / \delta \kappa^{*}_{jk}
= \delta \tilde{C}^{1,1}_q(\vec{r})  / \delta \kappa^{*}_{jk} 
= \delta \tilde{D}^{1,1*}_q(\vec{r}) / \delta \kappa_{jk}
= \delta \tilde{C}^{1,1*}_q(\vec{r}) / \delta \kappa_{jk}
= 0$. From the definitions \eqref{eq:def:Ddenpair} and
\eqref{eq:def:Cdenpair} follows directly that, in a gauge that
leads to real $\kappa_{ij} = \kappa_{ij}^*$, one also automatically
has real $\Delta_{jk} = \Delta^{*}_{jk}$, as all terms on the right-hand side
of Eqs.~\eqref{eq:Delta:def} and~\eqref{eq:DeltaS:def} become integrals 
over real-valued functions.


\section{Exploratory calculations}
\label{sec:calculations}


\subsection{Numerical choices}

We have extended the MOCCa code~\cite{RyssensPhD,MOCCa} to allow for the solution of 
the self-consistent mean-field problem with N2LO EDFs of the form presented in 
the previous section. This code represents the self-consistent mean-field 
problem in coordinate space, utilizing the properties of Lagrange 
meshes~\cite{Bay86a,Baye15}, for which the derivatives of a function 
are constructed from the values of said function on all mesh points.
Together with a rectangular quadrature rule, the Lagrange-mesh derivatives 
implicitly define an underlying basis of plane waves in a box~\cite{Ryssens15b}.
The Lagrange-mesh technique has the advantages over simpler finite-difference 
expressions for derivatives that integration by parts is exact up to machine 
precision, and that higher-order derivative matrices can be calculated as
products of matrices of first-order derivatives. For N2LO EDFs, these two 
features turn out to be even more important for the suppression of numerical 
noise brought by the higher number of repeated derivatives needed when
calculating the local densities and when applying the single-particle-Hamiltonian
for the Skyrme N2LO functional than for the standard NLO functional.

Using this representation, one obtains the total energy of nuclear 
configurations with an accuracy that is essentially independent of their 
deformation~\cite{Ryssens15b}. In addition to the stringent numerical tests mentioned 
in Sec.~\eqref{sec:rewritten} above, MOCCa was also benchmarked in detail against
the spherical code used in Ref.~\cite{Becker17}. For all calculations reported here,
we have set the mesh spacing $dx$ to $0.8$ fm while adapting the box size to 
the nucleus under consideration, as in Ref.~\cite{Ryssens15b}. 

The only presently available Skyrme parametrization for finite nuclei that 
includes N2LO terms is SN2LO1~\cite{Becker17}. For the purposes of comparison, 
we also report on calculations with two NLO parametrizations. 
For all nuclei considered below, we include results for the SLy5* 
parametrization~\cite{Pastore13}. This parametrization serves as a 
good point of comparison, as the fit protocol constructed for 
its adjustment was later also adopted for SN2LO1 with only minimal changes. 
When discussing very heavy nuclei, we also include results 
obtained with the NLO SLy5s1 parametrization that was adjusted with 
an extension of said fit protocol that incorporated an additional 
constraint on surface tension~\cite{Jodon16}. We include SLy5s1
because of its superior performance for deformation properties of 
heavier nuclei that results from its more realistic surface tension~\cite{Ryssens19a}. 

All three of these parametrizations were adjusted for doubly-magic nuclei 
for which pairing correlations vanish at the mean-field level. In order to 
include the effect of pairing correlations, we have added a pairing 
term~\eqref{eq:pair:EDF:ULB} to the EDF that corresponds to a 
density-dependent zero-range pairing interaction, supplemented with a 
smooth cutoff both above and below the Fermi energy, as originally 
proposed in Ref.~\cite{Terasaki95,Rigollet}. As the effective mass 
of all three parametrizations considered is almost identical to 
that of SLy4 for which this pairing EDF has been adjusted, 
we decided to use the same parameters $V_n = V_p = -1250 \, \text{MeV} \, \text{fm}^3$
and $\rho_c = 0.16 \, \text{fm}^{-3}$ of the pairing EDF and the
same cutoff parameters as adjusted for SLy4 in Refs.~\cite{Terasaki95,Rigollet}
in all calculations. This has 
the additional advantage of making the comparison between the three Skyrme
EDFs more straightforward. The resulting HFB equations were solved using 
the two-basis method \cite{Gall1994, Ryssens19b}. We have employed 
the Lipkin-Nogami procedure to avoid the collapse of pairing correlations~\cite{Gall1994}. 

If one wants to link the Skyrme EDF to an underlying pseudopotential, 
the coupling coefficients of the time-odd terms are completely determined when one 
has fixed the coupling coefficients of the time-even terms. For many Skyrme NLO
parametrizations, however, making such a choice is impossible in practice, 
as they exhibit finite-size instabilities with respect to spin polarization which 
show up in situations when time-reversal symmetry is broken. Examples 
are calculations with broken-pair blocked quasiparticles~\cite{Schunck10}, 
self-consistent cranked calculations~\cite{Hellemans12}, QRPA calculations
\cite{Pastore15}, or time-dependent 
calculations~\cite{Fracasso12}. When working with NLO parametrizations that exhibit 
such instability, one and/or the other of the coupling constants of the $\cc{A}^{(2,1)}_{t,\textrm{o}} \,  
\vec{D}^{1, \sigma}_{t}(\vec{r})  \cdot \big[ \Delta \vec{D}^{1, \sigma}_{t}(\vec{r}) \big]$
term in Eq.~\eqref{eq:SkTodd:2} (and/or the coupling constants of a similar term
proportional to $\big[ \vnabla \cdot \vec{D}^{1, \sigma}_{t}(\vec{r}) \big]^2$
that only contributes when including explicit tensor forces~\cite{Hellemans12} not considered here), 
has to be set to smaller value, which is typically chosen to be zero. 

The empirical criteria developed in Ref.~\cite{Hellemans13} for isospin instabilities 
have been sufficient to guard also against such unphysical spin instabilities in 
the cases of SLy5*~\cite{Pastore13} and SLy5s1~\cite{Ryssens19b}. For SN2LO1, the same 
techniques were used with the same empirical protocol to 
safeguard the adjustment of the parametrization, and we aim to validate their
effectiveness for the case of N2LO parametrizations. For this reason, we have 
kept all time-odd terms in the particle-hole channel as generated
by the underlying pseudopotential as considered in Ref.~\cite{Becker17,Pastore13,Jodon16}
for all three parametrizations.

As demonstrated in Ref.~\cite{Ryssens19b}, the presence of higher-order 
derivatives in the functional renders the convergence of the
iterative solution of the self-consistent problem more difficult 
to control numerically. We have employed the heavy-ball and 
preconditioning algorithms of that reference to reliably and quickly converge 
the calculations.

We will present results as a function of the dimensionless (mass) quadrupole 
multipole moment, defined as
\begin{equation}
\label{eq:beta20}
\beta_{20} 
= \frac{4 \pi}{3 R_0^2 A} \langle r^2 Y_{20} \rangle \, ,
\end{equation}
where $Y_{20}$ is a spherical harmonic and $R_0 = 1.2 \, A^{1/3}$ fm. 
Positive (negative) values of $\beta_{20}$
indicate that the nuclear configuration exhibits a prolate (oblate) shape. In 
order to compare to experimental data obtained from $E2$ transitions, we
also define the dimensionless charge deformation
\begin{equation}
\beta_{20,p} = \frac{4 \pi}{3 R_0^2 Z} \langle r^2_p Y_{20,p} \rangle \, ,
\end{equation}
which usually takes values that differ from those of $\beta_{20}$ by less
than a few percent.

Finally, we will also compare to experimental data on isotopic shifts. For
a nucleus with $N$ neutrons, the isotopic shift is defined as the difference 
between its mean-squared (ms) charge radius and the ms radius of a reference
isotope of the same element with $N_0$ neutrons as
\begin{equation}
\delta \langle r^2_p \rangle^{N, N_0} 
=  \langle r^2_p \rangle^{N} - \langle r^2_p \rangle^{N_0} \, .
\end{equation}
%


\subsection{Shape coexistence in the Krypton isotopes}
\label{sec:Krisotopes}

\begin{figure}[t!]
\includegraphics[width=.45\textwidth]{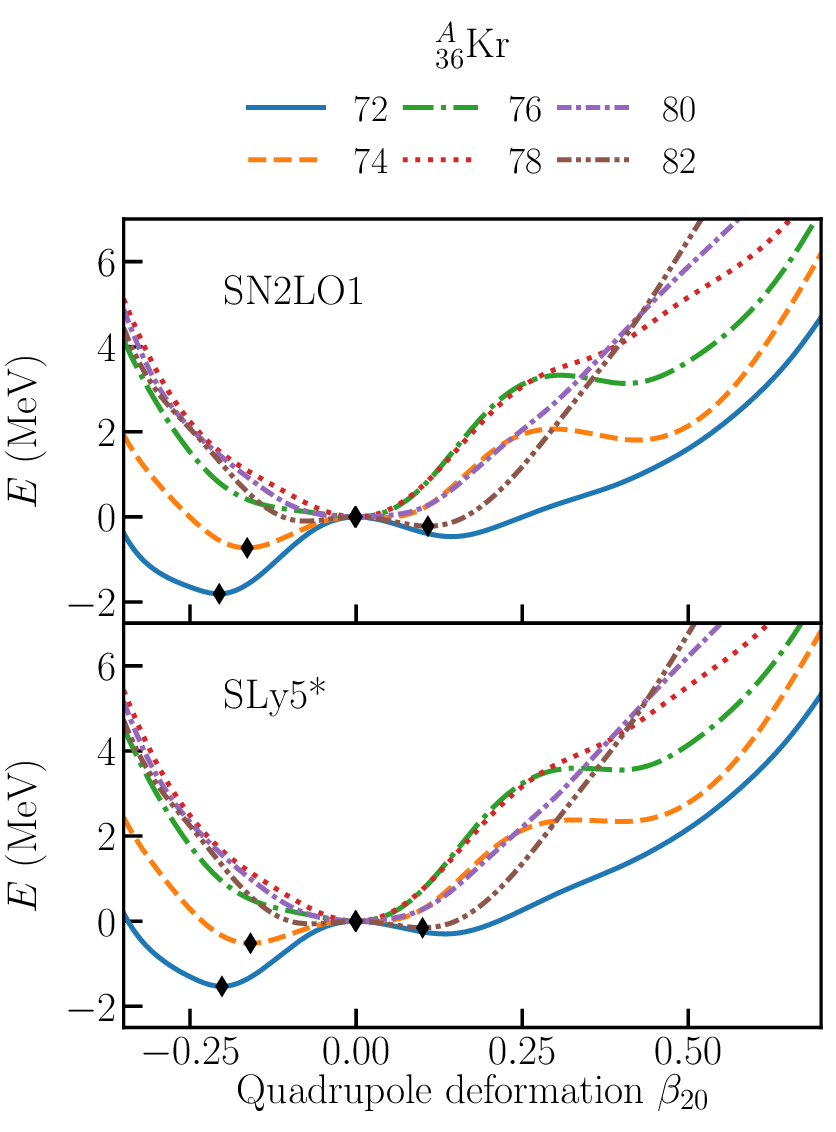}
\caption{Deformation energy curve for the lighter krypton 
         isotopes $^{72-82}$Kr as a function of the quadrupole moment $\beta_{20}$, 
         normalized to the spherical configurations. Top: results for SN2LO1, 
         bottom, results for SLy5*. Black diamond indicate the global minimum
         for each isotope. }
\label{fig:Kr_isotopes:light}
\end{figure}

\begin{figure}[t!]
\includegraphics[width=.45\textwidth]{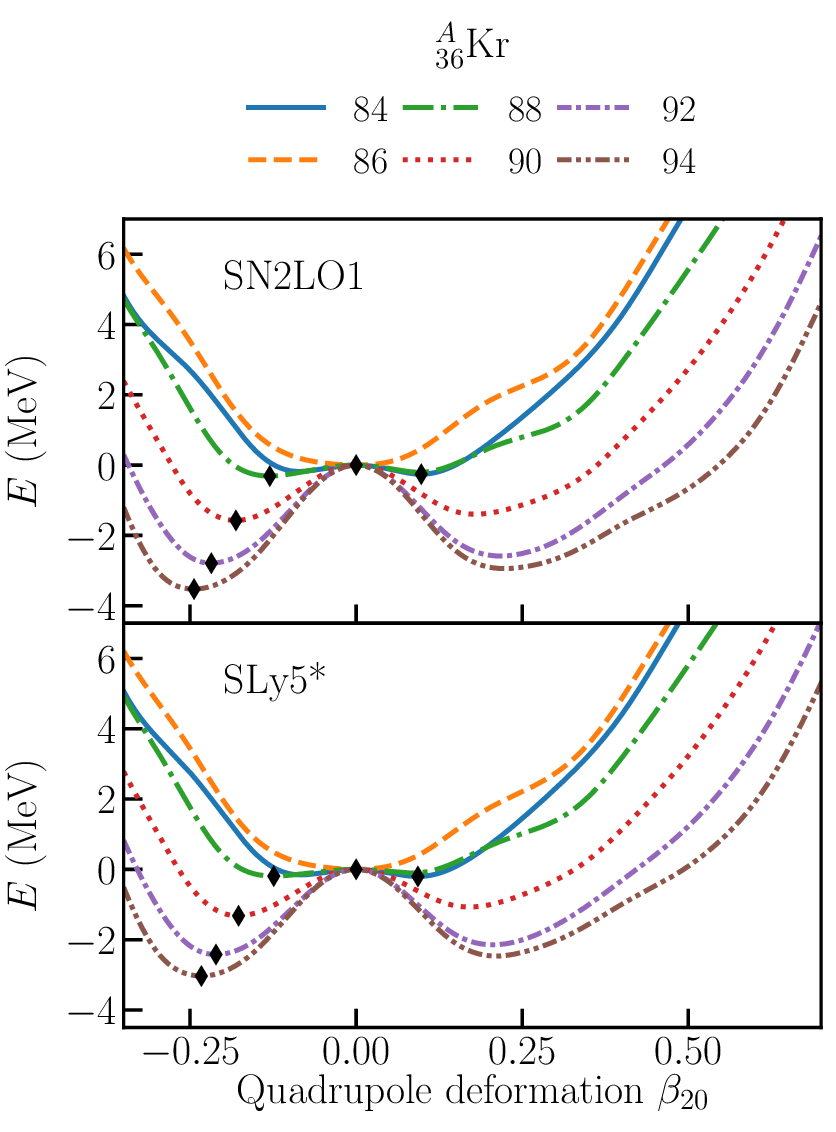}
\caption{Same as Fig.~\ref{fig:Kr_isotopes:light} but for the heavier 
        isotopes $^{84-94}$Kr.}
\label{fig:Kr_isotopes:heavy}
\end{figure}

We first consider the deformation energy curves of even-even krypton ($Z=34$)
isotopes,  spanning all isotopes from $N=36$ up to $N=58$. Figures~\ref{fig:Kr_isotopes:light}
and~\ref{fig:Kr_isotopes:heavy} show the energy curves of $^{72-82}$Kr and 
of $^{84-94}$Kr as calculated with the SN2LO1 parametrization (top panels) and 
calculated with the SLy5* parametrization (bottom panels). All curves have been
normalized to the energy of the spherical configuration.

The two parametrizations produce strikingly similar potential energy curves.
For both parametrizations the neutron-deficient $A=72$ and 74 isotopes exhibit a 
pronounced oblate minimum, which evolves to a spherical symmetric minimum for
$A=76$, $78$, $80$, and also $86$. $^{82}$Kr exhibits a very shallow 
prolate minimum, as does $^{84}$Kr. From $^{88}$Kr onwards, the isotopes develop 
a pronounced oblate minimum again. While minute differences between the results 
for both parametrizations exist, this general evolution is identical. This does 
not mean that the N2LO terms of SN2LO1 do not contribute to the energy: the 
contribution of $\mathcal{E}^{(4)}_{\rm e}(\vec{r})$ to the total energy of the 
spherical configuration of $^{76}$Kr integrates to roughly $-38$ MeV which is 
not negligible, but remains a small fraction of the total binding energy of
about $800$ MeV. The contribution of the N2LO terms varies on the level of
a few MeV as a function of deformation, but does not produce any meaningful 
shifts of the overall topography, as evidenced by Figs.~\ref{fig:Kr_isotopes:light}
and~\ref{fig:Kr_isotopes:heavy}. For the other nuclei discussed in 
what follows, the N2LO terms behave similarly.

The neutron-deficient $^{72-78}$Kr isotopes are a prime example of shape coexistence.
Detailed data from Coulomb excitation experiments for these nuclei can be 
interpreted in terms of a highly-deformed prolate and a lesser-deformed oblate (or 
possibly non-axial) structure that quickly change their relative position in the energy 
spectrum with neutron number. The situation is simplest for $^{76}$Kr and 
$^{78}$Kr, which have a strongly deformed prolate ground state~\cite{Becker06,Clement07} 
with $\beta_{2} \simeq 0.4$ that is the band-head of a rotational band built from
states with similar deformations. Going from $^{78}$Kr to $^{76}$Kr, the excitation 
energy of the $0^+_2$ state goes down from 1.017 MeV to 0.760 MeV, and even further 
down to 0.509 MeV in $^{74}$Kr. While the overall band structure of $^{74}$Kr 
is similar to the one of the heavier isotopes, the measured transition moments 
indicate that the two coexisting structures are strongly mixed in the low-lying
states~\cite{Clement07}. For the even lighter $N=Z$ isotope $^{72}$Kr, a recent
measurement finds that the ground-state has an oblate shape, while the yrast states
of higher spin remain prolate \cite{Wimmer20}. Beginning with $^{80}$Kr, the low-lying 
part of the spectrum of the heavier isotopes around $N=50$ can rather be interpreted 
in terms of anharmonic vibrations \cite{Doring95}.

This scenario is not reproduced by either of the Skyrme parametrizations.
Both of them predict either oblate-deformed or near-spherical shapes for all 
Kr isotopes in Figs.~\ref{fig:Kr_isotopes:light} and~\ref{fig:Kr_isotopes:heavy}. 
They both also produce a local minimum at large prolate deformation for $^{74-76}$Kr, 
but these configurations are still several MeV above the calculated oblate minimum. 
This is a problem that SLy5* and SN2LO1 share with many other modern parametrizations 
of the Skyrme EDF \cite{Ryssens19a,Bender06b}, while an early one like SIII correctly 
gives a prolate ground state for $^{74-76}$Kr \cite{Bonche85}. Similar problems are 
also found for the ground states of Zr isotopes \cite{Ryssens19a,Bender09}. It has 
been argued in Ref.~\cite{Bender06b} that this finding might be caused by incorrect 
relative distances between single-particle levels at spherical shape in this mass 
region. In any event, the differences between SN2LO1 and SLy5* are much smaller 
than what is typically found when comparing with results from other NLO Skyrme 
parametrizations for nuclei in this mass region \cite{Bender09,Ryssens19a}, 
which probably can be attributed to the similarity of their fit protocol.

On the other hand, using a Gogny interaction the shape transition and 
the overall structure of low-lying excited states in the neutron-deficient 
Kr isotopes has been successfully modeled in beyond-mean-field calculations 
when exactly or approximatively projecting on angular momentum and mixing states 
in the entire $\beta$-$\gamma$ plane \cite{Clement07,Wimmer20,Girod09,Rodriguez14}, 
which might hint at the insufficiency of pure mean-field 
calculations to describe shape coexistence phenomena in this mass region. 

%
%
\subsection{Deformation properties of the neodynium isotopes}
\label{sec:Ndisotopes}

\begin{figure}[t!]
\includegraphics[width=.45\textwidth]{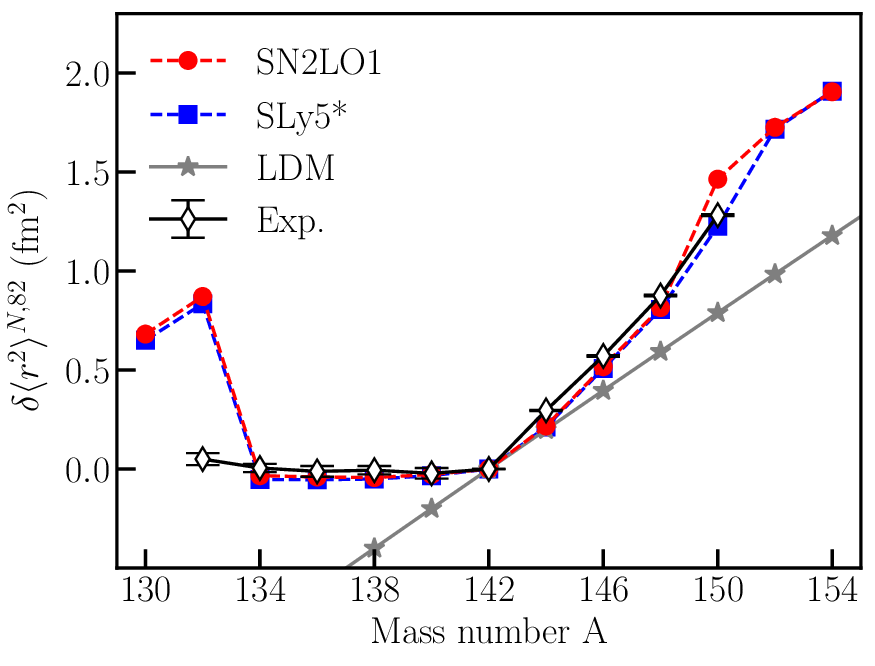}
\caption{Isotopic shifts $\delta \langle r^2 \rangle^{N, 82}$ of  
          mean-square charge radii for the Nd
          isotope chain calculated with SLy5* and SN2LO1. Experimental values 
          were taken from Ref.~\cite{Angeli}. The evolution of charge
          radii when assuming the global $A^{2/3}$ scaling of the spherical 
          liquid-drop model is also given for comparison.
          }
\label{fig:Nd_radii}
\end{figure}

\begin{figure}[t!]
\includegraphics[width=.45\textwidth]{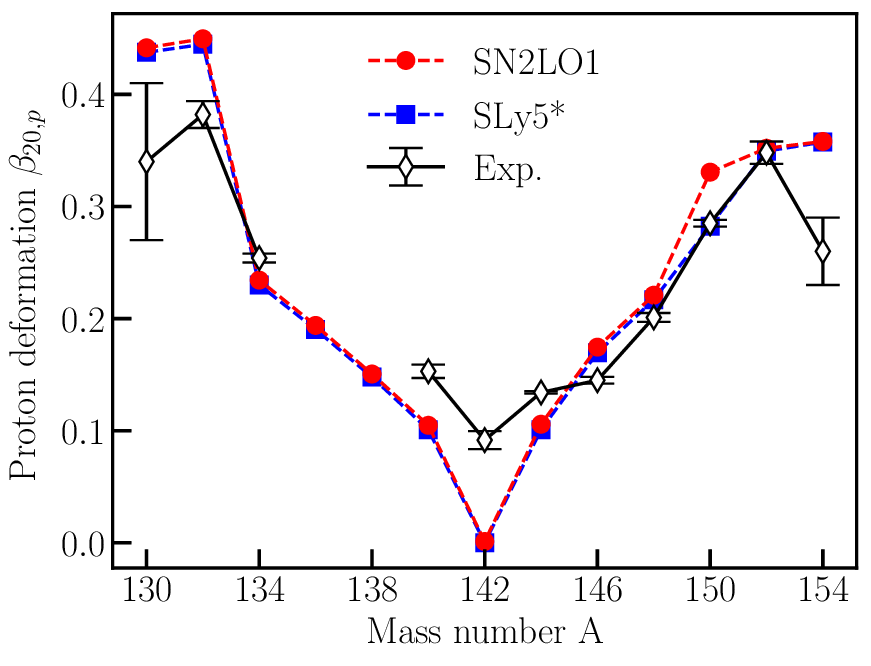}
\caption{Deformation $\beta_{20, p}$ of the proton distribution 
         in the Nd isotope chain calculated with SLy5* and SN2LO1, compared 
         to values of the transition charge deformation
         deduced from experimental $B(E2, 0^+_1 \to 2^+_1)$ values~\cite{Raman}
         under the assumption that the nuclei are rigid rotors.
         }
\label{fig:Nd_deformation}
\end{figure}

As a second example we consider the chain of Nd ($Z=60)$ isotopes, which 
present a medium-heavy set of nuclei that offer a large range of deformations 
that evolve from strongly deformed shapes of the neutron-deficient isotopes over 
spherical ones near $N=82$ to again strongly deformed neutron-rich isotopes.
Figure~\ref{fig:Nd_radii} confronts calculated isotopic shifts of charge radii 
obtained with the SN2LO1 and SLy5* parametrizations with the available data, 
whereas Fig.~\ref{fig:Nd_deformation} compares the deformation $\beta_{20,p}$ of 
protons in the calculated ground states of these nuclei with the experimental 
transition charge deformation as deduced from $B(E2,0^+_1 \to 2^+_1)$ values~\cite{Raman}.

The isotopic shifts of the Nd isotopes exhibit a clear kink at $A=142$, such 
that their overall growth does not follow the global scaling of ms charge radii 
with $A^{2/3}$. The latter is indicated on the figure by plotting the 
ms radius of a spherical liquid-drop $r^2 = (3/5) \, r_0^2 \, A^{2/3}$, where 
the surface radius constant $r_0$ is determined through 
$r_0^3 = 3/(4 \pi \rho_{\text{sat}})$ from the average of the very similar 
saturation densities of SLy5* and SN2LO1. The deviations from this smooth trend
can be attributed to the gradual onset of deformation on both sides of $N=82$ 
shell closure~\cite{Bender06}. From the structure of their excitation spectrum \cite{NuDat}, 
nuclei in the direct vicinity of $^{142}$Nd should be interpreted in terms of 
near-spherical soft anharmonic vibrators, such that their $B(E2,0^+_1 \to 2^+_1)$ 
values cannot be reliably linked to an intrinsic deformation through that 
assumption of a rigid rotor model. In fact, rotational bands built on the ground 
state  have been observed for $A \leq 136$ and $A \geq 146$, but their level spacing 
approaches the one of a rigid rotor only for $A \lesssim 136$ and $A \gtrsim 150$ 
\cite{NuDat}. The ms charge radii from the mean-field ground states nevertheless
follow closely the trend of the experimental ones.

The calculations overestimate the deformation and charge radii for the very 
neutron-deficient isotopes, but otherwise reproduce the experimental trend 
rather well across the entire chain. 

As for the Kr isotopes, the calculated ground-state deformations of all Nd 
isotopes are very similar for the two parametrizations, with the SN2LO1
ones being marginally larger for all nuclei. Consequently, the calculated
isotopic shifts are also very similar, with the SN2LO1 values always being  
slightly higher. The largest difference between parametrizations is found for
$^{150}$Nd: the deformation energy surface of 
this nucleus is very flat around $\beta_{20} \simeq 0.3$, such that a small 
difference in any one aspect of a parametrization can induce a rather large
change of deformation for this nucleus.

%
%
\subsection{The fission barrier of $^{240}$Pu}
\label{sec:fission}

\begin{figure}
\includegraphics[width=7.5cm]{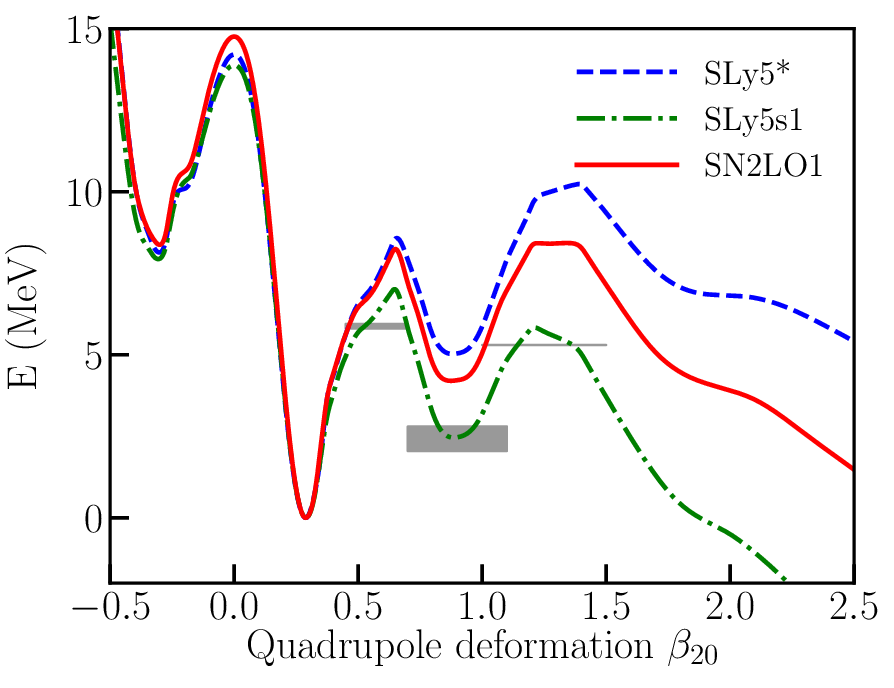}
\caption{Static fission path of $^{240}$Pu as a function of quadrupole deformation 
         $\beta_{20}$, as calculated with the SLy5s1, SLy5* and 
         SN2LO1 parametrizations. The energy curves are normalized to their
         respective prolate minimum. Along the path, nuclear shapes explore
         non-axial and/or reflection asymmetric deformations (see text). 
         The horizontal grey bars indicate experimental values for the height of the
         inner and outer barriers as well as the excitation energy of the fission isomer.
         taken from the same sources as in Ref.~\cite{Ryssens19a} 
         }
\label{fig:Pu240}
\end{figure}

Figure~\ref{fig:Pu240} shows the static fission barrier of $^{240}$Pu, 
calculated with the three parametrizations SN2LO1, SLy5* and SN2LO1, as a function of the 
dimensionless mass quadrupole moment $\beta_{20}$ as defined in Eq.~\eqref{eq:beta20}.
This nucleus provides one of the standard test benches for models of fission, two of the 
reasons being the availability of experimental data for several of the characteristic 
energies of its fission barrier, as well as the possibility to obtain a continuous
fission path by making calculations with a single constraint.
Moving from small to large deformations, the lowest mean-field 
configuration exhibits shapes with different intrinsic symmetries.
Shapes are axial around the local minima, but both around the inner and outer saddle 
points the nucleus takes non-axial shapes. Up to deformations slightly beyond
the fission isomer, shapes remain reflection symmetric, whereas beyond the 
configurations become increasingly reflection-asymmetric. Along the respective 
fission paths, the values of all multipole deformations up to at least $\ell = 8$ 
are near-identical for all three parametrizations. In particular, around the 
outer saddle point,  the favored nuclear shapes combine non-axiality 
and reflection asymmetry for all three parametrizations, 
a feature that was already reported on for SLy5s1 in Ref.~\cite{Ryssens19a},  
the UNEDF1 parametrization of the Skyrme EDF at NLO in Ref.~\cite{Ling20}, 
and for relativistic EDF approaches in Ref.~\cite{Lu2014}.

At small deformation, the overall profile of the barrier is very similar
for all three parametrizations. Only for $\beta_{20}\geq 0.5$
do we observe deviations larger than $1$ MeV. Between SLy5s1 and SLy5*, this
difference can be explained through the variation in surface tension of these 
parametrizations: with $a_{\text{surf}}^{\text{HF}} = 18.61 \, \text{MeV}$,
the surface energy coefficient
of SLy5* is much larger than the one of SLy5s1 that takes the value of
$a_{\text{surf}}^{\text{HF}} = 17.55 \, \text{MeV}$~\cite{Jodon16}, 
resulting in a larger loss in binding energy with increasing deformation. 
The correlation between the surface tension and the height of the 
fission barrier of $^{240}$Pu has been systematically studied in 
Refs.~\cite{Jodon16,Ryssens19a}, concluding that 
parametrizations with low surface tension like SLy5s1 in general provide a 
better description of the experimental data on fission barriers and deformation 
properties of heavy nuclei. Unfortunately, there is no tool available yet 
to calculate the surface energy coefficient of N2LO parametrizations like
SN2LO1. The observation that the deformation energy curve obtained with SN2LO1 
falls systematically between those for SLy5s1 and SLy5* suggests that also its
surface tension is in between the ones of these parametrizations. With that, 
SN2LO1 brings an improvement compared to SLy5*, but a further fine-tuning of 
its surface properties along the lines of Ref.~\cite{Jodon16} will be needed 
to arrive at realistic predictions for fission barriers.


\subsection{Superdeformed rotational band in $^{194}$Hg}
\label{sec:Hgband}

\begin{figure}
\includegraphics[width=7.5cm]{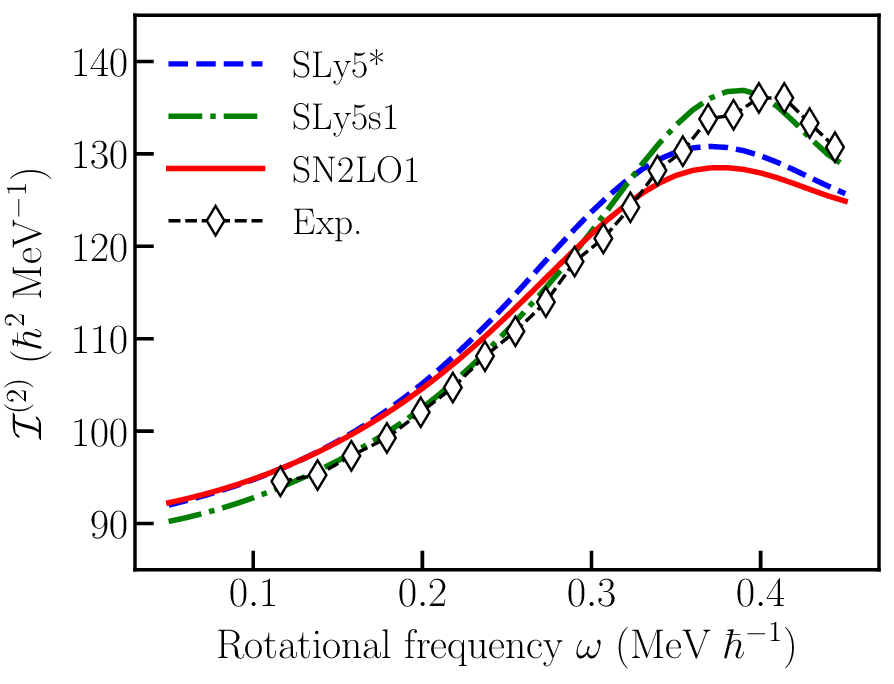}
\caption{Dynamical moment of inertia $\mathcal{I}^{(2)}$ of the superdeformed
         band in $^{194}$Hg as a function of the cranking frequency 
         $\omega$ for the SLy5s1, SLy5* and the 
         SN2LO1 parametrizations. As before, results for NLO (N2LO) parametrizations 
         are plotted with dashed (full) lines. The empty diamond symbols are experimental
         results for the yrast SD-1 band taken from Ref.~\cite{Singh02}.}
\label{fig:band}
\end{figure}

As a final example for yet another type of symmetry-breaking in 
mean-field calculations, we studied the properties of the yrast 
superdeformed rotational band of $^{194}$Hg, which is 
another often-used test bench for properties of the Skyrme EDF, in particular
its time-odd part \cite{Hellemans12,Ryssens19a}. We will focus
on the dynamical moment of inertia $\mathcal{I}^{(2)}$, which is very
sensitive to changes of the nuclear mean field in response to the nucleus'
rotation. Modelling this band as the semiclassical rotation of a 
deformed nuclear shape, this quantity is defined as
\begin{equation}
\label{eq:MOI}
\mathcal{I}^{(2)} 
\equiv \left( \frac{d^2 E}{d J^2} \right)^{-1} \, ,
\end{equation}
where $E$ is the total binding energy energy and $J$ is the size 
of the (classical) angular momentum. The moment of inertia $\mathcal{I}^{(2)}$ 
is usually studied as a function of the rotational frequency
\begin{equation}
\label{eq:freq}
\omega \equiv \frac{dE}{dJ} \, .
\end{equation}
To connect these quantities to experimentally accessible information, 
it is customary to calculate Eqs.~\eqref{eq:MOI} and \eqref{eq:freq} through 
finite differences of in-band $\gamma$-ray transition energies~\cite{Singh02}.

In a self-consistent mean-field model, the moment of inertia and rotational 
frequency can be accessed through cranked calculations, where, instead of the 
energy $E$, one minimizes the Routhian $R$
\begin{equation}
R(\omega) 
= E - \omega_{\rm crank} \langle \hat{J}_z\rangle
\label{eq:Routhian}
\end{equation}
as a function of the frequency $\omega_{\textrm{crank}}$.
In this procedure, the cranking frequency serves as the 
Lagrange parameter for a constraint on the angular momentum along the z-axis, 
which breaks time-reversal invariance. In such a calculation, all 
time-odd densities can take non-zero values, and the calculation reported 
on here is, to the best of our knowledge, the first that probes the time-odd 
terms of a N2LO parametrization in a finite nucleus. Shapes have been 
limited to triaxial ones as done in earlier calculations of this band
\cite{Gall1994,Terasaki95,Hellemans12,Ryssens19a}.

Figure~\ref{fig:band} confronts calculated values for the dynamical moment of inertia 
along the super-deformed band of $^{194}$Hg with experimental data on the 
band labelled ``SD-1'' from Ref.~\cite{Singh02}. In order to make the comparison, 
we have identified $\omega_{\rm crank}$ with the rotational frequency 
$\omega$ as deduced from observed $\gamma$-ray energies. 
Calculations with SN2LO1 and SLy5* produce almost identical curves, while 
the SLy5s1 moments of inertia are slightly lower at low spin, but 
then cross over and become slightly higher at high spin.
Overall, all three parametrizations provide a good descriptions of the 
experimental data, with a slight preference for SLy5s1.
The differences between them, however, are in fact on the same scale or 
even smaller than what is typically found when comparing different 
Skyrme NLO parameter sets~\cite{Hellemans12,Ryssens19a}, or different pairing 
models for the same parameter set~\cite{Terasaki95}, or when slightly 
varying pairing strengths~\cite{Terasaki95,Hellemans12} or specific 
coupling constants of NLO time-odd terms that are under-constrained 
by present fit protocols~\cite{Hellemans12}.

A more technical point implied by Figure~\ref{fig:band} is that the 
SN2LO1 parametrization is stable with respect to spurious finite-size 
instabilities in the spin channel even when all time-odd terms are taken into
account that are generated by the underlying two-body pseudopotential.
These calculations validate a posteriori also for N2LO EDFs the stability criteria proposed
in Ref.~\cite{Hellemans13} that were used during the parameter adjustment of
SLy5s1~\cite{Jodon16} and SN2LO1~\cite{Becker17}.
The earlier SLy5*~\cite{Pastore13} was adjusted along the same lines, 
which assured for the first time the stability of Skyrme EDFs against 
finite-size spin instabilities that are exhibited by many parameter sets of the 
Skyrme NLO EDF when using coupling constants $\cc{A}^{(2,1)}_{t,\textrm{o}}$ of the 
$\vec{D}^{1, \sigma}_{t}  (\vec{r}) \cdot \big[ \Delta \vec{D}^{1, \sigma}_{t} (\vec{r}) \big]$ 
terms in the NLO EDF \eqref{eq:SkTodd:2} as obtained from a generating two-body 
pseudopotential~\cite{Hellemans13,Pastore15}.


\section{Conclusions, Summary and Outlook}
\label{sec:conclusion}

We have presented an exploratory study of deformed nuclei using the 
N2LO Skyrme EDF proposed in Ref.~\cite{Becker17}, 
addressing the optimal formal and numerical representation of such extended 
Skyrme functionals in terms of local densities, and analyzing the performance 
of the SN2LO1 parameter set of Ref.~\cite{Becker17}.

Our main observations and conclusions concerning the formal representation
of the Skyrme EDF at N2LO are 
\begin{enumerate}
\item[(i)]
Going from NLO to N2LO requires the introduction of additional local 
densities~\cite{Carlsson08,Raimondi11a,Becker15,Becker17} that either contain additional 
gradients, or that have a higher-rank cartesian tensor structure. For the N2LO 
functional generated from a locally gauge-invariant central two-body pseudopotential 
with four gradients considered here, four new normal densities as well as two 
generalizations of previously defined densities to higher tensor rank 
are needed. The notation that was employed in the first exploratory studies 
of such EDF in Refs.~\cite{Becker15,Becker17} is not very transparent: it 
leads to an inflation in the number of symbols, none of which provides an 
indication of the operator structure of a given density and thereby limits the 
extensibility of this strategy to arbitrary order in gradients. 

\item[(ii)]
Several of the densities introduced in Refs.~\cite{Becker15,Becker17} to define
the N2LO functional turn out to be complex spatial functions and therefore 
do not exhibit definite behavior under time-reversal.

\item[(iii)]
As a solution to the problems and ambiguities concerning 
notation, we propose a new scheme for writing local densities and the 
corresponding mean fields. This notation indicates in a natural way the 
operator structure of densities and potentials at \textit{any} order in gradients.
Also, it produces by construction only densities that are real functions 
with definite behavior under time-reversal and thereby naturally separates
the parts of the N2LO functional that lead to time-even and time-odd mean 
fields.

\item[(iv)]
It turns out that some components of the higher-order densities 
introduced in Refs.~\cite{Becker15,Becker17} to define the N2LO functional 
can be expressed through gradients acting on lower-order densities.
As the latter are much simpler to handle formally and numerically, the 
use of such reducible densities should be avoided. The new notation 
introduced here helps to identify densities that are non-reducible
in that sense.

\item[(v)]
The definition of the Skyrme EDF in terms of local densities is not unique.
Beginning at NLO, a given EDF can be expressed through several different, 
but equivalent sets of densities that differ by a recoupling of gradients. 
To limit the number of densities and
their associated potentials to be calculated and stored, the use of redundant 
densities, that is densities that can be expressed as a linear combination 
of other densities and their gradients, should clearly be avoided.

\item[(vi)]
Profiting from the new notation and the freedom to recouple gradients, 
we propose a set of rules that guide 
choices for the construction of a set of non-redundant and non-reducible 
densities that are advantageous in terms of the computational cost of 
the floating-point operations necessary to construct a given local 
density and to apply the corresponding term in the single-particle Hamiltonian 
on a single-particle state, and also in terms of the memory size required to 
store the derivatives of single-particle states needed along the way. The set 
of densities proposed in Sec.~\ref{sec:choice} offers in our opinion the 
most straightforward and efficient way to represent the local, central 
N2LO functional of Ref.~\cite{Becker17}.

\end{enumerate}
We believe that the further exploration of degrees of freedom 
associated with higher-order gradients can benefit greatly from a 
more systematic notation like the one proposed here and its 
consequent categorization of densities. 
Not only does it allow to more readily identify reducibility and 
redundancies in any given set of local densities, it opens to the way to 
automated implementations of such complicated functionals. Already for the 
N2LO functional form of Ref.~\cite{Becker15,Becker17} that is limited to central 
and locally gauge-invariant terms, a manual implementation is 
time-consuming and error-prone, especially in the context of symmetry-breaking 
codes where every component of vector and tensor densities exhibits different 
symmetry properties. The future implementation of other higher-order terms such 
as the N2LO tensor pseudopotential and N3LO terms in general 
\cite{Davesne14,Davesne15,Davesne15b,Davesne16} in a deformed nuclear structure 
code will be even more laborious in this respect.

In that context, we recall that also the numerical treatment of N2LO terms in a self-consistent
mean-field solver is not entirely straightforward because of the contributions to 
the single-particle Hamiltonian that are of third and fourth order in gradients. 
Such terms cannot be treated with some of the widely-used algorithms to represent 
and solve the self-consistent mean-field equations~\cite{Becker17}. It turns out, 
however, that the N2LO terms can be very precisely handled in a coordinate-space 
representation based on a Lagrange-mesh, which consistently treats derivatives 
of all orders~\cite{Ryssens19b}, without any need for new developments. This 
is the technique we use here. While methods to directly integrate up a differential 
equation might face problems in the context of N2LO EDFs, iterative schemes 
in the spirit of the gradient-descent method to diagonalize the single-particle 
Hamiltonian can be applied without fundamental technical difficulties at N2LO. 
The convergence of such methods, however, can become an issue, but 
algorithms exist that offer comparable efficiency for NLO and N2LO 
functionals~\cite{Ryssens19b}. These are the techniques we use here.

We have also reported on the first symmetry-breaking self-consistent 
calculations with existing parametrizations of the  N2LO functional
beyond the convergence tests presented in Ref.~\cite{Ryssens19b}:
\begin{enumerate}
\item[(i)]
The examples cover typical applications for heavy nuclei: ground-state 
deformation and its impact on charge radii, rotational bands, and fission
barriers. For their description, we had to consider
a multitude of broken symmetries: breaking of rotational 
symmetry considering both axial and non-axial configurations, breaking
of reflection symmetry, and breaking of time-reversal symmetry. All of these are
frequently considered and explored in calculations of properties of finite 
nuclei.

\item[(ii)]
The example of the superdeformed band of $^{194}$Hg indicates 
that SN2LO1 is stable against finite-size spin-instabilities, such 
that time-reversal breaking calculations can be safely performed 
with this parametrization. 

\item[(iii)]
The results presented here, and others not mentioned, indicate
that the SN2LO1 parametrization shares both the successes and failures
of the NLO parametrizations: SN2LO1 describes the properties of 
deformed finite nuclei about as well as the NLO parameter set SLy5* that was
adjusted with the same fit protocol. Differences between these two 
parameter sets are much smaller than what is typically found when comparing
NLO parametrizations obtained with different fit protocols.

\item[(iv)]
While a careful covariance analysis has shown that the coupling constants 
of the N2LO terms considered for SN2LO1 are from a statistical point
of view significantly different from zero~\cite{Becker19}, they remain 
rather small and do neither significantly nor systematically improve the agreement 
with experiment compared to parametrizations of an NLO EDF adjusted with the 
same fit protocol, at least for the  set of results and observables studied here.

\end{enumerate}
Altogether, our results demonstrate that the N2LO functional of 
Ref.~\cite{Becker15,BeckerPhD,Becker17,Becker19} can be reliably used
to describe the structure of complex nuclei. There is, however, a clear 
need to better constrain the EDF during the parameter adjustment with 
observables that are particularly sensitive to the additional degrees of freedom
offered by the new N2LO terms and can discriminate them from the terms 
in the standard NLO functional. Work in that direction is underway, as is
the further extension of the Skyrme EDF through N2LO tensor terms 
as well as N3LO terms \cite{Davesne14,Davesne15,Davesne15b,Davesne16}.


\section*{Acknowledgments}

We wish to thank Pierre Becker, Karim Bennaceur, Dany Davesne and Jacques Meyer 
for many stimulating discussions that helped to shape the material presented in 
this paper, and Paul Proust for many fruitful discussions on its efficient
presentation.
The work of W.~R.\ was in part supported by the 
U.S.\ DOE grant No.\ DE-SC0019521, and in part by the FNRS (Belgium).  
The work of M.B. was supported by the french Agence Nationale de la Recherche 
under grant No.\ 19--CE31--0015--01 (NEWFUN). The computations were performed 
using HPC resources from the Consortium des \'Equipements de Calcul Intensif 
(CÉCI), funded by the Fonds de la Recherche Scientifique de Belgique (F.R.S.-FNRS) under
Grant No.~2.5020.11, the computing center of the IN2P3/CNRS as well as the 
computing resources of Yale University.


\begin{appendix}


\section{Transforming the functional from the original of Ref.~\cite{Becker17}}
\label{app:BeckerCD}

In this appendix, we show explicitly how to transform the N2LO 
energy density of Ref.~\cite{Becker17}, Eq.~\eqref{eq:originalN2LO}, into the 
form we propose, Eqs.~\eqref{eq:SkTeven:4} and \eqref{eq:SkTodd:4}. We start by 
rewriting Eq.~\eqref{eq:originalN2LO} in the new notation, employing the dictionary 
of subsection~\ref{sec:dictionary} and separating the energy density into
time-even and time-odd parts 
$\mathcal{E}^{(4)}_t (\vec{r}) =  \mathcal{E}^{(4)}_{t, \rm e} (\vec{r})  + \mathcal{E}^{(4)}_{t,\rm o} (\vec{r}) \, ,$
\begin{widetext}
\begin{align}
  \mathcal{E}^{(4)}_{t,\rm e} (\vec{r}) =
    +  C^{(4) \Delta \rho}_t &  \left[ \Delta D^{1,1}_t (\vec{r})\right]^2 
\nonumber \\
     + C^{(4) M\rho }_t  &      
      \left\{    
                       D^{1,1}_t(\vec{r}) \, D^{\Delta, \Delta}_t(\vec{r}) 
              +   \Big[ D^{(\nabla, \nabla)}_t(\vec{r}) \Big]^2
              + 2 \sum_{\mu \nu} D^{\nabla, \nabla}_{t,\mu \nu} (\vec{r}) \,
                                 D^{\nabla, \nabla}_{t,\mu \nu} (\vec{r})
              - 2 \sum_{\mu \nu} D^{\nabla, \nabla}_{t,\mu \nu} (\vec{r})
                                 \left[\nabla_{\mu} \nabla_{\nu} D^{1,1}_t(\vec{r}) \right]
      \right\} 
\nonumber \\
    + C^{(4) M s}_t &
       \left\{ 
            - 2 \sum_{\mu \nu \kappa} C^{\nabla, \nabla \sigma}_{t,\mu \nu \kappa}(\vec{r}) \,
                                      C^{\nabla, \nabla \sigma}_{t,\mu \nu \kappa}(\vec{r})
            -   \sum_{\nu} \left[\sum_{\mu}    \nabla_{\mu}    C^{1,\nabla \sigma}_{t, \mu \nu}    (\vec{r})\right]^2
            - 4 \sum_{\mu \nu} C^{1,\nabla\sigma}_{t,\mu \nu}(\vec{r}) \, C^{(\nabla, \nabla) \nabla \sigma}_{t,\mu\nu}(\vec{r})
       \right\}
\label{eq:app:original_N2LO_inDC_te}
 \, ,\\
 \mathcal{E}^{(4)}_{t,\rm o} (\vec{r}) =
     + C^{(4) \Delta s}_t   &   \left[ \Delta \vec{C}^{1,\sigma}_t (\vec{r})\right]^2
\nonumber \\
     + C^{(4) M\rho }_t & 
          \left\{
              - 2 \sum_{\mu \nu} C^{\nabla, \nabla}_{t,\mu \nu} (\vec{r}) \,
                                 C^{\nabla, \nabla}_{t,\mu \nu} (\vec{r})
              -  \left[ \vnabla \cdot \vec{C}^{1,\nabla}_t(\vec{r})\right]^2
              - 4 \, \vec{C}^{1,\nabla}_t(\vec{r}) \cdot \vec{C}^{(\nabla, \nabla) \nabla}_t(\vec{r})
          \right\}
\nonumber \\
   + C^{(4) M s}_t
   & \Bigg\{ 
              \vec{D}^{1,\sigma}_t(\vec{r}) \cdot \vec{D}^{\Delta, \Delta \sigma}_t(\vec{r}) 
            +   \left[\vec{D}^{(\nabla,\nabla) \sigma}_t(\vec{r}) \right]^2
            + 2 \sum_{\mu \nu \kappa} D^{\nabla, \nabla \sigma}_{t,\mu \nu \kappa}(\vec{r}) \,
                                      D^{\nabla, \nabla \sigma}_{t,\mu \nu \kappa}(\vec{r})
\nonumber \\
   & \;                                     
            - 2 \sum_{\mu \nu \kappa} D^{\nabla, \nabla \sigma}_{t,\mu \nu \kappa}(\vec{r})
                                      \left[ \nabla_{\mu} \nabla_{\nu} D^{1,\sigma}_{t,\kappa} (\vec{r})\right] 
     \Bigg\} \, .
\label{eq:app:original_N2LO_inDC_to}
\end{align}
\end{widetext}
We draw the readers attention to the minus signs in front of the terms 
bilinear in the reducible currents
$C^{\nabla,\nabla}_{t,\mu\nu}(\vec{r})$ and $C^{\nabla,\nabla\sigma}_{t,\mu\nu\kappa}(\vec{r})$,
which result from the square of the imaginary unit {\iunit}. 
We also remark that the term
$\sum_{\mu \nu} C^{\nabla, \nabla}_{t,\mu \nu} (\vec{r}) \, [\nabla_{\mu} \nabla_{\nu} D_t^{1,1}(\vec{r})]$
is absent: it vanishes identically as $C^{\nabla, \nabla}_{t,\mu \nu}(\vec{r})$
is skew-symmetric under the exchange $\mu \leftrightarrow  \nu$, {while the object
in square brackets is symmetric}. The term
$\sum_{\mu \nu} C^{\nabla, \nabla\sigma}_{t,\mu \nu} (\vec{r})[\nabla_{\mu} \nabla_{\nu} D_{t, \kappa}^{1,\sigma}(\vec{r})]$ 
vanishes as well for the same reason.

To obtain the new formulation of the functional, we will eliminate the two 
reducible currents of second order, $C^{\nabla,\nabla}_{t,\mu\nu}(\vec{r})$ and 
$C^{\nabla,\nabla\sigma}_{t,\mu\nu\kappa}(\vec{r})$, 
using Eqs.~\eqref{eq:reducetau} and \eqref{eq:reduceK}. Furthermore, 
we will rewrite two {currents}
of third order, $C^{(\nabla, \nabla) \nabla}_t(\vec{r})$ and 
$C^{(\nabla, \nabla) \nabla \sigma}_t(\vec{r})$ using Eqs.~\eqref{eq:rewriting_pi}
and ~\eqref{eq:laplacianexample_final}, respectively. Eliminating these four 
densities, the relevant terms of Eqs.~\eqref{eq:app:original_N2LO_inDC_te} and 
\eqref{eq:app:original_N2LO_inDC_to} become
\begin{widetext}
\begin{align}
              -2 \, C^{(4)M\rho}_t\sum_{\mu \nu} C^{\nabla, \nabla}_{t,\mu \nu} (\vec{r})  \,
                                 C^{\nabla, \nabla}_{t,\mu \nu} (\vec{r})
=  - C^{(4)M\rho}_t & \sum_{\mu \nu } \left\{ \left[ \nabla_{\mu} C^{1,\nabla}_{t,\nu}(\vec{r}) \right]^2 
-  \left[\nabla_{\mu} C^{1,\nabla}_{t,\nu}(\vec{r}\right]\left[\nabla_{\nu} C^{1,\nabla}_{t,\mu}(\vec{r})\right]\right\} \, ,
\label{eq:app:before_part1}
\\
             - 2 \, C^{(4)Ms}_t \sum_{\mu \nu \kappa}  C^{\nabla, \nabla \sigma}_{t,\mu \nu \kappa} (\vec{r}) \,
                                 C^{\nabla, \nabla\sigma}_{t,\mu \nu \kappa} (\vec{r})
= - C^{(4)Ms}_t & \sum_{\mu \nu } \left\{
\left[ \nabla_{\mu} C^{1,\nabla}_{t,\nu}(\vec{r}) \right]^2 
-  \left[\nabla_{\mu} C^{1,\nabla\sigma}_{t,\nu\kappa}(\vec{r})\right]\left[\nabla_{\nu} C^{1,\nabla\sigma}_{t,\mu\kappa}(\vec{r})\right] \right\}\, ,
\label{eq:app:before_part2} \\
-4 \, C^{(4) M \rho}_t \vec{C}^{1,\nabla}_t(\vec{r}) \cdot \vec{C}^{(\nabla, \nabla) \nabla}_t(\vec{r})
=
-C^{(4) M \rho}_t  & \Big\{
- 4 \,  \vec{C}^{1,\nabla}_t(\vec{r}) \cdot \vec{C}^{\Delta, \nabla}_t(\vec{r})
+2 \, \vec{C}^{1,\nabla}_t(\vec{r}) \cdot \big[ \Delta \vec{C}^{1, \nabla}_t(\vec{r}) \big] \nonumber \\
  & \qquad + 2  \sum_{\mu \nu } C^{1,\nabla}_{t,\mu}(\vec{r})\left[ \nabla_{\mu} \nabla_{\nu} C^{1, \nabla}_{t,\nu}(\vec{r})\right]\Big\}
\label{eq:app:before_part3}
\, , \\
-4 \, C^{(4)M s}_t\sum_{\mu \nu} C^{1,\nabla\sigma}_{t,\mu\nu}(\vec{r}) \, C^{(\nabla, \nabla) \nabla \sigma}_{t,\mu\nu}(\vec{r})
= - C^{(4) M s}_t &
 \sum_{\mu \nu}
 \Big\{
- 4 \, C^{1,\nabla\sigma}_{t,\mu\nu}(\vec{r}) \, C^{\Delta, \nabla\sigma}_{t,\mu \nu}(\vec{r})
+ 2 \, C^{1,\nabla\sigma}_{t,\mu\nu}(\vec{r}) \, \Big[ \Delta C^{1, \nabla\sigma}_{t,\mu\nu}(\vec{r}) \Big] 
 \nonumber \\
& \qquad + 2 \sum_{\kappa} C^{1,\nabla\sigma}_{t,\mu\nu}(\vec{r}) \left[\nabla_{\mu} \nabla_{\kappa} C^{1, \nabla\sigma}_{t,\kappa\nu}(\vec{r})\right]
\Big\}
\label{eq:app:before_part4}
\, .
\end{align}
%
Inserting Eqs.~\eqref{eq:app:before_part1}-\eqref{eq:app:before_part4} into 
Eqs.~\eqref{eq:app:original_N2LO_inDC_te} and \eqref{eq:app:original_N2LO_inDC_to} 
leads to an N2LO energy density $\mathcal{E}^{(4)}_{t}(\vec{r})$ that is 
equivalent (but not {yet} equal) to our formulation of the energy density in 
Eqs.~\eqref{eq:SkTeven:4} and \eqref{eq:SkTodd:4}. To complete the transformation, 
we note the following integral identities
%
\begin{align}
  -  \int d^3 r \,
  \sum_{\mu \nu}  \left[ \nabla_{\mu} C^{1,\nabla}_{t,\nu}(\vec{r}) \right]^2
  &=
  +\int d^3 r \,  
  \sum_{\mu}  C^{1,\nabla}_{t,\mu}(\vec{r}) \, \Delta \, C^{1,\nabla}_{t,\mu}(\vec{r})
  \label{eq:app:replace_1}
  \, , \\
%
  - \int d^3 r \, 
  \sum_{\mu \nu} \left[ \nabla_{\mu} C^{1,\nabla}_{t,\nu}(\vec{r}) \right] 
                                              \left[ \nabla_{\nu} C^{1,\nabla}_{t,\mu}(\vec{r}) \right]
  &=
  - \int d^3 r \, 
  \left[ \sum_{\mu }  \nabla_{\mu} C^{1,\nabla}_{t,\mu}(\vec{r}) \right]^2 \, , \\
 -  \int d^3 r \, 
 \sum_{\mu \nu \kappa} \left[ \nabla_{\mu} C^{1,\nabla \sigma}_{t,\nu \kappa}(\vec{r}) \right]^2
  &=
  + \int d^3 r \, 
  \sum_{\mu \nu}  C^{1,\nabla\sigma}_{t,\mu\nu}(\vec{r}) \, \Delta \, C^{1,\nabla \sigma}_{t,\mu \nu}(\vec{r})  \, , \\
%
  -\int d^3 r \, 
  \sum_{\mu \nu \kappa} \left[ \nabla_{\mu} C^{1,\nabla\sigma}_{t,\nu \kappa}(\vec{r}) \right] 
                                                  \left[ \nabla_{\nu} C^{1,\nabla\sigma}_{t,\mu \kappa}(\vec{r}) \right]
  &=
  -\int d^3 r \, 
   \sum_{\nu } \left[ \sum_{\mu} \nabla_{\mu} C^{1,\nabla \kappa}_{t,\mu \nu}(\vec{r}) \right]^2 \, , \\
  - \int d^3 r \, 
  \sum_{\mu \nu } C^{1,\nabla}_{t,\mu}(\vec{r})\left[ \nabla_{\mu} \nabla_{\nu} C^{1, \nabla}_{t,\nu}(\vec{r})\right]
  & =
  + \int d^3 r \, 
  \left[ \vnabla \cdot \vec{C}^{1,\nabla}_{t}(\vec{r}) \right]^2
  \, , \\
  - \int d^3 r \, 
   \sum_{\mu \nu\kappa} C^{1,\nabla\sigma}_{t,\mu\nu}(\vec{r}) \left[\nabla_{\mu} \nabla_{\kappa} C^{1, \nabla\sigma}_{t,\kappa\nu}(\vec{r})\right]
  & =
  + \int d^3 r \, 
  \sum_{\nu} \left[ \sum_{\mu}  \nabla_{\mu}  C^{1, \nabla\sigma}_{t,\mu\nu}(\vec{r})\right]^2 \, ,
  \label{eq:app:replace_last}
\end{align}
\end{widetext}
which can all be verified easily by partial integration. We are now 
allowed to replace the integrands on the left-hand-side of 
Eqs.~\eqref{eq:app:replace_1}-\eqref{eq:app:replace_last} in the formulation 
of the energy density by the expressions under the right-hand integral. 
While these replacements result in a different energy density 
$\mathcal{E}_{\rm Sk}^{(4)}$, they do not modify the total energy and hence
do not modify the physics of the EDF.

Combining all of the above, we obtain the final form of the N2LO energy 
density, provided we identify the N2LO coupling constants 
$\cc{A}^{(4,i)}_{\rm t,e/o}$ with the coupling constants of
Ref.~\cite{Becker17} as follows:
\begin{align}
\cc{A}_{t, \rm e}^{(4,1)} &= + \phantom{1} C^{(4) \Delta \rho}_{t} \, , &
\cc{A}_{t, \rm o}^{(4,1)} &= + \phantom{1} C^{(4) \Delta s   }_{t} \, , \nonumber \\
\cc{A}_{t, \rm e}^{(4,2)} &= + \phantom{1} C^{(4) M      \rho}_{t} \, , &
\cc{A}_{t, \rm o}^{(4,2)} &= + \phantom{1} C^{(4) M      s   }_{t} \, , \nonumber \\
\cc{A}_{t, \rm e}^{(4,3)} &= + \phantom{1} C^{(4) M      \rho}_{t} \, , &
\cc{A}_{t, \rm o}^{(4,3)} &= + \phantom{1} C^{(4) M      s   }_{t} \, , \nonumber \\
\cc{A}_{t, \rm e}^{(4,4)} &= +          2  C^{(4) M      \rho}_{t} \, , &
\cc{A}_{t, \rm o}^{(4,4)} &= +          2  C^{(4) M      s   }_{t} \, , \nonumber \\
\cc{A}_{t, \rm e}^{(4,5)} &= -          2  C^{(4) M      \rho}_{t} \, , &
\cc{A}_{t, \rm o}^{(4,5)} &= -          2  C^{(4) M      s   }_{t} \, , \nonumber \\
\cc{A}_{t, \rm e}^{(4,6)} &= - \phantom{1} C^{(4) M      s   }_{t} \, , &
\cc{A}_{t, \rm o}^{(4,6)} &= - \phantom{1} C^{(4) M      \rho}_{t} \, , \nonumber \\
\cc{A}_{t, \rm e}^{(4,7)} &= -          2  C^{(4) M      s   }_{t} \, , &
\cc{A}_{t, \rm o}^{(4,7)} &= -          2  C^{(4) M      \rho}_{t}\, ,  \nonumber \\
\cc{A}_{t, \rm e}^{(4,8)} &= +          4  C^{(4) M      s   }_{t}\, , &
\cc{A}_{t, \rm o}^{(4,8)} &= +          4  C^{(4) M      \rho}_{t}\, .
\end{align}

\section{Coupling constants}
\label{app:coupling}

In the expressions for the Skyrme EDF~\eqref{eq:SkTeven:0}--\eqref{eq:SkTodd:4}, 
we have given each term an individual coupling constant $\cc{A}^{(n, i)}_{t, \rm e/o}$, 
which is the most straightforward way for the implementation as well as subsequent 
testing and debugging of an EDF in a numerical code. Not all of these coupling 
constants, however, necessarily represent an independent degree of freedom of 
the EDF. While each individual term of the Skyrme EDF defined in  
Eqs.~\eqref{eq:SkTeven:0}-\eqref{eq:SkTodd:4} is invariant under spatial 
rotations, translations, space and time inversion of the coordinates used
to describe the nucleus,
there are a few more subtle invariances that in some cases can only be 
satisfied by specific combinations of several terms in the EDF, thereby 
reducing the number of independent coupling constants. The most prominent 
ones are Galilean invariance, local gauge invariance, and that the EDF
is generated by an underlying anti-symmetrized many-body interaction. 

Galilean invariance of the EDF, which signifies that only the kinetic 
energy changes in a specific way when going from one inertial frame of 
reference to another one, but not the interaction 
energy~\cite{Dobaczewski95,Dobaczewski96b,Carlsson08,Raimondi11a}, 
is a necessity for any meaningful dynamical EDF calculation. This concerns
time-dependent mean-field calculations and their limiting cases such 
as linear response theory and also cranked mean-field calculations 
such as the ones reported on in Sec.~\ref{sec:Hgband}.
Among the additional symmetries mentioned above, Galilean invariance 
introduces the smallest set of interdependencies among the coupling 
constants of the EDF~\cite{Carlsson08}.

According to Noether's second theorem, local gauge invariances are 
not at the origin of conservation laws, but impose constraints on the 
form of the equations of motion.
In the context of nuclear EDF methods, the requirement of local gauge 
invariance ensures that the continuity equations keep the simple form 
of Eqs.~\eqref{eq:continuity:normal} and~\eqref{eq:continuity:spin}. 
It has been pointed out that energy density functionals that are 
invariant under local gauge transformations are also automatically 
Galilean invariant \cite{Raimondi11a}, but not vice versa. Up to
NLO, however, these two symmetries impose the same relations 
among the coupling constants of the EDF \cite{Carlsson08,Raimondi11a}. 
Beginning with N2LO, one can construct Galilean-invariant terms that 
are not invariant under arbitrary local gauge transformations anymore.

Another requirement that can be imposed is that the EDF is generated
by an underlying effective many-body interaction, which signifies that
the EDF has the exchange symmetry of the Pauli principle~\cite{Bender09b,Stringari78a}.
The majority of parametrizations of the Skyrme NLO EDF does not 
satisfy such requirement, see for example 
Refs.~\cite{Bender03,Lesinski07,Bender09,Hellemans12,Chamel09,Ryssens15a}
and references therein for a discussion of motivations for, and consequences 
of, this practice. In recent years, there are efforts going in the opposite 
directions of either exploiting the additional freedom of lifting the constraints from
an underlying generating operator~\cite{Kortelainen10,Kortelainen12,Kortelainen14},
or of respecting those constraints as much as possible at least for the 
particle-hole part of the EDF, although such EDFs usually allow for 
density-dependent terms that also might be problematic in this 
context~\cite{Stringari78a,Robledo07a,Duguet09a,Robledo10a}. 
The necessary Galilean and optional local-gauge invariance of such operator
is then automatically transferred to the terms of the EDF that it generates. 

The construction of the SN2LO1 parametrization of Ref.~\cite{Becker17} has followed 
the latter strategy: the EDF has been generated from a locally-gauge-invariant
\footnote{Note
that Eq.~\eqref{eq:def:Vsk} only contains the locally-gauge-invariant central 
terms, but not the tensor terms discussed for example in
Refs.~\cite{Davesne13,Davesne14,Davesne15,Davesne15b,Davesne16}, which 
explains the presence of only four out of the six possible locally-gauge-invariant
terms at N2LO as identified in Ref.~\cite{Carlsson08,Raimondi11a}.  
} 
effective interaction $\hat{V}_{\rm Sk}$ of the form of Eq.~\eqref{eq:def:Vsk}
\cite{Davesne13,Davesne14,Davesne15,Becker15,Becker17,BeckerPhD}.
In this case, all 56 coupling constants of the Skyrme EDF
of Eqs.~\eqref{eq:SkTeven:0}--\eqref{eq:SkTodd:4}  can be 
related to the following much smaller set of thirteen parameters of 
$\hat{V}_{\rm Sk}$
\begin{align}
t_0, x_0, 
t_1^{(2)}, x_1^{(2)}, t_2^{(2)}, x_2^{(2)}, 
t_1^{(4)}, x_1^{(4)}, t_2^{(4)}, x_2^{(4)}, 
t_3, x_3, W_0 \, . \nonumber
\end{align}
The exponent $\alpha$ of the density dependence constitutes an additional
parameter of both sets.

As in Ref.~\cite{Becker17}, the NLO density-dependent interaction 
$\hat{V}^{\rm DD}_{\rm NLO} $ is parameterized by the parameters $t_3$, $x_3$, 
and $\alpha$,
while the parameter $W_0$ determines the NLO spin-orbit interaction 
$\hat{V}^{\rm SO}_{\rm NLO}$. The remaining parameters determine the central 
part of the interaction $\hat{V}^{C}_{\rm N2LO}$; this interaction reduces
to an NLO Skyrme interaction if the parameters 
$t_1^{(4)}, x_1^{(4)}, t_2^{(4)}, x_2^{(4)}$ are set to zero.

For the coupling constants of the central terms $\hat{V}^{\rm C}$, it
will be useful to use the following shorthand notation
%
%
\begin{alignat}{2}
C^{+}_{00} (t,x) &= +\tfrac{3}{8} t \, , \phantom{+\tfrac{1}{4} tx } & \quad
C^{+}_{01} (t,x) &= -\tfrac{1}{8} t - \tfrac{1}{4} tx           \, , \nonumber \\
C^{+}_{10} (t,x) &= -\tfrac{1}{8} t + \tfrac{1}{4} tx           \, , &
C^{+}_{11} (t,x) &= -\tfrac{1}{8} t \, , \phantom{+ \tfrac{1}{4} tx} \nonumber \\
C^{-}_{00} (t,x) &= +\tfrac{5}{8} t + \tfrac{1}{2} tx           \, , &
C^{-}_{01} (t,x) &= +\tfrac{1}{8} t + \tfrac{1}{4} tx           \, , \nonumber \\
C^{-}_{10} (t,x) &= +\tfrac{1}{8} t + \tfrac{1}{4} tx           \, , &
C^{-}_{11} (t,x) &= +\tfrac{1}{8} t \, .\phantom{+ \tfrac{1}{4} tx } 
\label{eq:CCshort:8}
\end{alignat}
The $C_{ST}^{\pi}(t,x)$ represent the generic coupling constants that 
are obtained when evaluating the fully anti-symmetrized Hartree-Fock expectation value 
of a generic central contact interaction\footnote{Similar generic combinations 
can also be constructed for spin-orbit and tensor interactions, but for the 
specific functional discussed here doing so will not further simplify the notation.}
\begin{align}
\langle \text{HF} & | \hat{V}^{\rm C}_i | \text{HF} \rangle
\nn \\
= & \, \langle \text{HF} | t_i \, ( 1 + x_i \hat{P}^{\sigma}_{12} ) \, 
\hat{O}^{\pi} (\vec{r}_1, \vec{r}_2, \vec{r}'_1, \vec{r}_2')
| \text{HF} \rangle
      \nn \\
  = & \,  \iiiint \! \rmd^3 r_1 \; \rmd^3 r_2 \; \rmd^3 r_1' \; \rmd^3 r_2' \;
      \hat{O}^{\pi} (\rvec_1, \rvec_2; \rvecp_1, \rvecp_2)
      \nn \\
&     \times \sum_{t=0,1} 
      \Big[
        C_{0t}^{\pi}(t_i,x_i) \,
        \rho_{t} (\rvec_1, \rvecp_1) \, \rho_{t} (\rvec_2, \rvecp_2)  
     \nn \\
&    \quad \quad  \quad     
       + C_{1t}^{\pi}(t_i,x_i) \,
        \vec{s}_{t} (\rvec_1, \rvecp_1) \cdot \vec{s}_{t} (\rvec_2, \rvecp_2)
      \Big] \, ,
\end{align}
where the $\hat{O}^{\pi} (\vec{r}_1, \vec{r}_2, \vec{r}'_1, \vec{r}_2')$
is the part of the operator of the $i$th term in the pseudopotential
$\hat{V}^{\rm C}$ that acts in position space and that has the parity 
$\pi = \pm$ under spatial inversion. For a central contact interaction, this part 
of the operator is by construction a scalar in coordinate space and does 
not act in spin or isospin space.

In terms of this shorthand, the expressions for the coupling constants of the 
time-even LO and NLO terms of the Skyrme functional, Eqs.~\eqref{eq:SkTeven:0} 
and \eqref{eq:SkTeven:2} are: 
\begin{alignat}{4}
\cc{A}_{t,\rm e}^{(0,1)} =&   
 + \phantom{\tfrac{1}{6}}  C^{+}_{0t}(t_0, x_0) \, , &
\nonumber \\
\cc{A}_{t,\rm e}^{(0,2)} =& 
 + \tfrac{1}{6}    C^{+}_{0t}(t_3, x_3) \, ,  &
\nonumber \\
\cc{A}_{t,\rm e}^{(2,1)} =&  
  -  \tfrac{3}{8} C^{+}_{0t}(t_1^{(2)}, x_1^{(2)}) 
  +  \tfrac{1}{8} C^{-}_{0t}(t_2^{(2)}, x_2^{(2)}) \, , &
\nonumber \\
\cc{A}_{t,\rm e}^{(2,2)} =&
  +   \tfrac{1}{2} C^{+}_{0t}(t_1^{(2)}, x_1^{(2)}) 
  +   \tfrac{1}{2} C^{-}_{0t}(t_2^{(2)}, x_2^{(2)}) \, , & 
\nonumber \\
\cc{A}_{t,\rm e}^{(2,3)} =&  
  -  \tfrac{1}{2} C_{1t}^{+}(t_1^{(2)},x_1^{(2)})   
  -  \tfrac{1}{2} C_{1t}^{-}(t_2^{(2)},x_2^{(2)})  \, , & 
\nonumber \\
\cc{A}_{0,\rm e}^{(2,4)} =& 
            - \tfrac{3}{4}  W_{\rm 0} \, , & \nonumber \\
\cc{A}_{1,\rm e}^{(2,4)} = &  
            - \tfrac{1}{4}  W_{\rm 0} \, . &
\end{alignat}
The coupling constants for the time-odd terms at LO and NLO are given by
\begin{alignat}{4}
\cc{A}_{t,\rm o}^{(0,1)} =& 
 + \phantom{\tfrac{1}{6}}  C^{+}_{1t}(t_0, x_0) \, , &
\nonumber \\
\cc{A}_{t,\rm o}^{(0,2)} =& 
 +  \tfrac{1}{6}   C^{+}_{1t}(t_3, x_3) \, , &
\nonumber \\
\cc{A}_{t,\rm o}^{(2,1)} =&
            - \tfrac{3}{8}  C_{1t}^+ (t_1^{(2)}, x_1^{(2)})      &
            + \tfrac{1}{8}  C_{1t}^- (t_2^{(2)}, x_2^{(2)})\, ,  &
\nonumber \\
\cc{A}_{t,\rm o}^{(2,2)} = &  
             + \tfrac{1}{2}  C_{1t}^{+}(t_1^{(2)},x_1^{(2)})  & 
             + \tfrac{1}{2}  C_{1t}^{-}(t_2^{(2)},x_2^{(2)})\, ,   &
\nonumber \\
\cc{A}_{t,\rm o}^{(2,3)} = & 
            - \tfrac{1}{2}   C^{+}_{0t}(t_1^{(2)}, x_1^{(2)}) &
            - \tfrac{1}{2}   C^{-}_{0t}(t_2^{(2)}, x_2^{(2)})\, ,   &
\nonumber \\
\cc{A}_{0,\rm o}^{(2,4)} =&
            - \tfrac{3}{4}  W_0 \, , & & 
\nonumber \\
\cc{A}_{1,\rm o}^{(2,4)} =&
            - \tfrac{1}{4}  W_0\, . & & 
\end{alignat}
The coupling constants appearing in the time-even, N2LO part of the functional 
$\mathcal{E}_{\rm Sk, e}^{(4)}$ of Eq.~\eqref{eq:SkTeven:4}, are
\begin{alignat}{5}
\cc{A}_{t,\rm e}^{(4,1)} = & 
  + \tfrac{3}{16} &  C^{+}_{0t}(t_1^{(4)}, x_1^{(4)})  &
  - \tfrac{1}{16} &  C^{-}_{0t}(t_2^{(4)}, x_2^{(4)}) \, , \nonumber\\
\cc{A}_{t,\rm e}^{(4,2)} =  
 &  + \tfrac{1}{4} & C^{+}_{0t}(t_1^{(4)}, x_1^{(4)})  &
    + \tfrac{1}{4} & C^{-}_{0t}(t_2^{(4)}, x_2^{(4)}) \, , \nonumber\\
\cc{A}_{t,\rm e}^{(4,3)} =  
 & + \tfrac{1}{4} & C^{+}_{0t}(t_1^{(4)}, x_1^{(4)}) &
   + \tfrac{1}{4} & C^{-}_{0t}(t_2^{(4)}, x_2^{(4)}) \, , \nonumber\\
\cc{A}_{t,\rm e}^{(4,4)} =
 & + \tfrac{1}{2} & C^{+}_{0t}(t_1^{(4)}, x_1^{(4)}) &
   + \tfrac{1}{2} & C^{-}_{0t}(t_2^{(4)}, x_2^{(4)}) \, , \nonumber\\
\cc{A}_{t,\rm e}^{(4,5)} =
 & -  \tfrac{1}{2} & C^{+}_{0t}(t_1^{(4)}, x_1^{(4)}) &
   -  \tfrac{1}{2} & C^{-}_{0t}(t_2^{(4)}, x_2^{(4)}) \, , \nonumber\\
\cc{A}_{t,\rm e}^{(4,6)} = 
 & -  \tfrac{1}{4} & C^{+}_{1t}(t_1^{(4)}, x_1^{(4)}) &
   -  \tfrac{1}{4} & C^{-}_{1t}(t_2^{(4)}, x_2^{(4)})\, , \nonumber \\
\cc{A}_{t,\rm e}^{(4,7)} = 
 & -  \tfrac{1}{2} & C^{+}_{1t}(t_1^{(4)}, x_1^{(4)})  &
   -  \tfrac{1}{2} & C^{-}_{1t}(t_2^{(4)}, x_2^{(4)})\, , \nonumber \\ 
\cc{A}_{t,\rm e}^{(4,8)} =  
&  + \phantom{\tfrac{1}{1}} & C^{+}_{1t}(t_1^{(4)}, x_1^{(4)})  &
   + \phantom{\tfrac{1}{1}} & C^{-}_{1t}(t_2^{(4)}, x_2^{(4)})     \,.
\end{alignat}
We have for the coupling constants appearing in the time-odd N2LO energy 
density $\mathcal{E}_{\rm Sk,o}^{(4)}$ of Eq.~\eqref{eq:SkTodd:4}
\begin{alignat}{5}
\cc{A}_{t,\rm o}^{(4,1)} = 
 &          +  \tfrac{3}{16} & C^{+}_{1t}(t_1^{(4)}, x_1^{(4)}) &
            -  \tfrac{1}{16} & C^{-}_{1t}(t_2^{(4)}, x_2^{(4)}) \, , \nonumber \\
\cc{A}_{t,\rm o}^{(4,2)} =  
 &          +  \tfrac{1}{4}  & C^{+}_{1t}(t_1^{(4)}, x_1^{(4)}) &
            +  \tfrac{1}{4}  & C^{-}_{1t}(t_2^{(4)}, x_2^{(4)}) \, , \nonumber\\
\cc{A}_{t,\rm o}^{(4,3)} =  
 &          +  \tfrac{1}{4}  & C^{+}_{1t}(t_1^{(4)}, x_1^{(4)}) &
            +  \tfrac{1}{4}  & C^{-}_{1t}(t_2^{(4)}, x_2^{(4)}) \, , \nonumber \\
\cc{A}_{t,\rm o}^{(4,4)} =
 &          + \tfrac{1}{2}   & C^{+}_{1t}(t_1^{(4)}, x_1^{(4)}) &
            + \tfrac{1}{2}   & C^{-}_{1t}(t_2^{(4)}, x_2^{(4)}) \, ,  \nonumber\\
\cc{A}_{t,\rm o}^{(4,5)} =
 &          - \tfrac{1}{2}   & C^{+}_{1t}(t_1^{(4)}, x_1^{(4)}) &
            - \tfrac{1}{2}   & C^{-}_{1t}(t_2^{(4)}, x_2^{(4)}) \, , \nonumber\\
\cc{A}_{t,\rm o}^{(4,6)} = 
 &          - \tfrac{1}{4}   & C^{+}_{0t}(t_1^{(4)}, x_1^{(4)}) &
            - \tfrac{1}{4}   & C^{-}_{0t}(t_2^{(4)}, x_2^{(4)}) \, , \nonumber\\
\cc{A}_{t,\rm o}^{(4,7)} = 
 &          - \tfrac{1}{2}   & C^{+}_{0t}(t_1^{(4)}, x_1^{(4)})  &
            - \tfrac{1}{2}   & C^{-}_{0t}(t_2^{(4)}, x_2^{(4)}) \, , \nonumber\\ 
\cc{A}_{t,\rm o}^{(4,8)} =  
 &          + \phantom{\tfrac{1}{1}}  & C^{+}_{0t}(t_1^{(4)}, x_1^{(4)})  &
            + \phantom{\tfrac{1}{1}}  & C^{-}_{0t}(t_2^{(4)}, x_2^{(4)})  \,.
\end{alignat}
A detailed analysis of Galilean invariance of particle-hole terms in the 
N2LO EDF that is hidden in some of the above relations will be given elsewhere.


\section{Mean-field potentials}
\label{app:potentials}

The mean-field potentials associated with the N2LO functional can be obtained 
in a straight-forward but slightly tedious way by varying the energy with 
respect to the local normal densities, Eq.~\eqref{eq:variation}. 
As a reference, we include their full expressions for the SN2LO1 
parametrization here. 
Defining a shorthand for the combinations of coupling constants that appear in contributions 
from the same ($q' = q$) and the other ($q' \neq q$) nucleon species
\begin{align}
\cc{A}^{(i,j)}_{qq',x} 
& =   \big( \cc{A}^{(i,j)}_{0,x} + \cc{A}^{(i,j)}_{1,x} \big) \, \delta_{qq'}
    +\big( \cc{A}^{(i,j)}_{0,x} - \cc{A}^{(i,j)}_{1,x} \big) \big( 1 - \delta_{qq'} \big) 
    \nn \\
& = \cc{A}^{(i,j)}_{0,x} 
   + \big( 2 \delta_{qq'} - 1 \big)  \cc{A}^{(i,j)}_{1,x} \, , 
\end{align}
the contribution from the Skyrme EDF to the potentials for the nucleon species 
$q = p$, $n$ that are associated with time-even densities are given by
\begin{widetext}
\begin{align}
F_{q}^{1,1} (\vec{r}) 
 = & \sum_{q' = p,n} \bigg\{ 
      2 \, \cc{A}^{(0,1)}_{qq',\text{e}} D^{1,1}_{q'} (\vec{r})
    + 2 \, \cc{A}^{(0,1)}_{qq',\text{e}} \big[ D^{1,1}_{0} (\vec{r}) \big]^\alpha \, D^{1,1}_{q'} (\vec{r})
     \nn \\
   & + 2 \cc{A}^{(2,1)}_{qq',\textrm{e}} \big[ \Delta D^{1,1}_{q'} (\vec{r}) \big]
     +   \cc{A}^{(2,2)}_{qq'\textrm{e}} D^{(\nabla,\nabla)}_{q'}
     +   \cc{A}^{(2,4)}_{qq',\textrm{e}} \big(  \vnabla \cdot \vec{C}^{1,\nabla \times \sigma}_{q'} \big)
     \nn \\
   & + 2 \cc{A}^{(4,1)}_{qq',\textrm{e}} \big[ \Delta \Delta D^{1,1}_{q'} (\vec{r}) \big]   
     +   \cc{A}^{(4,2)}_{qq',\textrm{e}} D^{\Delta, \Delta}_{q'} (\vec{r}) 
     +   \cc{A}^{(4,5)}_{qq',\textrm{e}}  
      \sum_{\mu,\nu} \big[ \nabla_\mu \nabla_\nu D^{\nabla,\nabla}_{q',\mu\nu} (\vec{r}) \big]
      \bigg\}
     \nn \\
   & + \alpha \, \cc{A}^{(0,2)}_{0,\text{e}} \, 
       \big[ D^{1,1}_{0} (\vec{r}) \big]^{\alpha-1} \, \big[ D^{1,1}_{0} (\vec{r}) \big]^2
     + \alpha \, \cc{A}^{(0,2)}_{1,\text{e}} \, 
       \big[ D^{1,1}_{0} (\vec{r}) \big]^{\alpha-1} \, \big[ D^{1,1}_{1} (\vec{r}) \big]^2
     \nn \\
   & + \alpha \, \cc{A}^{(0,2)}_{0,\text{o}} \, 
       \big[ D^{1,1}_{0} (\vec{r}) \big]^{\alpha-1} \, \big[ \vec{D}^{1,\sigma}_{0} (\vec{r}) \big]^2
     + \alpha \, \cc{A}^{(0,2)}_{1,\text{o}} \, 
       \big[ D^{1,1}_{0} (\vec{r}) \big]^{\alpha-1} \, \big[ \vec{D}^{1,\sigma}_{1} (\vec{r}) \big]^2  \, ,
     \\
G_{q, \mu\nu}^{1,\nabla\sigma}(\vec{r}) 
= & \sum_{q' = p,n} \Big\{
      2 \cc{A}^{(2,3)}_{qq',\textrm{e}} C^{1, \nabla \sigma}_{q', \mu \nu}
       - \cc{A}^{(2,4)}_{qq',\textrm{e}} 
       \sum_{\kappa} \epsilon_{\kappa \mu \nu} \big[ \nabla_{\kappa} D^{1,1}_{q'} (\vec{r}) \big]
      \nn \\
  & + 2 \cc{A}^{(4,6)}_{qq',\textrm{e}} 
      \big[ \Delta C^{1,\nabla\sigma}_{q',\mu \nu} \big]
   - 2 \cc{A}^{(4,7)}_{qq',\textrm{e}} 
      \sum_{\kappa} \big[ \nabla_\mu \nabla_\kappa C^{1,\nabla\sigma}_{q',\kappa \nu} \big]
   + \cc{A}^{(4,8)}_{qq',\textrm{e}} C^{\Delta,\nabla \sigma}_{q',\mu\nu} 
   \Big\} \, ,
      \\   
F_{q, \mu\nu}^{\nabla,\nabla} (\vec{r}) 
= & \sum_{q' = p,n} \Big\{ 
        \cc{A}^{(2,2)}_{qq',\textrm{e}} \, D^{1,1}_{q'} (\vec{r}) \, \delta_{\mu\nu} 
    + 2 \cc{A}^{(4,3)}_{qq',\textrm{e}} \, D^{(\nabla, \nabla)}_{q'} (\vec{r}) \, \delta_{\mu\nu} 
    + 2 \cc{A}^{(4,4)}_{qq',\textrm{e}} \, D^{\nabla,\nabla}_{q', \mu\nu} (\vec{r}) 
    +   \cc{A}^{(4,5)}_{qq',\textrm{e}} \, \big[ \nabla_\mu \nabla_\nu D^{1,1}_{q'} (\vec{r}) \big]
    \Big\} \, ,
     \\
G_{q, \mu\nu}^{\Delta, \nabla \sigma}(\vec{r}) 
= & \sum_{q' = p,n} \cc{A}^{(4,8)}_{qq',\textrm{e}} \, C^{1,\nabla \sigma}_{q',\mu\nu}
 \, ,
    \\
F_{q}^{\Delta,\Delta}(\vec{r}) 
= & \, \sum_{q' = p,n} \cc{A}^{(4,2)}_{qq',\textrm{e}} \, D^{1,1}_{q'} (\vec{r}) \, ,   
\intertext{whereas for the potentials associated with time-odd densities one has}
\vec{F}_{q}^{1,\sigma}(\vec{r})  
 = & \sum_{q' = p,n} 
      \bigg\{ 
       2 \cc{A}^{(0,1)}_{qq',\text{o}} \, \vec{D}^{1,\sigma}_{q'} (\vec{r})
     + 2 \cc{A}^{(0,2)}_{qq',\text{o}} 
       \big[ D^{1,1}_{0} (\vec{r}) \big]^\alpha \, \vec{D}^{1,\sigma}_{q'} (\vec{r})
     \nn \\
  &    + 2 \cc{A}^{(2,1)}_{qq',\text{o}} \big[ \Delta \vec{D}^{1,\sigma}_{q'}  (\vec{r}) \big]
       +   \cc{A}^{(2,2)}_{qq',\text{o}} \, \vec{D}^{(\nabla,\nabla)\sigma}_{q'}  (\vec{r})
       +   \cc{A}^{(2,4)}_{qq',\text{o}} 
       \big[ \vnabla \times \vec{C}^{1,\nabla}_{q'} (\vec{r}) \big]
     \nn \\
  &  + 2 \cc{A}^{(4,1)}_{qq',\text{o}} 
       \big[ \Delta \Delta \vec{D}^{1,\sigma}_{q'}  (\vec{r}) \big]
     + \cc{A}^{(4,2)}_{qq',\text{o}} \, \vec{D}^{\Delta,\Delta\sigma}_{q'}  (\vec{r})
     + \cc{A}^{(4,5)}_{qq',\text{o}} 
       \sum_{\mu \nu \kappa}
       \big[ \nabla_\mu \nabla_\nu D^{\nabla,\nabla \sigma}_{q',\mu \nu \kappa}  (\vec{r}) \big]
       \, \vec{e}_{\kappa} 
    \bigg\} \, ,
     \\
\vec{G}_{q}^{1, \nabla}(\vec{r})  
 = & \sum_{q' = p,n} 
      \bigg(
      2 \cc{A}^{(2,3)}_{qq',\text{o}} \, \vec{C}^{1,\nabla}_{q'} (\vec{r}) 
      + \cc{A}^{(2,4)}_{qq',\text{o}}
       \big[ \vnabla \times \vec{D}^{1,\sigma}_{q'} (\vec{r}) \big]
     \nn \\
  &  + 2 \cc{A}^{(4,6)}_{qq',\text{o}} \big[ \Delta \vec{C}^{1,\nabla}_{q'} (\vec{r}) \big] 
     + 2 \cc{A}^{(4,7)}_{qq',\text{o}} 
       \Big\{ \vnabla \big[ \vnabla \cdot \vec{C}^{1,\nabla}_{q'} (\vec{r}) \big] \Big\}
     + \cc{A}^{(4,8)}_{qq',\text{o}}\, \vec{C}^{\Delta,\nabla}_{q'} (\vec{r}) 
     \bigg) \, ,                
     \\
F_{q, \mu \nu \kappa}^{\nabla,\nabla\sigma} (\vec{r})  
 = & \sum_{q' = p,n}
     \Big\{ 
       \cc{A}^{(2,2)}_{qq',\text{o}} \,
       D^{1,\sigma}_{q',\kappa}  (\vec{r}) \, \delta_{\mu\nu}
     + 2 \cc{A}^{(4,3)}_{qq',\text{o}} \,
       D^{(\nabla,\nabla)\sigma}_{q',\kappa}  (\vec{r}) \, \delta_{\mu\nu}
     + 2 \cc{A}^{(4,4)}_{qq',\text{o}} \,
       D^{\nabla,\nabla\sigma}_{q',\mu \nu \kappa}  (\vec{r}) 
     + \cc{A}^{(4,5)}_{qq',\text{o}} \,
       \big[ \nabla_\mu \nabla_\nu D^{1,\sigma}_{q',\mu \nu \kappa}  (\vec{r}) \big]
    \Big\} \, ,
     \\
\vec{G}_{q}^{\Delta,\nabla} (\vec{r})  
 = & \sum_{q' = p,n}
       \cc{A}^{(4,8)}_{qq',\text{o}} \, \vec{C}^{1,\nabla}_{q'} (\vec{r}) \, ,
     \\
\vec{F}_{q}^{\Delta, \Delta \sigma}(\vec{r})  
 = & \sum_{q' = p,n} 
      \cc{A}^{(4,2)}_{qq',\text{o}} \, \vec{D}^{1,\sigma}_{q'} (\vec{r}) \, ,
\end{align}
\end{widetext}
where $ \vec{e}_{\kappa}$ is a Cartesian unit vector. The contributions from the kinetic energy, 
c.m.\ correction and Coulomb energy that are also contained in Eq.~\eqref{eq:Etot} still have to
be added to these expressions.


\section{Point-group transformations and point-group symmetries of local densities}
\label{app:symmetries}

In what follows, we will sketch how the symmetries of single-particle states, 
of the densities as defined in Sec.~\ref{sec:definition}, and of the corresponding 
potentials as defined in Sections~\ref{sec:singleh} and~\ref{sec:singleDelta}
can be used to reduce the numerical cost of symmetry-restricted HFB
calculations with N2LO EDFs.
We will concentrate on symmetries under the point-group transformation of the 
coordinate system as defined in Ref.~\cite{Doba00} that provides the symmetries
that are of relevance in a 3d Cartesian representation. Elements $\hat{U}$ of that
group that are relevant for our discussion are parity $\hat{P}$, the signature 
operators $\hat{R}_{\mu} = e^{i \pi \hat{J}_{\mu}}$, the simplex operators 
$\hat{S}_{\mu} = \hat{R}_{\mu} \hat{P}$, all of which are linear, and the anti-linear 
time-reversal operator $\check{T}$. Note that the properties of these operators 
and the corresponding group are slightly different when applied in the spaces of
many-body states representing systems with even or odd particle number, and also
in the space of single-particle states~\cite{Doba00}. We will only need the latter for
our discussion.

The point-group transformations can be used to set up symmetry-restricted HFB
calculations. To this end, the quasi-particle vacuum is constructed such that
the single-particle states
\begin{align}
\label{eq:Psi:decomp}
\Psi_k(\vec{r})
& = \left(
    \begin{array}{l}
    \Re \{ \psi_{k}(\vec{r},+) \} + \iunit \, \Im \{ \psi_{k}(\vec{r},+) \} \\
    \Re \{ \psi_{k}(\vec{r},-) \} + \iunit \, \Im \{ \psi_{k}(\vec{r},-) \}
    \end{array}
    \right)
\end{align}
are eigenstates of some transformation operators. For discrete point-group 
symmetries as considered here, the group 
elements can be directly used as symmetry operators. Not every combination of
point-group transformations can be chosen as conserved symmetries, though~\cite{Doba00b}: 
the maximal set comprises a linear hermitian, a linear anti-hermitian, an anti-linear
hermitian and an anti-linear anti-hermitian operator that have to fulfill
specific commutation relations~\cite{RyssensPhD}.

The action of the basic point-group operators on the
spinor components can be summarized as
\begin{align}
\label{eq:sp:parity}
\psi_k^{\hat{P}} (x,y,z,\sigma)
& =  \psi_k (-x,-y,-z,\sigma) \, ,
      \\
\label{eq:sp:tinv}
\psi_k^{\check{T}}(x,y,z,\sigma)
& =  -\sigma \, \psi_k^* (x,y,z,-\sigma) \, ,
      \\
\label{eq:sp:rotx}
\psi_k^{\hat{R}_x} (x,y,z,\sigma)
& =  - \iunit \, \psi_k (x,-y,-z,-\sigma) \, ,
      \\
\label{eq:sp:roty}
\psi_k^{\hat{R}_y}(x,y,z,\sigma)
& =  \sigma \, \psi_k (-x,y,-z,-\sigma) \, ,
      \\
\label{eq:sp:rotz}
\psi_k^{\hat{R}_z} (x,y,z,\sigma)
& =  - \iunit \, \sigma \psi_k (-x,-y,z,\sigma) \, ,
\end{align}
from which the behavior of the four real functions needed to represent a single-particle
state~\eqref{eq:Psi:decomp} under point-group transformations can be
easily deduced. For all calculations reported in this paper, we assumed that single-particle states 
are eigenstates of a subset of the following operators: (linear hermitian) parity $\hat{P}$,
(linear anti-hermitian) $z$ signature $\hat{R}_z$, and the (anti-linear hermitian) 
time $y$ simplex $\check{S}^T_y$
\begin{align}
\label{eq:spwf:P}
\hat{P} \Psi_k(\vec{r})
& = \pi_k \, \Psi_k(\vec{r}) \, ,
    \\
\label{eq:spwf:Rz}
\hat{R}_z \Psi_k(\vec{r})
& = \iunit \, \eta_k \, \Psi_k(\vec{r}) \, ,
    \\
\label{eq:spwf:STy}
\check{S}^T_y \Psi_k(\vec{r})
  = \, \check{T} \hat{P} \hat{R}_y \Psi_k(\vec{r})
& = + \Psi_k(\vec{r}) \, ,
\end{align}
where $\pi_k = \pm 1 $ and $\eta_k = \pm 1$ are the parity and signature quantum 
numbers.\footnote{Note that there is an alternative convention where the factor $-\iunit$
is not taken out from the eigenvalue, such that the single-particle states are labeled by 
the imaginary values $\iunit \eta_k = \pm \iunit$.
} 
The relation for $\check{S}^T_y$ establishes a spatial symmetry, but because of its
anti-linearity there is no associated quantum number. Relation~\eqref{eq:spwf:STy}
for $\check{S}^T_y$, however, fixes the relative phase between the single-particle 
states~\cite{Doba00b}.

In addition, for some of the calculations reported in Sec.~\ref{sec:calculations}
we impose time-reversal symmetry. As an anti-linear anti-hermitian operator, $\check{T}$
does not provide an eigenvalue equation, but instead establishes a relation between two 
single-particle states in the canonical basis that we will label as $k$ and $ \bar{k} $ 
as customarily done
\begin{align}
\Psi_{\bar{k}} (\vec{r}) 
& = \Psi_k^{\check{T}}  (\vec{r}) 
  = \check{T} \Psi_k  (\vec{r}) \, ,
\end{align}
meaning that only one out of the two states needs to be explicitly calculated and
kept in storage. (When time-reversal symmetry is broken, the time-reversed
of a given state in the canonical basis is in general not a state in the 
canonical basis, but a superposition of such states.)
As time-reversal changes the sign of the imaginary eigenvalues of linear 
anti-hermitian operators, the states to be treated explicitly can be chosen as the 
single-particle states for which the signature quantum number $\eta_k$ is positive. 
Note that applying time-reversal twice on a Fermionic state yields the 
negative of the original state, $\check{T}^2 = -1$, {such that} one has 
$\Psi_k (\vec{r}) = - \check{T} \Psi_{\bar{k}} (\vec{r})$. 

From Eqs.~\eqref{eq:sp:parity}--\eqref{eq:sp:rotz} follows that the four
real functions out of which the eigenstates of Eqs.~\eqref{eq:spwf:P}--\eqref{eq:spwf:STy}
are built satisfy the relations
\begin{widetext}
\begin{align}
\label{eq:cr8:symmetries:1}
\Re \{ \psi_k(x,y,z,+) \} & = +               \, \Re \{ \psi_k( x,-y, z,+) \} 
                            = +        \eta_k \, \Re \{ \psi_k(-x,-y, z,+) \} 
                            =             p_k \, \Re \{ \psi_k(-x,-y,-z,+) \} \, , \nn \\
\Im \{ \psi_k(x,y,z,+) \} & = -               \, \Im \{ \psi_k( x,-y, z,+) \} 
                            = +        \eta_k \, \Im \{ \psi_k(-x,-y, z,+) \} 
                            =             p_k \, \Im \{ \psi_k(-x,-y,-z,+) \} \, , \nn \\
\Re \{ \psi_k(x,y,z,-) \} & = +               \, \Re \{ \psi_k( x,-y, z,-) \} 
                            = -        \eta_k \, \Re \{ \psi_k(-x,-y, z,-) \}
                            =             p_k \, \Re \{ \psi_k(-x,-y,-z,-) \} \, , \nn \\
\Im \{ \psi_k(x,y,z,-) \} & = -               \, \Im \{ \psi_k( x,-y, z,-) \}
                            = -        \eta_k \, \Im \{ \psi_k(-x,-y, z,-) \} 
                            =             p_k \, \Im \{ \psi_k(-x,-y,-z,-) \} \, .
\end{align}  
\end{widetext}
From suitable combinations of these one can then construct the symmetries of
each of the four real functions under plane reflection as detailed in 
Table~\ref{tab:spwf:simplexes}. These imply that for 
eigenstates of $\big\{ \hat{P}, \hat{R}_z, \check{S}^T_y \big\}$ it is 
sufficient to have a numerical representation of the single-particle
states in $1/8$ of the box. The same choice has been used earlier in
Refs.~\cite{Bonche85,Hellemans12,Ryssens15a}.

\begin{table}[b!]
\caption{\label{tab:spwf:simplexes}
Plane reflection symmetries of the real and imaginary parts of the upper and
lower spinor components of eigenstates of $\big\{ \hat{P}, \hat{R}_z, \check{S}^T_y \big\}$
(see text). 
}
\begin{tabular}{c@{\qquad}c@{\quad}c@{\quad}c}
\hline\noalign{\smallskip}
                                  &  $(-x,y,z)$   & $(x,-y,z)$  & $(x,y,-z)$    \\
\noalign{\smallskip}\hline\noalign{\smallskip}
  $\Re \{ \psi_{k}(\vec{r},+) \}$ &  $+\eta_k$  &   $+$  & $+\eta_k \, p_p$ \\
  $\Im \{ \psi_{k}(\vec{r},+) \}$ &  $-\eta_k$  &   $-$  & $+\eta_k \, p_k$ \\
  $\Re \{ \psi_{k}(\vec{r},-) \}$ &  $-\eta_k$  &   $+$  & $-\eta_k \, p_k$ \\
  $\Im \{ \psi_{k}(\vec{r},-) \}$ &  $+\eta_k$  &   $-$  & $-\eta_k \, p_k$ \\
\noalign{\smallskip}\hline
\end{tabular}
\end{table}

The same triaxial shape degrees of freedom can also be described choosing signature 
and time-simplex operators that refer to other cartesian directions~\cite{Doba00b}. 
For reasons that ultimately are related to the special role played by the $z$
direction for the quantization of spin and orbital angular momenta, choosing a
different set of point-group symmetries to describe states with triaxial shapes
leads in general to different symmetries of the four real functions needed to 
represent the single-particle states, and will also lead to different symmetries 
for some of the local densities.

Among all possible choices, however, ours has the advantage that each of the 
four functions needed to represent a single-particle state~\eqref{eq:Psi:decomp} 
individually takes three plane reflection symmetries. This is specific to working 
with eigenstates of $\hat{R}_z$ and $\check{S}^T_y$. The other signature and 
time-simplex all establish relations between different spinor components and/or 
their real and imaginary parts, see Eqs.~\eqref{eq:sp:parity}--\eqref{eq:sp:rotz}. 
Choosing $\big\{ \hat{P}, \hat{R}_x, \check{S}^T_y \big\}$ instead, as done in
many other implementations of triaxial HFB~\cite{Doba00b}, leads to a situation where 
the continuation of each of the four functions needed to represent $\Psi_k(\vec{r})$ 
into other octants is provided by a different one among these functions.  

The symmetry properties of the single-particle states are then transferred on to 
the two-body and many-body states, the local densities and the local potentials, 
meaning that the numerical representation can be limited to some sector of the 
full volume in which the nucleus is placed, and the information about the eliminated 
sectors is reconstructed through the conserved 
symmetries~\cite{Doba00b,Rohozinski10a,Ryssens15a,Hellemans12}.

\begin{table}[b!]
\caption{Plane reflection symmetries of the time-even normal densities
defined in Sec.~\ref{sec:choice} when working with eigenstates of  
$\big\{ \hat{P}, \hat{R}_z, \check{S}^T_y \big\}$ 
(see text). 
}
\label{tab:densymmetries:normalte}
\begin{ruledtabular}
\begin{tabular}[t]{lc@{\hskip 10pt}ccc}
Density                     & $\mu\nu$  
                            & $(-x,y,z)$   & $(x,-y,z)$  & $(x,y,-z)$  \\
\noalign{\smallskip} \hline \noalign{\smallskip}
 $D^{1,1}_q (\vec{r})$, $D^{\Delta, \Delta}_q(\vec{r})$ 
  &                    &  $+$ & $+$ & $+$ \\
\noalign{\smallskip} \hline \noalign{\smallskip} 
\multirow{4}{*}{$D^{\nabla, \nabla}_{q,\mu \nu}(\vec{r})$ }
  &  $xx$, $yy$, $zz$  &   $+$ &  $+$ &  $+$ \\
  &  $xy$, $yx$        &   $-$ &  $-$ &  $+$ \\
  &  $xz$, $zx$        &   $-$ &  $+$ &  $-$ \\
  &  $yz$, $zy$        &   $+$ &  $-$ &  $-$ \\
\noalign{\smallskip} \hline \noalign{\smallskip} 
\multirow{4}{*}{
\begin{tabular}{c}
$C^{1, \nabla \sigma}_{q, \mu \nu}(\vec{r})$ \\[1mm] 
$C^{\Delta, \nabla \sigma}_{q, \mu \nu}(\vec{r})$
\end{tabular}
}
  & $xx$, $yy,zz$ &  $-$ &  $-$ &  $-$ \\
  & $xy$, $yx$    &  $+$ &  $+$ &  $-$ \\
  & $xz$, $zx$    &  $+$ &  $-$ &  $+$ \\
  & $yz$, $zy$    &  $-$ &  $+$ &  $+$ \\
\end{tabular}
\end{ruledtabular}
\end{table}

\begin{table}[t]
\caption{\label{tab:densymmetries:normalto}
Same as Tab.~\ref{tab:densymmetries:normalte}, but for 
time-odd normal densities.}
\begin{ruledtabular}
\begin{tabular}[t]{lc@{\hskip 10pt}ccc}
Density                     & $\mu\nu$  
                            & $(-x,y,z)$   & $(x,-y,z)$  & $(x,y,-z)$  \\
\noalign{\smallskip} \hline \noalign{\smallskip}
\multirow{3}{*}{
\begin{tabular}{l}
$D^{1, \sigma}_{q, \mu}(\vec{r})$ \\[1mm]
$D^{\Delta, \Delta \sigma}_{q, \mu}(\vec{r})$ 
\end{tabular}}
          & $x$ &  $-$ & $+$ & $-$ \\
          & $y$ &  $+$ & $-$ & $-$ \\
          & $z$ &  $+$ & $+$ & $+$ \\
\noalign{\smallskip} \hline \noalign{\smallskip}
\multirow{12}{*}{$D^{\nabla, \nabla \sigma}_{q, \mu \nu \kappa}(\vec{r})$ } 
         &  $xxx$, $yyx$, $zzx$, & $-$ & $+$ & $-$ \\
         &  $xxy$, $yyy$, $zzy$  & $+$ & $-$ & $-$ \\
         &  $xxz$, $yyz$, $zzz$  & $+$ & $+$ & $+$ \\
         &  $xyx$, $yxx$         & $+$ & $-$ & $-$ \\
         &  $xyy$, $yxy$         & $-$ & $+$ & $-$ \\        
         &  $xyz$, $yxz$         & $-$ & $-$ & $+$ \\
         &  $xzx$, $zxx$         & $+$ & $+$ & $+$ \\
         &  $xzy$, $zxy$         & $-$ & $-$ & $+$ \\
         &  $xzz$, $zxz$         & $-$ & $+$ & $-$ \\
         &  $yzx$, $zyx$         & $-$ & $-$ & $+$ \\
         &  $yzy$, $zyy$         & $+$ & $+$ & $+$ \\
         &  $yzz$, $zyz$         & $+$ & $-$ & $-$ \\
\noalign{\smallskip} \hline
\noalign{\smallskip}
\multirow{3}{*}{
\begin{tabular}{l}
$C^{1 , \nabla}_{q, \mu}(\vec{r})$ \\[1mm]
$C^{\Delta, \nabla }_{q, \mu}(\vec{r})$ 
\end{tabular}}
    &  $x$ & $+$ & $-$ & $+$ \\
    &  $y$ & $-$ & $+$ & $+$ \\
    &  $z$ & $-$ & $-$ & $-$ \\
\end{tabular}
\end{ruledtabular}
\end{table}


\begin{table}[t]
\caption{\label{tab:densymmetries:pairte}
Same as Tab.~\ref{tab:densymmetries:normalte}, but for
local pair densities that are time-even in a gauge where $\kappa$
is real. The reflection symmetries of the real and imaginary 
parts of these possibly complex pair densities are the same.
}
\begin{ruledtabular}
\begin{tabular}[t]{lc@{\hskip 10pt}ccc}
Density                     & $\mu\nu$  
                            & $(-x,y,z)$   & $(x,-y,z)$  & $(x,y,-z)$  \\
\noalign{\smallskip} \hline \noalign{\smallskip}
$\tilde{D}^{1, 1}_{q}(\vec{r})$, $\tilde{D}^{\Delta, \Delta}_{q}(\vec{r})$ 
              &    & $+$ &  $+$  & $+$ \\
\noalign{\smallskip} \hline
\noalign{\smallskip}
\multirow{4}{*}{$\tilde{D}^{\nabla, \nabla}_{q, \mu\nu}(\vec{r})$ } 
           & $xx$, $yy$, $zz$  &  $+$ & $+$ & $+$ \\
           & $xy$, $yx$        &  $-$ & $-$ & $+$ \\
           & $xz$, $zx$        &  $-$ & $+$ & $-$ \\
           & $yz$, $zy$        &  $+$ & $-$ & $-$ \\
\noalign{\smallskip} \hline
\noalign{\smallskip}
\multirow{4}{*}{
\begin{tabular}{c}
$\tilde{C}^{1, \nabla \sigma}_{q, \mu\nu}(\vec{r})$ \\[1mm]
$\tilde{C}^{\Delta, \nabla \sigma}_{q, \mu\nu}(\vec{r}) $
\end{tabular} 
} 
    &    $xx$, $yy$, $zz$  &  $-$ & $-$ & $-$ \\
    &    $xy$, $yx$        &  $+$ & $+$ & $-$ \\
    &    $xz$, $zx$        &  $+$ & $-$ & $+$ \\
    &    $yz$, $zy$        &  $-$ & $+$ & $+$ \\
\noalign{\smallskip} 
\noalign{\smallskip}
\end{tabular}
\end{ruledtabular}
\end{table}


\begin{table}[t]
\caption{\label{tab:densymmetries:pairto}
Same as Tab.~\ref{tab:densymmetries:normalte}, but for
local pair densities that are time-odd in a gauge where $\kappa$
is real. Again, the reflection symmetries of the real and imaginary 
parts of these possibly complex pair densities are the same.
}
\begin{ruledtabular}
\begin{tabular}[t]{lc@{\hskip 10pt}ccc}
Density                     & $\mu\nu$  
                            & $(-x,y,z)$   & $(x,-y,z)$  & $(x,y,-z)$  \\
\noalign{\smallskip} \hline \noalign{\smallskip}
\multirow{1}{*}{$\tilde{C}^{1, 1}_{q}(\vec{r}) $, $\tilde{C}^{\Delta, \Delta}_{q}(\vec{r})$} 
              &   &  $-$ & $-$ & $+$ \\
\noalign{\smallskip}  \hline
\noalign{\smallskip}
\multirow{4}{*}{$\tilde{C}^{\nabla, \nabla}_{q, \mu\nu}(\vec{r})$} 
    &   $xx$, $yy$, $zz$ &  $-$ & $-$ & $+$ \\
    &   $xy$, $yx$       &  $+$ & $+$ & $+$ \\
    &   $xz$, $zx$       &  $+$ & $-$ & $-$ \\
    &   $yz$, $zy$       &  $-$ & $+$ & $-$ \\
\noalign{\smallskip}  \hline
\noalign{\smallskip}
\multirow{4}{*}{
\begin{tabular}{c}
$\tilde{D}^{1, \nabla\sigma}_{q, \mu\nu}(\vec{r})$ \\[1mm]
$\tilde{D}^{\Delta, \nabla\sigma}_{q, \mu\nu}(\vec{r})$
\end{tabular}
}  
&  $xx$, $yy$, $zz$ &  $+$ & $+$ & $-$ \\
&  $xy$, $yx$       &  $-$ & $-$ & $-$ \\
&  $xz$, $zx$       &  $-$ & $+$ & $+$ \\
&  $yz$, $zy$       &  $+$ & $-$ & $+$ \\
\noalign{\smallskip} 
\noalign{\smallskip}
\end{tabular}
\end{ruledtabular}
\end{table}


Tables~\ref{tab:densymmetries:normalte} and~\ref{tab:densymmetries:normalto}
detail the behavior of all cartesian components of the normal time-even densities 
entering our choice of representation of the N2LO Skyrme EDF of 
Eqs.~\eqref{eq:SkTeven:0}--\eqref{eq:SkTodd:4} under plane reflection reflection in 
the three cartesian directions. These are obtained writing out the expressions
of Eqs.~\eqref{eq:def:Dden:3}--\eqref{eq:def:Csvec:3} in terms of the four real
functions on the right-hand-side of Eq.~\eqref{eq:Psi:decomp} and using that the
component $\nabla_\mu$ of the nabla operator changes sign when replacing $r_\mu$
by $-r_\mu$. For example, having the values of the $x$-component of the current
$C^{\Delta, \nabla}_{q, x}(\vec{r})$ in the octant where $x$, $y$, and $z$ are
all positive, from Table~\ref{tab:densymmetries:normalto} it can be deduced 
that its values in the three adjacent octants are obtained as
\begin{align}
C^{\Delta, \nabla}_{q, x}(-x,y,z) 
& = + C^{\Delta, \nabla}_{q, x}(x,y,z) \, ,
    \\
C^{\Delta, \nabla}_{q, x}(x,-y,z) 
& = - C^{\Delta, \nabla}_{q, x}(x,y,z) \, ,
    \\
C^{\Delta, \nabla}_{q, x}(x,y,-z) 
& = + C^{\Delta, \nabla}_{q, x}(x,y,z) \, .
\end{align}
The value of $C^{\Delta, \nabla}_{q, x}(\vec{r})$ in the other four octants can 
be obtained from suitable combinations of these relations. Note that the Tables 
with the point-group symmetries of normal densities provided by Ref.~\cite{Doba00} 
represent something different: they represent the sign change of a given density 
when going to the \textit{transformed} coordinate system, and the entries 
in the tables provided there can only used for the actually conserved symmetries
of the many-body states. By contrast, 
our Tables represent the continuation of the densities when going from one octant
to another in the \textit{same} coordinate system. None of the plane reflection symmetries 
listed in Tables~\ref{tab:densymmetries:normalte} and~\ref{tab:densymmetries:normalto} 
corresponds to a conserved symmetry of the actual single-particle wave 
functions and therefore the properties of the functions might disagree with the 
symmetries of densities under a simplex transformation $\hat{S}_\mu$ as 
given in Ref.~\cite{Doba00}.

Tables~\ref{tab:densymmetries:pairte} and \ref{tab:densymmetries:pairto} provide 
the same information for the time-even pair and time-odd pair 
densities, respectively.
When time-reversal is a conserved symmetry, all densities listed in 
Tables~\ref{tab:densymmetries:normalto} and~\ref{tab:densymmetries:pairto}
are strictly zero.

The energy density $\mathcal{E}(\vec{r})$ is a scalar under all spatial
similarity transformations, meaning that it has to have the same symmetries 
as the local normal matter density $D^{1,1}_q(\vec{r})$. This necessitates 
that each individual term in Eqs.~\eqref{eq:SkTeven:0}--\eqref{eq:SkTodd:4} 
is the product of two densities (or of a density and the derivative of a 
density) that have the same behavior under all plane reflection symmetries. 
From Eq.~\eqref{eq:def:potentials:F} follows then that the potentials 
$\mathsf{F}_{q, {\rm a}}(\vec{r})$ take exactly the same symmetries as the 
density $\mathsf{R}_{q, {\rm a}}(\vec{r})$ they are associated with.

When breaking parity among the symmetries discussed above, such that only 
$\big\{\hat{R}_z, \check{S}^T_y \big\}$ remain conserved, one loses the 
reflection symmetry in the $z=0$ plane in 
Tables~\ref{tab:spwf:simplexes}--\ref{tab:densymmetries:pairto}.
Keeping parity conserved while breaking one of the other spatial symmetries, 
however, will lead to a situation with a point reflection symmetry instead 
of plane reflection symmetries that cannot be deduced from these Tables anymore.

The calculations of Kr and Nd isotopes discussed in Sec.~\ref{sec:calculations} were 
performed imposing the full $\{ \hat{P}, \hat{R_z}, \check{S}_y^T, \check{T}  \}$ symmetry
on the single-particle states. This corresponds to time-reversal invariant triaxial 
many-body states, i.e.\ states whose densities take three plane reflection symmetries, 
and which necessarily have an average angular momentum of zero, see 
Ref.~\cite{Ryssens15a} for a detailed discussion of this choice. The calculations of the SD band of \nuc{194}{Hg} 
used the {single-particle symmetries} $\{ \hat{P}, \hat{R_z}, \check{S}_y^T \}$ instead, which also 
corresponds to triaxial many-body states, but the broken time-reversal invariance now 
allows for finite average angular momentum in $z$ direction $\langle \hat{J}_z \rangle$, 
see for example \cite{Bonche87a,Hellemans12} for further discussion of this choice. 
Finally, the fission barrier of \nuc{240}{Pu} was calculated with conserved 
$\{ \hat{R_z}, \check{S}_y^T, \check{T} \}$, where broken parity lifts one of the 
plane reflection symmetries of the densities and allows for describing reflection-asymmetric
shapes.

%
%
\section{Properties of local pair densities that motivate our definitions}
\label{app:pairdens}

In the following, we further detail the motivations behind our choice 
of local pair densities as defined in Eqs.~\eqref{eq:def:Ddenpair}--\eqref{eq:def:Csvecpair}.

Despite usually being evaluated for densities obtained from a symmetry-breaking
reference state, any meaningful EDF nevertheless has to be constructed such that it is
invariant under all symmetry transformations that correspond to the empirical conservation 
laws of nuclear physics. Among these, most relevant for the purpose of defining local 
pair densities are global gauge transformations.
Global gauge-invariance is the global symmetry related to the conservation 
of particle number~\cite{Bally21}. When not mixing protons and neutrons, the many-body 
state has two separate global gauge invariances $\mathrm{U}(1)_N \times \mathrm{U}(1)_Z$, 
one with respect to gauge rotation of neutrons, the other with respect to gauge rotations of 
protons~\cite{Bally21}.

A global $\mathrm{U}(1)$ gauge transformation $e^{\iunit \varphi \hat{N}}$ changes the 
phase of every single-particle state by the angle $\varphi$. On the one hand, the gauge 
transformation of a given many-body state $| \Phi \rangle$ can be represented through 
the change of the phase of all single-particle wave functions 
$\Psi_k(\vec{r}) \to \Psi_k(\vec{r}) \, e^{\iunit \varphi}$, while  
the creation and annihilation operators $\hat{a}^\dagger_k$ and $\hat{a}_k $ still refer 
then to the original single-particle basis. On the other hand, as an alternative 
one can keep the single-particle wave functions unchanged and transform the 
corresponding creation and annihilation operators instead
\begin{align}
\label{eq:gauge:a}
e^{+\iunit \varphi \hat{N}} \hat{a}_k \, e^{-\iunit \varphi \hat{N}} 
& = \hat{a}_k \, e^{-\iunit \varphi} \, ,
    \\
\label{eq:gauge:a+}
e^{+\iunit \varphi \hat{N}} \hat{a}^\dagger_k \, e^{-\iunit \varphi \hat{N}}
& = \hat{a}^\dagger_k \, e^{+\iunit \varphi} \, ,
\end{align}
such that the phase factors are treated as separate entities throughout the 
calculation. For formal and computational reasons that will become clear in 
what follows, the latter convention for gauge transformations is more practical 
when considering the definition of local pair densities.

As we are interested here in the transformations of density matrices
\eqref{eq:def:nonlocaldensity} and~\eqref{eq:def:nonlocalpair}
for a given particle species $q$, we will limit our discussion to the 
global gauge transformation of the auxiliary Bogoliubov quasiparticle 
vacuum for one generic nucleon species.

In static calculations and when conserving an anti-linear anti-hermitian 
symmetry \cite{RyssensPhD}, the amplitudes $v_k$ and $u_k$ can be chosen 
to be real in the canonical basis when not mixing protons and neutrons.
This is the choice that is usually made in numerical codes, but it is 
not a necessity. For such specific choice of canonical basis, the corresponding 
quasiparticle vacuum can be written in the simple BCS form
\begin{equation}
\label{eq:BCS}
| \Phi(0) \rangle
\equiv \prod_{f} \hat{a}^\dagger_f 
       \prod_{\stackrel{k > 0}{ k, \bar{k} \neq f}} 
       \big( u_k + v_k \hat{a}^\dagger_k \hat{a}^\dagger_{\bar{k}} \big)
      | - \rangle \, ,
\end{equation}
where $f$ labels completely filled single-particle states and $k$ the paired ones \cite{Schunck2019}.
A gauge-rotated quasiparticle vacuum $| \Phi(\varphi) \rangle$ is then obtained as~\cite{Bender09b}
\begin{align}
\label{eq:BCS:gauge}
| \Phi(\varphi) \rangle
& \equiv e^{\iunit \varphi \hat{N}} | \Phi(0) \rangle 
      \nn \\
& = \prod_{f} \hat{a}^\dagger_f \, e^{\iunit \varphi}
    \prod_{\stackrel{k > 0}{ k, \bar{k} \neq f}} 
    \big( u_k + v_k \, e^{2\iunit \varphi} \, \hat{a}^\dagger_k \hat{a}^\dagger_{\bar{k}} \big)
    | - \rangle \, .
\end{align}
As paired Bogoliubov quasiparticle vacua are not eigenstates of particle number, the global
gauge rotation does \textit{not} simply correspond to a change of phase of the 
many-body state by $\varphi N_0$ as it would be the case for a Slater determinant and any other
eigenstate of the particle-number operator  $\hat{N}$ with eigenvalue $N_0$. Also, there are no simple 
transformation rules for Bogoliubov quasiparticle creation and annihilation operators, 
as these mix single-particle creation and annihilation operators that according to 
Eqs.~\eqref{eq:gauge:a} and~\eqref{eq:gauge:a+} transform differently.

From the relations~\eqref{eq:gauge:a} and~\eqref{eq:gauge:a+}
follows that the normal and anomalous density matrices in some single-particle basis transform as 
\begin{alignat}{3}
\label{eq:rho:ggauge}
\rho^{\varphi \varphi}_{k \ell}
& = \langle \Phi(\varphi) |  \hat{a}^\dagger_\ell \hat{a}_k \, | \Phi(\varphi) \rangle 
& \; = \; & \rho^{00}_{k \ell} \, ,
    \\
\label{eq:kappa:ggauge}
\kappa^{\varphi \varphi}_{k \ell}
& = \langle \Phi(\varphi) |  \hat{a}_\ell \hat{a}_k \,  | \Phi(\varphi) \rangle
& \; = \; & \kappa^{00}_{k \ell} \, e^{+2\iunit \varphi} \, ,
    \\
\label{eq:kappas:ggauge}
\kappa^{\varphi \varphi*}_{k \ell}
& = \langle \Phi(\varphi) |  \hat{a}^\dagger_k \hat{a}^\dagger_\ell   | \Phi(\varphi) \rangle \; 
& \; = \; & \kappa^{00*}_{k \ell} \, e^{-2\iunit \varphi} \, ,
\end{alignat}
when changing the gauge of the quasiparticle vacuum~\eqref{eq:BCS} they are constructed 
from. These relations imply that the choice of a real anomalous density matrix
$\kappa^{*}_{k \ell} = \kappa_{k \ell}$ requires a specific choice for the 
global gauge angle $\varphi$. By contrast, the normal one-body density matrix 
is not affected by a global gauge transformation of the many-body state it 
is constructed from, which implies that all local normal densities as defined in 
Eqs.~\eqref{eq:def:Dden}--\eqref{eq:def:Csvec} are automatically invariant under
global gauge transformations.
Even when working with symmetries and a gauge that permit $\kappa^{*} = \kappa$,
the distinction between these two matrices should be kept throughout the 
formalism as it distinguishes the independent degrees of freedom when deriving 
the HFB equation, see Sec.~\ref{sec:singleDelta}.

Yet another point to be taken into consideration is that the 
definition of local densities has to ensure that any possible symmetry 
of the single-particle wave functions translates into a symmetry of the 
densities. This is a pre-requisite for using such symmetries to 
reduce the numerical cost of calculations in a symmetry-restricted HFB 
code as explained in Appendix~\ref{app:symmetries}. 
When doing so, the numerical representation is limited to some sector 
of the full volume in which the nucleus is placed, and the information
about the sectors eliminated from the numerical representation 
is reconstructed through the conserved symmetries. From a practical
point of view, in a coordinate-space representation this particularly 
concerns the calculation of numerical derivatives that are to be 
calculated with different expressions when the function they are 
applied to is even or odd under some reflection symmetry~\cite{Ryssens15a}. 
Similarly, the conserved symmetries are then used to determine a priori 
that integrals over products of functions that have different symmetries 
are zero. That such integrals vanish exactly 
can in most cases not be deduced from numerically summing up the
integrand in the restricted volume; instead, the integral being zero usually
results from the cancellation between contributions of same absolute value 
but different sign from the various sectors of space that are connected 
by the symmetries.

For normal densities defined along the lines of Eqs.~\eqref{eq:def:Dden}--\eqref{eq:def:Csvec},
the symmetries of the single-particle wave functions are indeed 
automatically transferred to the densities, which follows from the hermiticity 
of normal density matrices. For pair densities, the situation is different as the 
non-local pair densities defined through Eqs.~\eqref{eq:def:rhot:nonlocal:2} 
and~\eqref{eq:def:st:nonlocal:2} transform as tensors of rank 2, as do the 
objects they are calculated from these relations. This has a number of consequences 
for our discussion. For example, the analysis of their behavior under similarity 
transformations in space-time reveals that the real and imaginary parts of 
the two-body wave function $\tilde{\varrho}_{jk}(\vec{r}',\vec{r})$ and 
$\tilde{\varsigma}_{jk,\mu}(\vec{r}',\vec{r})$ in general transform differently,
which should not be a surprise as the same is found for the real and imaginary 
parts of the upper and lower components of single-particle wave functions, 
see the discussion in Appendix~\ref{app:symmetries}.
For $\tilde{\varrho}_{jk}(\vec{r}',\vec{r})$ this can be seen from the 
symmetries of $\tilde{D}^{1,1}_q(\vec{r})$ and $\tilde{C}^{1,1}_q(\vec{r})$ 
as given in Tables~\ref{tab:densymmetries:pairte} and~\ref{tab:densymmetries:pairto} 
in Appendix~\ref{app:symmetries}, as these densities are simply a weighted sum of its 
real and imaginary parts at $\vec{r} = \vec{r}'$.

As argued above, objects that have different spatial symmetries need to 
be represented separately in a symmetry-restricted HFB code. This, however, 
cannot be done in a transparent manner in the traditional notation of local 
pair densities, i.e.\ using $\tilde{\rho}_q(\vec{r})$, $\tilde{\tau}_{q}(\vec{r})$, 
$\tilde{J}_{q, \mu \nu}(\vec{r})$, etc,
when $\kappa_{ij}$ is complex: the symmetries of real and imaginary parts of 
$\tilde{\varrho}_{jk}(\vec{r},\vec{r})$ and $\tilde{\varsigma}_{jk,\mu}(\vec{r},\vec{r})$ 
are lost when calculating the real and imaginary parts of these pair densities.
This becomes different when separating the traditional pair densities into a 
$\tilde{D}$ and a $\tilde{C}$ object that only depend on either the real
or the imaginary part of $\tilde{\varrho}_{jk}(\vec{r},\vec{r})$ and 
$\tilde{\varsigma}_{jk,\mu}(\vec{r},\vec{r})$ as done in the definitions of 
Eqs.~\eqref{eq:def:Ddenpair}--\eqref{eq:def:Csvecpair}. While the $\tilde{D}$ 
and a $\tilde{C}$ might still be complex, their real and imaginary parts now
have the same spatial symmetries.

Under a global gauge transformation of the reference state, the local pair 
densities as defined in Eqs.~\eqref{eq:def:Ddenpair}--\eqref{eq:def:Csvecpair} 
transform as
\begin{align}
\label{eq:Ddenpair:ggauge} 
\tilde{D}_q^{A, B} (\vec{r})
& \to \tilde{D}_q^{A, B} (\vec{r}) \, e^{2 \iunit \varphi} \, ,
  \\
\label{eq:Cdenpair:ggauge} 
\tilde{C}_q^{A, B} (\vec{r})
& \to \tilde{C}_q^{A, B} (\vec{r}) \, e^{2 \iunit \varphi} \, ,
  \\
\label{eq:Dsvecpair:ggauge} 
\tilde{D}_q^{A, B\sigma} (\vec{r})
& \to \tilde{D}_q^{A, B\sigma} (\vec{r}) \, e^{2 \iunit \varphi} \, ,
  \\
\label{eq:Csvecpair:ggauge} 
\tilde{C}_q^{A, B\sigma} (\vec{r})
& \to \tilde{C}_q^{A, B\sigma} (\vec{r}) \, e^{2 \iunit \varphi} \, ,
\end{align}
which is a special case of the transformation rules of pair densities under 
more general local gauge transformations discussed in Ref.~\cite{Perlinska04}, 
$\phi(\vec{r}) \to \varphi$. As repeatedly mentioned, the ingredients of the pair 
densities transform as two-body wave functions, so the pair densities transform 
with the square of the phase factor caught by each single-particle state. 
By contrast, one can easily show that all normal densities are invariant 
under global gauge transformations (but not necessarily local gauge transformations
\cite{Dobaczewski95,Dobaczewski96b,Perlinska04,Carlsson08,Raimondi11a}), 
which simply follows from Eq.~\eqref{eq:rho:ggauge} and originates in the 
hermiticity of the normal density matrices: the phase factors of a
single-particle state and its hermitian conjugate always cancel.

The same problem of clearly separating the parts of the local pair densities 
that have different symmetries also emerges when projecting symmetry-restricted HFB 
states on particle number, in which case only one of the two HFB states is
globally gauge rotated~\cite{Robledo07a,Bender09b}. Again, the weights that 
multiply $\tilde{\varrho}_{jk}(\vec{r},\vec{r})$ and $\tilde{\varsigma}_{jk,\mu}(\vec{r},\vec{r})$ 
when summing up the local mixed pair densities become complex.

Another similarity transformation that constrains the possible 
combinations of terms in the EDF is Galilean 
invariance~\cite{Dobaczewski95,Dobaczewski96b,Carlsson08,Raimondi11a},
see also Appendix~\ref{app:coupling}. From the transformation of 
the pair density matrix under a Galilean boost transformation~\cite{Perlinska04}
\begin{equation}
\tilde{\rho}_q (\vec{r}\sigma, \vec{r}' \sigma')
\to \tilde{\rho}_q (\vec{r}\sigma, \vec{r}' \sigma') \, 
  e^{\iunit \vec{P} \cdot ( \vec{r} + \vec{r}') / \hbar} \, ,
\end{equation}
where $\vec{P}$ is the vector of the center-of-mass momentum in the boosted frame,
follows for example that
\begin{align}
\label{eq:Ddenpair:Galilei}
\tilde{D}_q^{1, 1} (\vec{r})
& \to \tilde{D}_q^{1, 1} (\vec{r}) \, e^{2 \iunit \vec{P} \cdot \vec{r} / \hbar} \, ,
  \\
\label{eq:Cdenpair:Galilei}
\tilde{C}_q^{1, 1} (\vec{r})
& \to \tilde{C}_q^{1, 1} (\vec{r}) \, e^{2 \iunit \vec{P} \cdot \vec{r} / \hbar} \, ,
\end{align}
meaning that pair densities that are constructed to be real-valued functions
in the rest frame necessarily become complex-valued functions in the boosted 
frame. 
This behavior is again very different from the one of normal densities,
that remain real also in the boosted frame~\cite{Dobaczewski95,Dobaczewski96b,Perlinska04}.
Pair densities containing gradients become complex in the boosted frame 
as well, but their transformation usually involves also other densities and/or 
gradients of other densities (as is the case for their normal counterparts), 
see Ref.~\cite{Perlinska04} for a discussion in traditional notation. The same
difference in behavior of normal and pair densities is also found for the arbitrary 
local gauge transformations considered in Refs.~\cite{Dobaczewski95,Dobaczewski96b,Perlinska04}.

The relations above show that pair densities cannot be chosen to be real under all circumstances, but
that there is an interest to separate the traditionally used expressions for pair 
densities into two different objects whose real and imaginary parts take the {\emph{same}}
symmetries in symmetry-restricted HFB calculations. This is achieved by the 
definitions~\eqref{eq:def:Ddenpair}--\eqref{eq:def:Csvecpair}. A pairing EDF
that is constructed to be manifestly gauge- and Galilean-invariant is then necessarily 
composed of suitable bilinear forms of a pair density and the complex conjugate of 
a possibly different pair density as is the case for the pairing EDF of Eq.~\eqref{eq:pair:EDF:ULB},
such that the complex phase factors that arise from gauge or Galilean transformations 
cancel out. A more detailed analysis of Galilean invariance of 
particle-hole and possible particle-particle terms in the N2LO EDF using the conventions 
introduced here will be given elsewhere.


\section{Further comments on alternative densities at NLO}
\label{app:funfact}

The issues with alternative local densities, possible redundancies and 
reducibilities sketched in Sec.~\ref{sec:problems} are much less of a problem
for the standard NLO EDF than at higher orders. For the LO densities 
$\rho_q(\vec{r}) = \boldsymbol{D}^{1,1}_q(\vec{r})$ and 
$\vec{s}_q(\vec{r}) = D^{1,1 \sigma}_q(\vec{r}) $ this is even trivially 
the case as they do not contain gradients. Containing just one gradient, 
because of Eqs.~\eqref{eq:symmetry} and~\eqref{eq:recoupleC}, neither of 
the two currents $\vec{j}_q(\vec{r}) = \vec{C}^{1,\nabla}_q(\vec{r})$ 
and $J_{q,\mu\nu}(\vec{r}) = C^{1,\nabla\sigma}_{q,\mu\nu}(\vec{r})$ that 
are customarily used to express the NLO EDF can be re-expressed through 
a different current of reduced to the gradient of some other current.

From a computational point of view, it is in general advantageous to 
construct the EDF from densities that can also be used to calculate 
frequently-used one-body observables. In this way, the number of densities
that need to be constructed and stored can be kept as small as possible.
Doing so gives a natural preference for the currents $\vec{j}_q(\vec{r})$ and 
$J_{q,\mu\nu}(\vec{r})$, which together with $\rho_q(\vec{r})$ and 
$\vec{s}_q(\vec{r})$ enter the expectation values of angular momenta,
electromagnetic observables and the expectation values of the operators that 
are frequently used as constraints in the HFB equation.
The same four local densities enter also the 
continuity equations for the transport of matter with given spin and 
isospin~\cite{Raimondi11b}. The only ambiguity in their definition
concerns constant prefactors, where the choice provided by Eqs.~\eqref{eq:def:rho},
\eqref{eq:def:vecs}, \eqref{eq:def:vecj}, and \eqref{eq:def:Jmunu} is 
the most symmetric one.

For the kinetic densities $\tau_q(\vec{r}) = D^{(\nabla,\nabla)}_q(\vec{r})$, 
$\vec{T}_q(\vec{r}) = D^{(\nabla,\nabla)\sigma}_q(\vec{r}) $,
and $\vec{F}_q(\vec{r}) = D^{\nabla,(\nabla\sigma)}_q(\vec{r})$ 
entering the NLO EDF, however, the situation is less straightforward. While 
$\tau_q(\vec{r}) = D^{(\nabla,\nabla)}_q(\vec{r}) $ as defined in 
Eq.~\eqref{eq:def:tau} can be used without ambiguities to evaluate the 
one-body operator of kinetic energy
\begin{align}
\label{eq:ekin:1}
E_{\text{kin}}
& =  \sum_{q=p,n} \frac{\hbar^2}{2m_q} \int \! d^3r \, \sum_{jk} \rho_{kj} \, 
     \big[ \nabla \Psi^\dagger_j(\vec{r}) \big] \cdot
     \big[ \nabla \Psi_k(\vec{r}) \big]
     \nn \\
& =  \sum_{q=p,n} \int \! d^3r \,  \frac{\hbar^2}{2m_q} \, D^{(\nabla,\nabla)}_q(\vec{r}) \, ,
\end{align}
this definition of kinetic energy is not unique. An alternative form 
is given by
\begin{align}
\label{eq:ekin:2}
E_{\text{kin}}
=  & - \sum_{q=p,n} \frac{\hbar^2}{2m_q} \int \! d^3r \, \sum_{jk} \rho_{kj} \, 
       \nonumber \\
   &  \times \tfrac{1}{2}
      \Big\{ \Psi^\dagger_j(\vec{r})  \, \big[ \Delta \Psi_k(\vec{r}) \big]
             + \big[ \Delta \Psi^\dagger_j(\vec{r}) \big] \, \Psi_k(\vec{r}) 
      \Big\}
      \nonumber \\
= & \sum_{q=p,n} \int \! d^3r \, \bigg[ - \frac{\hbar^2}{2m_q}  D^{1,\Delta}_q(\vec{r}) \bigg] \, ,
\end{align}
which is also frequently used in a non-hermitian form without the symmetrization in the first 
line on the right-hand side of this equation. All these expressions give the same value for the 
kinetic energy.

While the calculation of $D^{1,\Delta}_q(\vec{r})$ and {$\vec{D}^{1,\Delta\sigma}_q(\vec{r})$}
only requires the storage and application of the Laplacian on the single-particle wave
functions, which is less costly than the application and storage of the three components
of the nabla, there are several reasons in favor of using $D^{(\nabla,\nabla)}_q(\vec{r})$
instead of $D^{1,\Delta}_q(\vec{r})$ and using $\vec{D}^{(\nabla,\nabla)\sigma}_q(\vec{r})$ instead 
of $\vec{D}^{1,\Delta\sigma}_q(\vec{r})$. On the one hand, the first-order derivatives of the 
single-particle wave functions are required anyway to construct other densities and currents needed
at NLO that cannot be re-expressed solely through Laplacians. On the other hand, as argued in 
Secs.~\ref{sec:computational} and~\ref{sec:singleh}, there are computational reasons to 
privilege densities for which the operators $\hat{A}$ and $\hat{B}$ are as balanced as possible.

The kinetic energy \textit{densities} $\mathcal{E}_{\text{kin}}(\vec{r})$, which are provided in 
each case by the integrand in the second line of Eqs.~\eqref{eq:ekin:1} and~\eqref{eq:ekin:2}, 
respectively, are however different because of \cite{Wong76a,Graf80a,Lombard91,Lombard94}
\begin{equation}
D^{(\nabla,\nabla)}_q(\vec{r}) 
= - D^{1,\Delta}_q(\vec{r})  + \tfrac{1}{2} \, \big[ \Delta D^{1,1}_q (\vec{r}) \big] \, ,
\end{equation}
which can be easily shown applying twice Eq.~\eqref{eq:eqredundant} and using the
symmetry $D_q^{A, B}(\vec{r}) = D_q^{B, A}(\vec{r})$,
Eq.~\eqref{eq:symmetry}. Note that Refs.~\cite{Wong76a,Graf80a,Lombard94,Lombard91,Prakash81a} 
work with an object that equals $-D^{1,\Delta}_q(\vec{r})$, and for which some 
authors~\cite{Prakash81a} actually use the symbol $\tau_q(\vec{r})$.
As demonstrated in these references, there are subtle differences when using one or the 
other expression for the kinetic energy as the starting point for semi-classical 
approximations to the kinetic energy density, such as gradient corrections in extended Thomas-Fermi 
methods. As has been argued in Refs.~\cite{Lombard91,Prakash81a}, it is the average of the 
two different kinetic energy densities
$\tfrac{1}{2} \, \big[ D^{(\nabla,\nabla)}_q(\vec{r}) - D^{1,\Delta}_q(\vec{r})\big]$
that establishes the connection to the average classical kinetic energy density, and not 
the kinetic density $D^{(\nabla,\nabla)}_q(\vec{r})$ that is traditionally used with the
Skyrme EDF. After integration, both give of course the same value for the kinetic energy.

\end{appendix}

\newpage{\pagestyle{empty}\cleardoublepage}

\begin{thebibliography}{100}

\bibitem{Bender03} M. Bender, P.-H. Heenen, and P.-G. Reinhard, 
{Self-consistent mean-field models for nuclear structure}, 
Rev. Mod. Phys. \textbf{75}, 121, (2003).

\bibitem{Schunck2019} M. Bender, N. Schunck, J. P. Ebran, and T. Duguet,
Single-Reference and Multi-Reference Formulation,
Chapter 3 of \textit{Energy Density Functional Methods for Atomic Nuclei}, 
edited by N. Schunck (IOP, Bristol 2019), pp 3-1 to 3-78.

\bibitem{Goriely13} S. Goriely, N. Chamel, and J. M. Pearson, 
{Hartree-Fock-Bogoliubov nuclear mass model with 0.50 MeV accuracy based on
 standard forms of Skyrme and pairing functionals.}
Phys. Rev. C \textbf{88}, 061302(R) (2013).

\bibitem{Bonneau07} L. Bonneau, P. Quentin, and P. M{\"o}ller,
{Global microscopic calculations of ground-state spins and parities for odd-mass nuclei.}
Phys. Rev. C \textbf{76}, 024320 (2007).

\bibitem{Dobaczewski15} J. Dobaczewski, A. V. Afanasjev, M. Bender, L. M. Robledo, and Y. Shi,
{Properties of nuclei in the nobelium region studied within the covariant, 
Skyrme, and Gogny energy density functionals.},
Nucl. Phys. A \textbf{944}, 388 (2015).

\bibitem{Kortelainen10} M. Kortelainen, T. Lesinski, J. Mor{\'e},
W. Nazarewicz, J. Sarich, N. Schunck, M. Stoitsov, and S. Wild, 
{Nuclear energy density optimization},
Phys. Rev. C \textbf{82}, 024313 (2010).

\bibitem{Kortelainen12} M. Kortelainen, J. McDonnell, W. Nazarewicz, 
P.-G. Reinhard, J. Sarich, N. Schunck, M. Stoitsov, and S. Wild, 
{Nuclear energy density optimization: Large deformations},
Phys. Rev. C \textbf{85}, 024304 (2012).

\bibitem{Kortelainen14} M. Kortelainen, J. McDonnell, W. Nazarewicz, E. Olsen,
P.-G. Reinhard, J. Sarich, N. Schunck, and S. Wild
{Nuclear energy density optimization: Shell structure},
Phys. Rev. C \textbf{89}, 054314 (2014).

\bibitem{Lesinski07} T. Lesinski, M. Bender, K. Bennaceur, T. Duguet, and J. Meyer, 
{Tensor part of the Skyrme energy density functional: Spherical nuclei},
Phys. Rev. C \textbf{76}, 014312 (2007).

\bibitem{Bender09} M. Bender, K. Bennaceur, T. Duguet, P.-H. Heenen, T. Lesinski, and J. Meyer,
{Tensor part of the Skyrme energy density functional. II. Deformation 
properties of magic and semi-magic nuclei},
Phys. Rev. C \textbf{80}, 064302 (2009).

\bibitem{Hellemans12} V. Hellemans, P.-H. Heenen, and M. Bender,
{Tensor part of the Skyrme energy density functional. III. Time-odd parts at high spin},
Phys. Rev. C \textbf{85}, 014326 (2012).

\bibitem{Colo07} G. Col\`o,  H. Sagawa, S. Fracasso, and P. F. Bortignon,
{Spin-orbit splitting and the tensor component of the Skyrme interaction}, year:2010
Phys. Lett. B \textbf{646}, 227 (2007).

\bibitem{Zalewski08} M. Zalewski, J. Dobaczewski, W. Satu{\l}a, and T. R. Werner, 
{Spin-orbit and tensor mean-field effects on spin-orbit splitting including
        self-consistent core polarizations}, 
Phys. Rev. C \textbf{77}, 024316 (2008).

\bibitem{Lesinski06} T. Lesinski, K. Bennaceur, T. Duguet, and J. Meyer, 
{Isovector splitting of nucleon effective masses, ab initio benchmarks and 
      extended stability criteria for Skyrme energy functionals}, 
Phys. Rev. C \textbf{74}, 044315 (2006).

\bibitem{Erler2010} J. Erler, P. Kl{\"u}pfel, and P.-G. Reinhard, 
{Exploration of a modified density dependence in the Skyrme functional}, 
Phys. Rev. C \textbf{82}, 044307 (2010).

\bibitem{Krewald1977} S. Krewald, V. Klemt, J. Speth, and A. Faessler, 
{Limits on the use of Skyrme forces}, 
Nucl. Phys. A \textbf{281}, 166 (1977).

\bibitem{Pearson1994} J. M. Pearson and M. Farine, 
{Relativistic mean field theory and a density dependent spin orbit Skyrme force},
 Phys. Rev. C \textbf{50}, 185 (1994).

\bibitem{Chamel09} N. Chamel, S. Goriely, and J. M. Pearson, 
{Further explorations of Skyrme-Hartree-Fock-Bogoliubov mass formulas. 
XI. Stabilizing neutron stars against a ferromagnetic collapse},
Phys. Rev. C \textbf{80}, 065804 (2009).

\bibitem{Sadoudi13a} J. Sadoudi, T. Duguet, J. Meyer, and M. Bender, 
{Skyrme functional from a three-body pseudo-potential of second
order in gradients. Formalism for central terms}, 
Phys. Rev. C  \textbf{88}, 064326 (2013).

\bibitem{Sadoudi13b} J. Sadoudi, M. Bender, K. Bennaceur, D. Davesne, R. Jodon, and T. Duguet, 
{Skyrme pseudo-potential-based EDF parametrization for spuriosity-free 
MR-EDF calculations},
Phys. Scr.  \textbf{T154}, 014013 (2013).

\bibitem{Carlsson08} B. G. Carlsson, J. Dobaczewski, and M. Kortelainen,
{Local nuclear energy density functional at next-to-next-to-next-to-leading order},
Phys. Rev. C \textbf{78}, 044326 (2008).

\bibitem{Carlsson10a} B. G. Carlsson, J. Dobaczewski, J. Toivanen and P. Vesel\'y,
{ Solution of self-consistent equations for the N3LO nuclear energy density functional in spherical symmetry. The program hosphe (v1.02).} 
Comp. Phys. Comm.  \textbf{181}, 1641-1657 (2010).

\bibitem{Carlsson10b} B. G. Carlsson and J. Dobaczewski,
{Convergence of Density-Matrix Expansions for Nuclear Interactions},
Phys. Rev. Lett. \textbf{105}, 122501 (2010).

\bibitem{Raimondi11a} F. Raimondi, B. G. Carlsson, and J. Dobaczewski,
{Effective pseudopotential for energy density functionals with
higher-order derivatives},
Phys. Rev. C \textbf{83}, 054311 (2011).

\bibitem{Skyrme56} T. H. R. Skyrme, Philos. Mag. \textbf{1}, 1043 (1956).

\bibitem{Bell56} J. S. Bell and T. H. R. Skyrme, Philos. Mag. \textbf{1}, 1055 (1956).

\bibitem{Skyrme59a} T. H. R. Skyrme, Nucl. Phys. \textbf{9}, 615 (1958/59).

\bibitem{Skyrme59b} T. H. R. Skyrme, Nucl. Phys. \textbf{9}, 635 (1958/59).

\bibitem{Davesne13} D. Davesne, A. Pastore, and J. Navarro,
{Skyrme effective pseudopotential up to the next-to-next-to-leading order},
J. Phys. G \textbf{40}, 095104 (2013).

\bibitem{Davesne14} D. Davesne, A. Pastore, and J. Navarro, 
{Fitting N3LO pseudo-potentials through central plus tensor Landau
parameters},
J. Phys. G \textbf{41}, 065104 (2014).

\bibitem{Davesne15} D. Davesne, J. Navarro, P. Becker, R. Jodon, J. Meyer, and A. Pastore,
{Extended Skyrme pseudopotential deduced from infinite nuclear matter properties},
Phys. Rev. C \textbf{91}, 064303 (2015).

\bibitem{Davesne15b} D. Davesne, J. Meyer, A. Pastore, and J. Navarro,
{Partial wave decomposition of the N3LO equation of state},
Phys. Scr. \textbf{90}, 114002  (2015).

\bibitem{Davesne16} D. Davesne, P. Becker, A. Pastore, and J. Navarro,
{Infinite matter properties and zero-range limit of non-relativistic finite-range interactions},
Ann. Phys. \textbf{375}, 288 (2016).

\bibitem{Becker15} P. Becker, D. Davesne, J. Meyer, J. Navarro, and A. Pastore, 
{Tools for incorporating a D-wave contribution in Skyrme energy density functionals},
J. Phys. G  \textbf{42}, 034001 (2015).

\bibitem{BeckerPhD} P. Becker, 
{D{\'e}veloppement d'une interaction nucl{\'e}aire effective de nouvelle
         g\'en\'eration}, 
PhD Thesis, Universit{\'e} Claude Bernard Lyon 1 (2017).

\bibitem{Becker17} P. Becker, D. Davesne, J. Meyer, J. Navarro, and A. Pastore,
{Solution of Hartree-Fock-Bogoliubov equations and fitting procedure 
      using the N2LO Skyrme pseudopotential in spherical symmetry}, 
Phys. Rev. C \textbf{96}, 044330 (2017).

\bibitem{Becker19} P. Becker, A. Pastore, D. Davesne, and J. Navarro,
{Error analysis of the parameters of the Skyrme N2LO pseudo-potential},
Il Nuovo Cimento \textbf{42} C, 88 (2019).

\bibitem{ChabanatA} E. Chabanat, P. Bonche, P. Haensel, J. Meyer, and R. Schaeffer,
{ A Skyrme parametrization from subnuclear to neutron star densities}, 
Nucl. Phys. A \textbf{627}, 710 (1997).

\bibitem{ChabanatB} E. Chabanat, P. Bonche, P. Haensel, J. Meyer, and R. Schaeffer,
{A Skyrme parametrization from subnuclear to neutron star densities 
Part II. Nuclei far from stabilities},
Nucl. Phys. A \textbf{635}, 231 (1998); 
{Erratum}  Nucl. Phys. A \textbf{643}, 441 (1998).

\bibitem{Pastore13} A. Pastore, D. Davesne, K. Bennaceur, J. Meyer, and V. Hellemans, 
{Fitting Skyrme functionals using linear response theory},
Physica Scripta \textbf{T154}, 014014 (2013).

\bibitem{RyssensPhD} W. Ryssens,
{Symmetry breaking in nuclear mean-field models},
PhD Thesis, Universit\'e Libre de Bruxelles (2016).

\bibitem{MOCCa} W. Ryssens, M. Bender, and P.-H. Heenen, 
MOCCa code
(unpublished).

\bibitem{Pastore15} A. Pastore, D. Tarpanov, D. Davesne, and J. Navarro,
{Spurious finite-size instabilities in nuclear energy density functionals: Spin channel},
 Phys. Rev. C \textbf{92}, 024305 (2015).

\bibitem{Perlinska04} E. Perli{\'n}ska, S. G. Rohozi{\'n}ski, J. Dobaczewski, and W. Nazarewicz,
{Local density approximation for proton-neutron pairing correlations: Formalism},
Phys. Rev. C \textbf{69}, 014316 (2004).

\bibitem{Ryssens15a} W. Ryssens, V. Hellemans, M. Bender, and P.-H. Heenen,
{Solution of the Skyrme-HF+BCS equation on a 3D mesh, II: a new version of the EV8 code},
Comp. Phys. Comm. \textbf{187}, 175 (2015).

\bibitem{Ryssens19b} W. Ryssens, M. Bender, and P.-H. Heenen,
{Iterative approaches to the self-consistent nuclear energy density
functional problem. Heavy ball dynamics and potential preconditioning}, 
Eur. Phys. J. A \textbf{55}, 93 (2019).

\bibitem{Dobaczewski84} J. Dobaczewski, H. Flocard, and J. Treiner, 
{Hartree-Fock-Bogolyubov description of nuclei near the neutron-drip line},
Nucl. Phys.  A \textbf{422}, 103 (1984).

\bibitem{Doba00} J. Dobaczewski, J. Dudek, S. G. Rohozi{\'n}ski, and T. R. Werner,
{Point symmetries in the Hartree-Fock approach. 
      I. Densities, shapes and currents},
Phys. Rev. C \textbf{62}, 014310 (2000).

\bibitem{Dobaczewski96} J. Dobaczewski, W. Nazarewicz, T. R. Werner, J. F. Berger, C. R. Chinn, and J. Decharg{\'e},
{Mean-field description of ground-state properties of drip-line nuclei: Pairing and continuum effects},
Phys. Rev. C \textbf{53}, 2809 (1996).

\bibitem{RotivalPhD} V. Rotival,
{Fonctionnelles d'{\'e}nergie non-empiriques pour la structure nucl{\'e}aire}, 
PhD Thesis, Universit{\'e} Paris Diderot (2009).

\bibitem{Rohozinski10a} S. G. Rohozi{\'n}ski, J. Dobaczewski, and W. Nazarewicz,
{Self-consistent symmetries in the proton-neutron Hartree-Fock-Bogoliubov approach},
Phys. Rev. C \textbf{81}, 014313 (2010).

\bibitem{Dobaczewski96b} J. Dobaczewski and J. Dudek,
{Time-odd components in the rotating mean field and identical bands},
Acta Phys. Pol. B \textbf{27}, 45 (1996).

\bibitem{KBJMnotes} K. Bennaceur and J. Meyer,
discussion notes (unpublished).

\bibitem{Engel75} Y. M. Engel, D. M. Brink, K. Goeke, S. J. Krieger, and D. Vautherin,
{Time-dependent Hartree-Fock theory with Skyrme's interaction},
Nucl. Phys. A \textbf{249}, 215 (1975).

\bibitem{Dobaczewski95} J. Dobaczewski and J. Dudek,
Time-odd components in the mean field of rotating superdeformed nuclei,
Phys. Rev. C \textbf{52}, 1827 (1995);
Erratum \textit{ibid} \textbf{55}, 3177 (1997).

\bibitem{Raimondi11b} F. Raimondi, B. G. Carlsson, J. Dobaczewski, and J. Toivanen,
{Continuity equations and local gauge invariance for the N3LO nuclear 
energy density functionals},
Phys. Rev. C \textbf{84}, 064303 (2011).

\bibitem{Bender09b} M. Bender, T. Duguet, and D. Lacroix,
Particle-number restoration within the energy density functional formalism
Phys. Rev. C \textbf{79}, 044319 (2009).

\bibitem{Stringari78a} S. Stringari and D. M. Brink,
Constraints on effective interactions imposed by antisymmetry and charge independence,
Nucl. Phys. A \textbf{304}, 307 (1978).

\bibitem{Terasaki95} J. Terasaki, P.-H. Heenen, P. Bonche, J. Dobaczewski, and H. Flocard,
{Superdeformed rotational bands with density dependent pairing interactions},
Nucl. Phys. A \textbf{593}, 1 (1995).

\bibitem{Rigollet} C. Rigollet, P. Bonche, H. Flocard, and P.-H. Heenen, 
{Microscopic study of the properties of identical bands in the $A = 150$ mass region}, 
Phys. Rev. C 59, 3120 (1999).

\bibitem{Gall1994} B. Gall, P. Bonche, J. Dobaczewski, H. Flocard, and P.-H. Heenen,
{Superdeformed rotational bands in the mercury region.
A cranked Skyrme-Hartree-Fock-Bogoliubov study},
Z. Phys. A \textbf{348}, 183 (1994).

\bibitem{Bay86a} D. Baye and P.-H. Heenen,
{Generalised meshes for quantum mechanical problems}, 
J. Phys. A \textbf{19}, 2041 (1986).

\bibitem{Baye15} D. Baye,
{The Lagrange-mesh method}, 
Phys. Rep. \textbf{565}, 1 (2015).

\bibitem{Ryssens15b} W. Ryssens, P.-H. Heenen, and M. Bender, 
{Numerical accuracy of mean-field calculations in coordinate space},
Phys. Rev. C \textbf{92}, 064318 (2015).

\bibitem{Jodon16} R. Jodon, M. Bender, K. Bennaceur, and J. Meyer, 
{Constraining the surface properties of effective Skyrme interactions},
Phys. Rev. C \textbf{94}, 024335 (2016).

\bibitem{Ryssens19a} W. Ryssens, M. Bender, K. Bennaceur, P.-H. Heenen, and J. Meyer, 
{The impact of the surface energy coefficient on the deformation properties
of atomic nuclei as predicted by Skyrme energy density functionals}, 
Phys. Rev. C \textbf{99}, 044315 (2019).

\bibitem{Schunck10} N. Schunck,  J. Dobaczewski, J. McDonnell, J. Mor{\'e}, W. Nazarewicz, J. Sarich,
and M. V. Stoitsov,
{One-quasiparticle states in the nuclear energy density functional theory},
Phys. Rev. C \textbf{81}, 24316 (2010).

\bibitem{Fracasso12} S. Fracasso, E. B. Suckling, and P. D. Stevenson, 
{Unrestricted Skyrme-tensor time-dependent Hartree-Fock model and its 
       application to the nuclear response from spherical to triaxial nuclei},
Phys. Rev. C \textbf{86}, 044303 (2012).

\bibitem{Hellemans13} V. Hellemans, A. Pastore, T. Duguet, K. Bennaceur, D. Davesne, J. Meyer, M. Bender, and P.-H. Heenen, 
{Spurious finite-size instabilities in nuclear energy density functionals},
Phys. Rev. C \textbf{88}, 064323 (2013).

\bibitem{Becker06} F. Becker \etal,
{Coulomb excitation of $^{78}$Kr},
Nucl. Phys. A \textbf{770}, 107 (2006).

\bibitem{Clement07} E. Cl\'{e}ment \etal, 
{Shape coexistence in neutron-deficient krypton isotopes}, 
 Phys. Rev. C  \textbf{75}, 054313 (2007).

\bibitem{Wimmer20} K. Wimmer \etal, 
{Shape coexistence revealed in the $N=Z$ isotope $^{72}$Kr through inelastic scattering},
Eur. Phys. J. A \textbf{56}, 159 (2020).

\bibitem{Doring95} J. D{\"o}ring \etal,
{High-spin bands in $^{80}$Kr},
Phys. Rev. C \textbf{52}, 76 (1995).

\bibitem{Bender06b} M. Bender, P. Bonche, and P.-H. Heenen,
{Shape coexistence in neutron-deficient Kr isotopes: Constraints on the 
single-particle spectrum of self-consistent mean-field models from collective excitations},
Phys. Rev. C \textbf{74}, 024312 (2006).

\bibitem{Bonche85} P. Bonche, H. Flocard, P.-H. Heenen, S. J. Krieger, and M. S. Weiss,
{Self-consistent study of triaxial deformations: Application to the isotopes of Kr, Sr, Zr and Mo},
Nucl. Phys. A \textbf{443}, 39 (1985).

\bibitem{Girod09} M. Girod, J. P. Delaroche, A. G{\"o}rgen, and A. Obertelli, 
{The role of triaxiality for the coexistence and evolution of shapes in light krypton isotopes},
Phys. Lett. B \textbf{676}, 39 (2009).

\bibitem{Rodriguez14} T. R. Rodr{\'i}guez, 
{Structure of krypton isotopes calculated with symmetry-conserving configuration-mixing methods},
Phys. Rev. C \textbf{90}, 034306 (2014).

\bibitem{Angeli} I. Angeli and K. P. Marinova, 
{Table of experimental nuclear ground state charge radii: An update},
At. Data Nucl. Data Tables \textbf{99}, 69 (2013).

\bibitem{Raman} S. Raman, C. W. Nestor, and P. Tikkanen,
{Transition probability from the ground to the first-excited $2^+$ state of 
         even-even nuclides}, 
At. Data Nucl. Data Tables \textbf{78}, 1 (2001).

\bibitem{Bender06} M. Bender, G. F. Bertsch, and P.-H. Heenen,
{Global study of quadrupole correlation effects}, 
Phys. Rev. C \textbf{73}, 034322 (2006).

\bibitem{NuDat} National Nuclear Data Center, 
information extracted from the NuDat 2 database, 
https:\/\/www.nndc.bnl.gov\/nudat2\/

\bibitem{Ling20} C. Ling, C. Zhou, and Y. Shi,
{Fission barriers of actinide nuclei with nuclear density functional theory: influence of the triaxial deformation},
Eur. Phys. J. A \textbf{56}, 180 (2020).

\bibitem{Lu2014} B. N. Lu, J. Zhao, E. G. Zhao, and S. G. Zhou, 
{Multidimensionally-constrained relativistic mean-field models and 
potential-energy surfaces of actinide nuclei}, 
Phys. Rev. C \textbf{89}, 014323 (2014).

\bibitem{Singh02} B. Singh, R. Zywina, and R. B. Firestone, 
{Table of superdeformed nuclear bands and fission isomers},  
Nucl. Data Sheets \textbf{97}, 241 (2002).

\bibitem{Robledo07a} L. M. Robledo, 
Particle number restoration: Its implementation and impact in nuclear structure calculations,
Int. J. Mod. Phys. E \textbf{16}, 337 (2007).

\bibitem{Duguet09a} T. Duguet, M. Bender, K. Bennaceur, D. Lacroix, and T. Lesinski,
Particle-number restoration within the energy density functional formalism: 
Non-vialibility of terms depending on non-integer powers of the density matrices,
Phys. Rev. C \textbf{79}, 044320 (2009).

\bibitem{Robledo10a} L. M. Robledo, 
Remarks on the use of projected densities in the density-dependent part of Skyrme or Gogny functionals,
J. Phys. G \textbf{37}, 064020 (2010).

\bibitem{Doba00b} J. Dobaczewski, J. Dudek, S. G. Rohozi{\'n}ski, and T. R. Werner,
{Point symmetries in the Hartree-Fock approach. II. Symmetry-breaking schemes},
Phys. Rev. C \textbf{62}, 014311 (2000).

\bibitem{Bonche87a} P. Bonche, H. Flocard, and P.-H. Heenen, 
{Self-consistent calculation of nuclear rotations: The complete yrast line of $^{24}$Mg},
Nucl. Phys. A \textbf{467}, 115 (1987).

\bibitem{Bally21} B. Bally and M. Bender,
Projection on particle number and angular momentum: Example of triaxial Bogoliubov quasiparticle states,
Phys. Rev. C \textbf{103} 024315 (2021).

\bibitem{Wong76a} C. Y. Wong,
{On the Thomas-Fermi approximation of the kinetic energy density},
Phys. Lett. B \textbf{63}, 395 (1976).

\bibitem{Graf80a} H. Gr{\"a}f,
{Thomas-Fermi kinetic-energy density with gradient corrections},
Nucl. Phys. A \textbf{343}, 91 (1980).

\bibitem{Lombard91} R. J. Lombard, D.  Mas, and S. A. Moszkowski, 
{On the kinetic energy density}, 
J. Phys. G \textbf{17}, 455 (1991).

\bibitem{Lombard94} R. J. Lombard and S. A. Moszkowski, 
{On the positivity of the kinetic-energy density}, 
Il Nuovo Cim.  \textbf{109}, 1291 (1994).

\bibitem{Prakash81a} M. Prakash, S. Shlomo, and V. M. Kolomietz,
{Shell and surface effects in the static Wigner phase-space distribution function of nuclei},
Nucl. Phys. A \textbf{370}, 30 (1981).

\end{thebibliography}
\end{document}